%% file: Huxor_Grebel_Final.tex
\documentclass[useAMS,usenatbib]{mn2e}
\usepackage{epsfig,fp}
\usepackage{amsmath}

\newcommand{\mwRadiusKpc}{18}
\newcommand{\mwHalfHeight}{5}

\title[Tracing MW features with carbon-rich LPVs]{Tracing the tidal streams of the Sagittarius dSph, and halo Milky Way features, with carbon-rich long-period variables}
\author[A. P. Huxor and E. K. Grebel]{A. P. Huxor $^{1}$\thanks{E-mail:
avon@ari.uni-heidelberg.de} and E. K. Grebel$^{1}$\\
$^{1}$Astronomisches Rechen-Institut, Zentrum f\"{u}r Astronomie der Universit\"{a}t Heidelberg, M\"{o}nchhofstra{\ss}e 12 - 14, \\
69120 Heidelberg, Germany.\\
}

\begin{document}

\date{Accepted ... Received ...  in original form ...}

\pagerange{\pageref{firstpage}--\pageref{lastpage}} \pubyear{2002}

\maketitle

\label{firstpage}

\begin{abstract} 

We assemble 121 spectroscopically-confirmed halo carbon stars, drawn from the literature, exhibiting measurable variability in the Catalina Surveys. We present their periods and amplitudes, which  are used to estimate distances from period-luminosity relationships.  The location of the carbon stars -- and their velocities when available --  allow us to trace the streams of the Sagittarius (Sgr) dwarf  spheroidal  galaxy. These are compared to a canonical numerical simulation of the accretion of Sgr. We find that the data match this model well for heliocentric distances of 15--50 kpc, except for a virtual lack of carbon stars in the trailing arm just north of the Galactic Plane, and there is only tentative evidence of the leading arm south of the Plane. The majority of the sample can be attributed to the Sgr accretion. We also find  groups of carbon stars which are not part of Sgr; most of which are associated with known halo substructures.  A few have no obvious attribution and may indicate new substructure.  We find evidence that there may be a structure behind the Sgr leading stream apocentre, at $\sim$100 kpc, and a more distant extension to the Pisces Overdensity also at $\sim$100 kpc. We study a further 75 carbon stars for which no good period data could be obtained, and for which NIR magnitudes and  colours are used to estimate distances. These data add support for the features found at distances beyond 100 kpc.

\end{abstract}


\section{Introduction}
\label{introduction}

In the standard $\Lambda$CDM cosmology \citep{Bullocketal01,WhiteFrenk91} large galaxies form through the merger and accretion of smaller galaxies. We are fortunate to be witness to  this process with the ongoing accretion of the Sagittarius (Sgr) dwarf galaxy  \citep{Ibataetal94, Majewskietal03} onto  the  Milky Way (MW). Its proximity provides a unique opportunity to study  accretion in detail, through the tidally stripped streams that straddle the sky.

There have been considerable efforts to create simulation models  of the Sgr accretion, including \citet{Penarrubiaetal10,Lokasetal10,Purcelletal11,Ibataetal13,VeraCiroHelmi13}. However, the model most used by the community is that of  \citet{LawMajewski10a} [hereafter LM10], designed to fit much of the observational data known to that date.

A recurring theme in the study of the tidal tails of Sgr is their use as a probe of the mass and shape of the MW dark matter halo, for knowledge of the orbit of the Sgr dSph is invaluable in constraining such models \cite[e.g.][]{Ibataetal01, MartinezDelgadoetal04,Johnstonetal05,Carlinetal12a,DegWidrow13,Ibataetal13,VeraCiroHelmi13}. But no consistent and agreed solution has been achieved. Further observational data on the actual orbit, especially at larger Galactocentric radii, will provide additional constraints on models of its accretion, and hence the dark matter profile of the MW. 

Many stellar tracers have been used to study  the Sgr Streams. An early study used carbon stars \citep{Ibataetal01} but significant progress came with \citet{Majewskietal03} who found that M-giants revealed details of the leading and trailing tidal streams (arms) -- data that informed the LM10 model. Since its publication, the LM10 model has been tested against observations of stars -- their positions, proper motions and velocities. It captures many of the main features of the streams, and has become a benchmark for such studies. Indeed, many regions of the streams are now relatively well characterised up to a distance of $\sim$50 kpc by a variety of other tracers including horizontal branch stars \citep{Ruhlandetal11,Shietal12}, RR Lyrae  \citep{Vivasetal06, Drakeetal13a}, red clump stars \citep{Correntietal10, Carrelletal12}, upper main-sequence and main-sequence turn-off stars \citep{Belokurovetal06,Koposovetal12,Slateretal13}.

Despite its success in reproducing many features of the observed data, the LM10 model has some limitations. For example, it does not explain the ``bifurcation" seen in the actual stream \citep{Belokurovetal06,Koposovetal12,Slateretal13}, prompting suggestions \citep{Koposovetal12,Newbyetal13} that there may actually be two progenitors, with separate streams emanating from each.  Other deviations from the LM10 model have also been found: Using RR Lyrae, \citet{Sesaretal11b} find no evidence of the leading arm at low right ascension -- although expected from the model. Similarly, \cite{Drakeetal13a}, see no suggestion of the Sgr trailing arm in RR Lyrae in the sky just north of the Galactic Plane.

Beyond about 50 kpc, the data are more ambiguous.  For example, \citet{Newbergetal03,Drakeetal13b,Belokurovetal14} each describe a halo feature, located at a distance of $\sim$90 kpc (the ``Gemini Arm"), which has been recently attributed to the trailing arm of the Sagittarius stream. If the Gemini Arm were part of the trailing arm of the Sgr stream, it will have a great impact on the possible orbit of the progenitor. These data are far from the model, as seen on the sky and in distance, indicating that the LM10 model  requires serious revision.
 
Clearly there is much work still to be done to understand both the nature of the progenitor of Sgr, and its fateful journey around the Galaxy during accretion.

 \subsection{Carbon stars}

In this paper we return to the study of carbon stars as a tracer population, the first  to reveal the extent of the streams \citep{Ibataetal01}. Specifically, we focus of those carbon stars that exhibit clear variability. There are a number of reasons for doing so. Firstly, there are newly published lists of halo carbon stars, enlarging the sample. Secondly, the ability of the period of any variability to give a good estimate of distance is now  better understood thanks to studies of the Magellanic Clouds. And, finally there is now an excellent set of `almost" all-sky, variable star, data available in the Catalina Surveys \citep{Drakeetal13a}. 
Carbon stars possess properties that make them good probes. They have photospheric C/O ratios that exceed one, and so show distinct features of C$_{2}$ and CN in their spectra, making their identification simple even in relatively low-resolution spectra.
 The  strong  band heads also aid velocity determination \citep{Aaronson83}, making them good kinematic probes. For example, they have been used to study the kinematics of the outer regions of the Galactic disk \citep{DemersBattinelli07},  the Galactic halo \citep{HartwickCowley85,Mouldetal85, Bothunetal91,Greenetal94}, the LMC \citep{Graffetal00,AlvesNelson00, OlsenMassey07}, and the SMC \citep{Kunkeletal00}.

The population of carbon stars is composed of a complex "zoo'' of many types and with a variety of origins \citep{Abiaetal03}, and whose original classification was determined by the presence of certain spectroscopic features. N-type stars have substantial absorption at the blue end of the spectrum, while  R-types stars still show the blue end of the spectrum, and CH stars have strong CH absorption. The J stars exhibit very strong isotopic bands of C$_{2}$ and CN.

There are two types of explanation as to how carbon finds itself on the surface of the star: intrinsic and extrinsic. The intrinsic carbon stars are  formed by internal processes,  in which the carbon  products of helium burning mix into the convective envelope and are brought to the surface. One major process is the third dredge-up (TDU), which occurs on the asymptotic giant branch (AGB), creating the classic N-type carbon stars, which typically pulsate and are losing mass.  Recent work  \citep{Zamoraetal09,Knappetal01} suggests that the cooler, late-type R carbon stars are the same as the N-type, although possibly sitting nearer to the bottom of the AGB phase. The warmer, early-type R stars are, by contrast, near the red clump, and may be formed through an anomalous He-flash. 

Extrinsic carbon stars are formed by the accretion of carbon-rich material from a donor star within a binary system. These are normally identified (in stars at high Galactic latitudes) by high radial velocities --   CH stars, for example --   \citep{McClure84}. Similar are the nearby so-called "carbon dwarfs'' (dCs). The first was discovered by \citet{Dahnetal77} and was subsequently found to be a binary system  \citep{Dearbornetal86}, in which mass transfer has taken place on to a dwarf star. These dCs can, however, be identified by their high proper motion.

The AGB carbon stars are the main focus of this paper. They are luminous and so can be detected to large distances \citep{Grebel07}, and our primary interest is in the outer halo of the MW.  Due to a minimum stellar mass required for the TDU to occur \citep[e.g.][]{IbenRenzini83,Stranieroetal03}, such carbon stars are also seen as representatives of young to intermediate-age populations -- and hence of the accretion of younger stars than is typical of the halo. It should be noted, however, there is still some uncertainty about the minimum mass with new models allowing carbon stars at lower mass, at low metallicity \citep{Karakas10}. 

Unsurprisingly, given the variety of carbon star types, early work using carbon stars to probe the MW halo constantly encountered the problem that they possess a wide variety of intrinsic luminosities in a range of passbands \citep{Alksnisetal98, Abia08}.  Using AGB (N and late-R type) stars alone might be expected to reduce the problem.  Thus, \citet{TottenIrwin98} assumed a single value of $M_{\rm R} = -3.5$ mag. But even N-types have a wide range of absolute visual magnitudes as determined from Hipparcos distances  \citep{Alksnisetal98}. \citet{Mauronetal04} use a standard  $M_{K_{\rm s}}$ calibrated against LMC carbon stars, but with the additional refinement of including ($J-K_{\rm s}$) colour. This approach gives better results, as NIR data take the mass-loss that is characteristic of N-type carbon stars into account.

The uncertainties of distance determination using carbon stars can be considerably reduced if the carbon star is a long-period variable (LPV) -- as most AGB stars are -- as we can exploit  period-luminosity relationships (e.g. \citealt{Itaetal04b,Whitelocketal08, Soszynskietal09a, Riebeletal10}). 
But carbon stars are rare -- and those with clear and measurable variability more so. The challenge is to garner a statistically significant sample of such variable carbon stars. This paper tries to do so by drawing together all carbon stars in the halo that are to be found in the literature. 

Being giants, and  often suffering significant mass-loss, the stars of interest are typically very red. For this reason, the largest sample we use is drawn from the large list of 2MASS-selected carbon stars by Mauron and his collaborators, having colours most consistent with being AGB carbon stars. We also draw on other samples, but many of these have used  spectra at the blue end to identify carbon stars, and so preferentially select the bluer early-type R stars, CH stars and dwarf carbon stars. Using long-period variables also allows us to be more confident that we are dealing with AGB stars, and we adopt a minimum period of $\sim$80 days for our selection.

With AGB carbon stars, we should be able to trace the Sgr dwarf tidal features to large distances. Carbon stars that do not lie on the streams may indicate the presence of previously undetected wraps of the Sagittarius stream, helping constrain its orbit. Alternatively they could point to new, or confirm previously proposed, substructure.

This paper proceeds as follows: the data we use are described in \S \ref{data}.  The methods used to derive distances are given in \S \ref{methods}. The results are presented in \S \ref{results}. In \S \ref{noPeriods}, we study those carbon stars for which we do not have period data. The results are discussed in \S \ref{discussion}, and we summarise in \S \ref{summary}.

\section{Data}
\label{data}

Our sample of carbon stars are  drawn from a variety of sources in the literature  and are all spectroscopically-confirmed. 
They have very different means of detection, selection criteria and spatial coverage. The Galactic disk, being young, contains many carbon stars, but in this paper we are concerned with potential halo structures, hence we focus on those at high Galactic latitudes. 

The final selection of carbon stars is divided into two catalogues. There are those stars for which we have measurable periods from the Catalina Surveys (hereafter called the ``first sample"). Note that these stars are listed in Table \ref{tab:first_sample_catalog}, and are identified throughout this paper by the prefix ``HG" followed by a running number.  A ``second sample" contains those carbon stars for which we do not have good variability data. These stars are listed in Table \ref{tab:second_sample_catalog} and have the prefix ``HG2-".

For our study, we are interested in the AGB stars, the N-type and late-R types. Our samples below rely to a large extent on the classifications given by the authors of the various papers. In some cases (such as \citet{TottenIrwin98}) each star is specifically classified from its spectrum. In others (such as the many Mauron group papers), a general comment is made that properties of the sample are consistent with them all being AGB stars. And in a few cases (in particular those carbon stars found during searches for other types of stars), there is no indication of which sub-type they belong to. However, we use the colours of the stars to mostly assist in attributing them to AGB stars. In the few cases where the stars lie outside this colour selection, we use available spectra or other information to decide whether to include them in our sample.

For our purposes we require that the stars sit on the luminosity-period relationships. 
Note that we accept ``candidate" LPV carbon stars, and so may have some CH stars in our sample, for example, which may mimic low-amplitude AGB stars. However, as we are primarily concerned with tracing distant halo structures, we are willing to accept a few contaminants. One aim of this study is to locate good candidate carbon stars that may trace previously unknown halo substructure. These stars can be used as targets for deep imaging to reveal any features in the fainter, but more numerous, tracer populations. Given that such features may be found over the whole sky, having such targets is a very cost-effective means of seeking new substructure.

\subsection{Mauron  et al. sample}

Our primary source, providing just over half of all our variable carbon stars, comes from the lists  published by the Mauron  group
\citep{Mauronetal04, Mauronetal05, Mauronetal07b, Mauron08,Mauronetal14}. These were selected from candidates based on 2MASS \citep{Skrutskieetal06} photometry and then spectroscopically confirmed. Those from \citet{Mauronetal04} and \citet{Mauronetal05} have radial velocities determined, while the later papers do not give velocities.
For \citet{Mauronetal04} they had access to the second incremental data release of 2MASS, only about half the sky; their subsequent work, however, exploited the all-sky 2MASS data. 

To identify carbon AGB star candidates they employed the 2MASS colours of known carbon stars, which lie in a known linear feature in the ($H-K_{\rm s})/(J-H)$ colour plot, selecting stars $\pm$0.15 mag from the median line. Initially they limit this sample  with the cuts on colour  ($H-K_{\rm s}$) $> 0.4$ and  ($J-H$) $> 0.95$, and a faint magnitude limit ($K_{\rm s} < 13.0$). They exclude a region either side of the Galactic Plane, only going down to $|$b$|$ $>$ 25$^\circ$. For later papers in the series they enlarged the selection criteria to $K_{\rm s} <  13.5$,  $|$b$|$ $>$ 20$^\circ$ and ($J-K_{\rm s}$) $> 1.3$ mag. One star in the Mauron sample (\#80 in their numbering scheme) is actually an S-type, and is removed from our list. We also exclude their M56, M57 and M58, which they note are known ``warm carbon stars" (all with a ($J-K_{\rm s}$) $ <  1$), and not AGB stars, and which they observed to provide comparison spectra across a range of carbon star types.

The published errors on heliocentric velocities, where given, are  $\pm$10--12 kms$^{-1}$ \citep{Mauronetal04}. Certain stars have entries that require comment:  \citet{Mauronetal05} report a typographic error  for star  HG2-51 (M15 in the Mauron numbering system, which uses a preceeding 'M') in their earlier paper \citep{Mauronetal04}. Its heliocentric radial velocity is --158 and not +158 kms$^{-1}$. HG75 (M43) has a greater heliocentric velocity (+54 kms$^{-1}$) in \citet{Mauronetal05}  compared to the more recent observation (+15 kms$^{-1}$) of  \citet{Deasonetal12}. Such differences are not totally unexpected, however. Significant differences  30 kms$^{-1}$ have been found between the velocities obtained by absorption and emission lines  and between optical and radio lines \citep{Menziesetal06}. The reasons are not clear, but they recommend using absorption lines rather than Balmer emission lines for velocity determination.
In our table we use the Mauron value of HG75 for consistency with the rest of the table sample, as the velocity difference does not significantly affect our conclusions. For HG89 and HG92 we obtain radial velocities from \citet{Lagadecetal10}.

\citet{Mauronetal04} believe the majority of the sample in that paper to be AGB stars on account of: their colours, the presence in half of H$\alpha$ emission, and the fact that two-thirds have spectra (in which the wavelength range permits) that  rises to the red side. In their later papers they only note that the spectra are typical of N-type stars.

We also include the carbon stars from the First Byurakan Survey (FBS) for which the Mauron group undertook additional spectroscopy \citep{Mauronetal07b}. For a few stars, the new spectra clearly show them to be N-type.

 \subsection{Other Sources}
 
 The Mauron sample provides the largest number of carbon stars, but a review of the literature reveals the following additional samples. Many of these are also duplicated between themselves, or with the large Mauron sample. In these cases, the reference for the star given in Table \ref{tab:first_sample_catalog} either indicates the first report of the star, or the reference that provided us with the most data.
  
\subsubsection{Additional 2MASS-selected stars}
 
A number of carbon stars have also been found serendipitously during searches for late-type dwarf stars -- based on their 2MASS colours --  by \citet{Cruzetal03,Cruzetal07,Kirkpatricketal08,Reidetal08}, 
and for whom  carbon stars are contaminants. Fortunately, they published the status of these.   As contaminants, the authors only state that they are carbon stars, and give no indication of the sub-class, so they may include CH, dC and early-R type carbon stars in addition to the late-R and N-types.

 \citet{Cruzetal03} use the following selections: 
 
 \begin{equation}
\begin{split}
|b|>10^{\circ}  \\
 \mbox{} (J-K_{\rm s}) >1 \\
 \mbox{} (J-H) > 1 - (J-K_{\rm s}) \\
 \mbox{and } J\leq1.5 \times (J-K_{\rm s})+10.5 
 \end{split}
 \label{eqn:cruz0}
\end{equation}   

With additional cuts in the NIR  colour-colour diagram of:
 
\begin{equation}
\begin{split}
 (J-H) \leq 0.8  \\
 \mbox{ for } 0.30 \leq (H-K_{S}) \leq 0.35 
 \end{split}
 \label{eqn:cruz1}
\end{equation}

  and

\begin{equation}
\begin{split}
(J-H) \leq 1.75 \times (H-K_{S})+01875 \\
 \mbox{ and }(J-H) \geq 1.75 \times (H-K_{S})-0.4750 \\
\mbox{ for } 0.35 \leq (H-K_{S}) \leq 1.20
\end{split}
 \label{eqn:cruz2}
\end{equation}

They obtained spectroscopy from facilities in both the northern (Kitt Peak Observatory) and southern (Cerro Tololo Inter-American Observatory) hemispheres. Hence, their sample covers much of the sky except for the Galactic Plane.
\citet{Cruzetal07} and \citet{Reidetal08} essentially use the same selections as \citet{Cruzetal03}. 

\citet{Kirkpatricketal08} selected candidates with the following constraints:

 \begin{equation}
\begin{split}
|b| > 25^{\circ}  \\
 \mbox{} --50^{\circ}< Dec <+5^{\circ}1 \\
 \mbox{} 8 < K_{\rm s} < 13.6 \\
 \mbox{} (J-K_{\rm s}) >  1.3 \\ 
 \mbox{and } (R-K_{\rm s}) > 6.5
 \end{split}
 \label{eqn:kirkpatrick1}
\end{equation}   

Where $R$ magnitudes are obtained  from USNO-A\footnote{http://tdc-www.harvard.edu/catalogs/ua2.html}).  That is, they focus on the southern sky. Later in their programme, they reduced the  magnitude and colour cuts to: $9.5 < K_{\rm s} < 13.6$ and $(R-K_{\rm s}) > 5.5$ mags respectively.

 A similar, but much smaller, 2MASS-based sample of carbon stars was published by \citet{Liebertetal00}, possessing ``very red" NIR colours, and high Galactic latitudes (they too were interested in objects in the halo, not the disk). These stars are heavily dust enshrouded. Of these, there is only a velocity for HG44 (IRAS 08546+1732) from \citet{Groenewegenetal97}.  They present spectra, and note that all are consistent with being N-type.
 
\subsubsection{Totten and Irwin}

\citet{TottenIrwin98} list a sample of carbon stars that were serendipitously discovered in an APM (Automatic Plate Measuring) survey for high redshift quasars (which also appear as very red, faint stellar-like objects). 
The APM survey covered the whole of the northern sky away from the Galactic Plane($|b|>$30$^{\circ}$), and much of the southern sky (see Figure 1 of \citealt{Ibataetal01}). Candidates were selected on optical colour $(B_{J}-R  < 2.5)$ and magnitude $(11 < R < 17)$ and the equivalent POSS1 O and E  colours and magnitudes.
They undertook follow-up spectroscopy and give classifications for their sample, based on these spectra. We exclude those stars which they classify as CH-type, but include HG21 (their 0351-1127) which they classify as borderline CH/N-type.

The radial velocities also come from  \citet{TottenIrwin98}, except  for  HG60, where  a more recent value is given in \citet{Lagadecetal10}.

\subsubsection{SDSS}

A number of authors have searched through the SDSS database and generated catalogues of carbon stars, identified from the SDSS's own spectroscopy \citep{Margonetal02, Downesetal04,Green13}. Most of these carbon stars are rather blue in colour, due to the SDSS passbands, but a handful of AGB stars have been picked up. The SDSS catalogues also provide velocities, but -- as \citet{Green13} notes -- these are derived from fitting to templates that do not include carbon stars. Hence we do not use these values.
For example, in one case (HG45)  \citet{Deasonetal12} report  a heliocentric velocity of 200 $\pm$10 kms$^{-1}$, which is very different from the SDSS catalogue value of --313 kms$^{-1}$. We do, however,  adopt the  \citet{Deasonetal12} value.

 We have spectra for all these, and include those that clearly have N-type spectra, and one carbon star that is a candidate late-R type (see Section \ref{R_star}). Many of these are among the most distant carbon stars and so we discuss these later in Section \ref{distant_stars}.

\subsubsection{Others}

The few stars that do not appear in these surveys (and mostly listed in \citet{Alksnisetal01}), include: HG2-35 (CGCS 6480) from \citet{Moodyetal97}, found serendipitously during a search for emission line galaxies;  and HG90 (CGCS 3801) from \citet{MeusingerBrunzendorf01} found during a search for AGN in the fields around two Galactic globular clusters.  \citet{LeeChen09}  provide one, HG28.  
\citet{Christliebetal01} created a catalog of 403 faint high latitude carbon stars from a region of 6,400 deg$^{2}$, from the Hamburg/ESO  (HE) objective prism survey. This covers Dec$< +2.5^{\circ}$ and $|b|>30$$^{\circ}$.
The vast majority of these are not variable, and of those few that are, most are found in other lists described above. One exception is HG121 (HE 2319-1534).

\subsubsection{Completeness of sample}  
\label{completeness}
 
It is difficult to be certain about completeness for a sample that is assembled from a set of heterogeneous lists. But one limitation we have arises from our use of Catalina Surveys data (see \S \ref{variability_data}) which excludes the regions around the Galactic Plane, Magellanic Clouds, and the Equatorial Poles.

\citet{Mauronetal04} discuss their sample selection in some detail, and we can extrapolate their findings to estimate how well we recover the AGB carbon star population within the Catalina footprint. Using the colour cuts that we described above, they have an initial list of $\sim$1200 objects in the 2MASS second incremental data release, which covers 19,600 square degrees of sky. However many of these are, for example, distant galaxies. After cleaning this initial list, they finally have $\sim$200 ``good" candidates. The Catalina Surveys cover 33,000 square degrees, so we would expect $\sim$340 comparable carbon star candidates in this region. \citet{Mauronetal04} also report that some 30\% of the 97 stars for which they actually obtained spectra were carbon-rich stars, the remainder being oxygen-rich. Thus naively one might expect that just over 100 stars will be found in the Catalina Surveys. We actually find 121 in our LPV sample, with a few dozen other stars that although carbon stars, do not have good light curves. Thus we will have found all carbon stars that match the \citet{Mauronetal04} ``good" candidates. Of course, there may be many AGB carbon stars that did not match their ``good" criteria, which are not given, and where the criteria may have been too stringent.

\begin{figure} 
  \centering
 \includegraphics[angle=0,width=85mm]{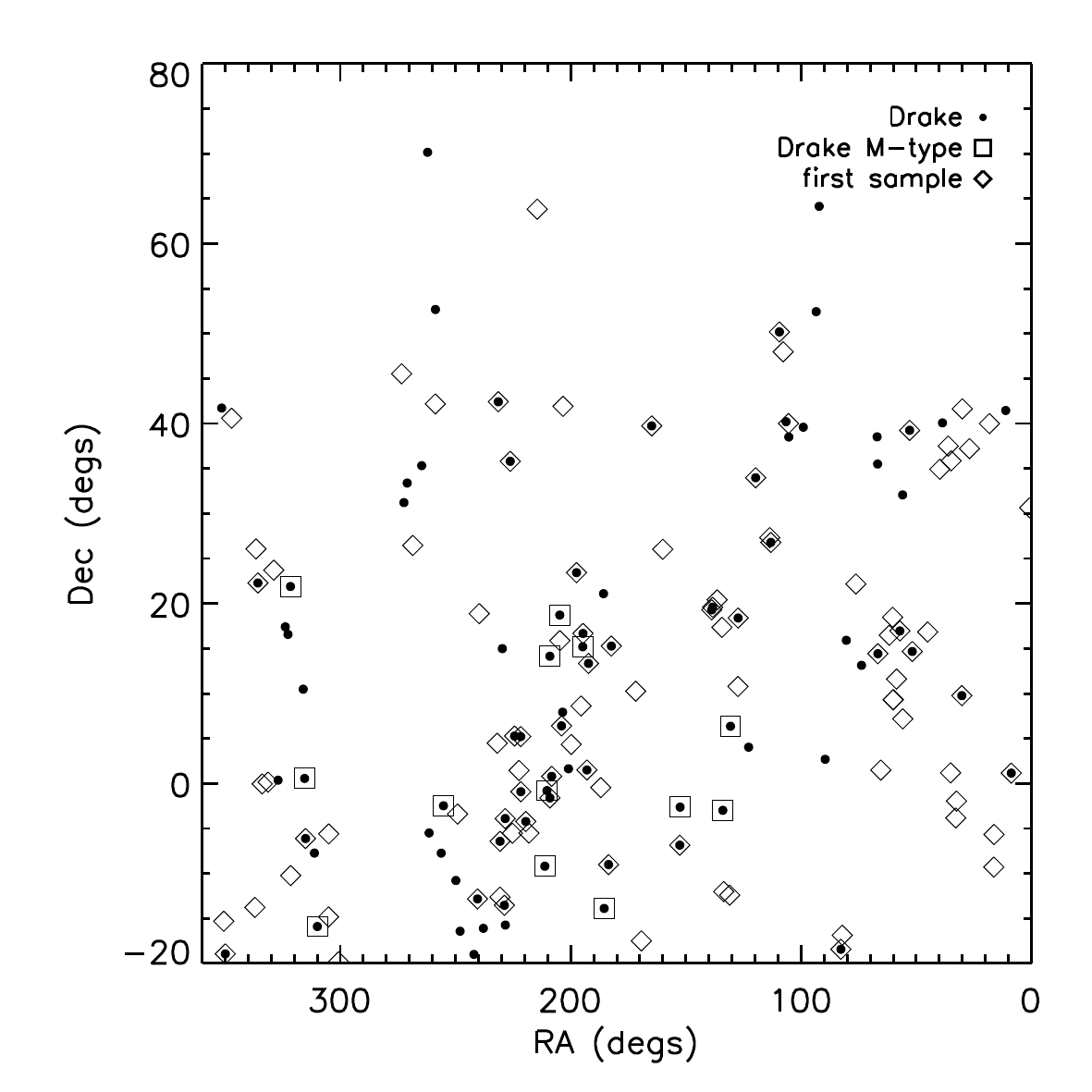}
 \caption{Comparison of \citet{Drakeetal14} sub-sample of LPV stars with our  comparable sub-sample drawn from the ``first sample". Stars known to be oxygen-rich (M stars) are also marked. }\label{Fi:compare_with_drake14_sample}
\end{figure}

Another measure of the coverage of our sample is to compare it to the sample of the LPVs from \citet{Drakeetal14}. They use Data Release 1 of the Catalina Surveys, which only reached down to a declination of --20$^{\circ}$, and so does not cover the region south of this (nor the region of the northern equatorial pole). We create a test sub-sample from our LPV (first) sample by applying these cuts for sky location.
We also apply  appropriate cuts to the Drake et al. sample with $(H-K_{\rm s}) > 0.4$, and $|$b$| >$20, to make it comparable to ours. This leaves us with two sub-samples (Fig. \ref{Fi:compare_with_drake14_sample}).
Of the 87 Drake et al.  stars (filled circles), which include both carbon-rich and oxygen-rich stars, our sample of carbon stars (diamonds) matches to 37, some 42\%. 
We can compare this match to \citet{Mauronetal04} where some 30\% were shown to be carbon stars. So we believe that we are recovering most of the carbon stars in the Drake et al. sub-sample, the remainder being oxygen-rich. (Those Drake et al. stars that are known to be M stars (oxygen-rich), from the SIMBAD or Vizier databases, are marked out with open squares in Fig. \ref{Fi:compare_with_drake14_sample}.) 

\begin{figure*} 
  \centering
 \includegraphics[angle=0,width=180mm]{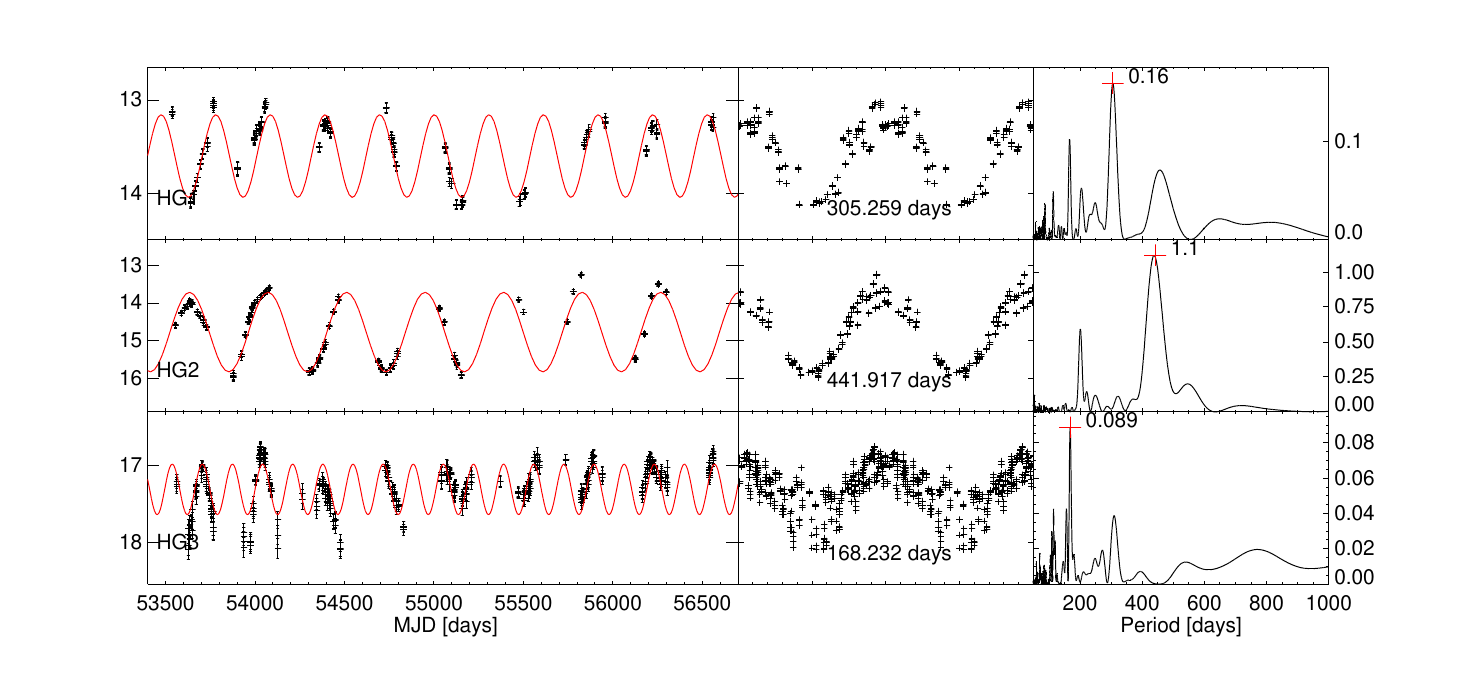}
 \caption{Light curves from  the  Catalina Sky Survey for the carbon stars HG1, HG2 and HG3. Left: The x-axis represents the Modified Julian Date (MJD) from 53400 to 56700 days. The y-axis represents magnitude in the Catalina system, where each tick indicates an integer value, allowing an estimate of the amplitude of the period. Centre: A folded light curve over two periods, with the period labelled. Right: A plot of the power spectrum (scaled to near to the maximum power), from 50 to 1000 days. The chosen period is indicated by a red cross, and the value of the power is labelled.}\label{Fi:LC_phase_power_plotting_HG1_HG3}
\end{figure*}

However, what is notable is that we have many LPVs in our sample that are not found in the Drake et al. sample. We believe a number of reasons may explain our greater success rate. Firstly, Drake et al. fail to detect many low amplitude stars. They compared their sample to the LINEAR  (which we describe below in \S\ref{variability_data}) variability survey and found that they missed those stars with an amplitude $<$ 0.1 mag.  Secondly, they appear to have only used the folded light curve when visually selecting their variable stars. Many of our stars have odd folded light-curves, but are more convincing in the full light-curve (e.g HG24, HG92 and HG104). Thirdly, the Drake et al. DR1 data covered a shorter duration (to June 2011: about MJD 55,700 compared to DR2 which extends to $\sim$MJD 56,500). If a light curve is sparse, our longer period will reveal periodicity that may be less obvious in DR1 (e.g. HG12,HG14, HG91, HG109 and HG119). These light curves are shown in the Online Material accompanying this paper. However the light-curves and period fits for three stars (HG1, HG2 and HG3) are shown in Fig. \ref{Fi:LC_phase_power_plotting_HG1_HG3}, to show typical examples.

Nonetheless, there are still many stars in our sample and that have good light curves in Catalina which did not make it into the Drake et al. sample. The lead author of that paper (AJD) visually inspected a remarkable $\sim$112,000 stars, so it is not unexpected that some true LPV stars might slip through the process. So, although we recover the expected proportion of their sample, there remains some uncertainty as to the completeness of that sample, and by extension to ours. 

The fact that our input lists of carbon stars are so heterogeneous can partially help here. They combine data found in both the optical and NIR, from both the northern and southern hemispheres etc. Thus selection biases inherent in one will be compensated by others.
In summary, it is extremely difficult to have a true measure of completeness for our sample. However, given the arguments above we believe that we are finding the majority of the variable AGB carbon stars in the region of sky covered by the Catalina Surveys.

\subsection{Carbon stars in MW satellites}  

We also investigate the variability properties, if any, of known carbon stars in satellite galaxies of the Milky Way. These galaxies have well-established distances, and can be used to check our procedures. The stars used are listed in Table \ref{tab:satellite_catalog}.

\subsubsection{Carbon stars in the Sgr dwarf spheroidal (dSph) galaxy}

There are 26 previously published spectroscopically-confirmed carbon stars in the Sgr main body \citep{Whitelocketal99}. We find that three  have detectable periods in the Catalina Surveys data; those at the highest right ascension. The others are not in the Catalina Surveys as they lie too close to the Galactic plane. We also have two of the variable stars listed in \citet{BattinelliDemers13}, giving a total of five known Sgr carbon stars.

None of the carbon stars from the Sgr main body given in \citet{McDonaldetal12} are in the Catalina Surveys, as they also are too close to the Galactic plane, and outside its footprint.

\subsubsection{Carbon stars in the Fornax dSph galaxy}

We use the  \citet{Whitelocketal09} list of carbon stars in the Fornax dwarf. We can identify long-period variable (LPV) stars more effectively than \citet{Whitelocketal09}, as the Catalina data are in the optical whilst they used NIR photometry, and the amplitude of variability is known to be greater in the optical than in the NIR \citep{Smithetal02,Nowotnyetal11b}. Also they have 15 epochs over three years, whilst we have many more epochs over a seven year period. Thus we get periods for stars that they do not. Note that one star (star \#105 in their table 2) is listed as having a period of 340 days, but of 270 days in the caption to their Fig 9. We obtain a period of 280 days, so we assume the shorter value was intended. 
One of the  \citet{Whitelocketal09} stars is also in the Mauron sample (M30 in their numbering system --  our S19). We also include any additional new carbon stars from the Fornax dSph, into our input list,  given in \citet{Mauronetal14}.

\subsubsection{Carbon stars in other dSph  satellites of the MW}

\citet{Azzopardietal85} provide lists of carbon stars  for the Sculptor, Carina and Leo II dSph galaxies.
 Sculptor carbon stars are also given in \citet{Menziesetal11}. Certain  Mauron  stars are already known to be within MW dSph galaxies: their star M4 (our S25) is in the Sextans dwarf galaxy, and is discussed more detail below. Their M29 (our S1) is in Sculptor, and discussed in \citet{Menziesetal11} and \citet{Sloanetal12}. As with the Fornax dSph star from the Mauron sample, we consider them as part of the satellite carbon star sample. 

\subsection{Photometric properties of our sample}
\label{properties_of_sample}

We can  characterise the ensemble properties of the full set of our sample, and compare their properties to our examples of MW satellite carbon stars.  The stars from the MW satellites are not included in the ``first" sample, but are used only for calibration and comparison.

\begin{figure}
 \centering
 \includegraphics[angle=0,width=85mm]{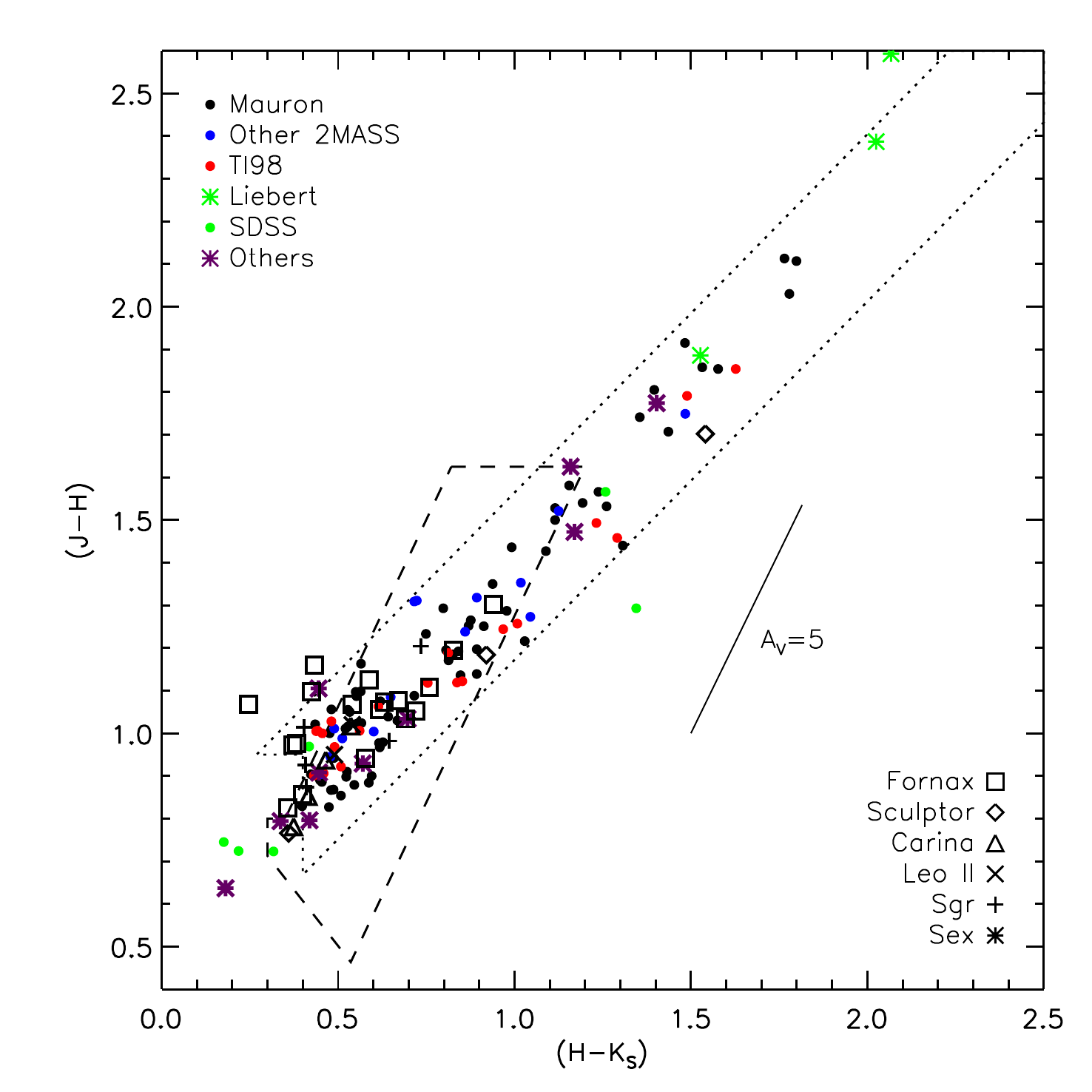}
 \caption{Colour-colour diagram of $(J-H)$ against $(H-K_{\rm s})$, uncorrected for reddening. The points are the variable carbon stars in Table \ref{tab:first_sample_catalog}. The different symbols represent either the source of the carbon star, or in the case of the MW satellite carbon stars, the host galaxy. The dotted region illustrates the approximate region selected by the Mauron group. They defined their cuts from known (but unnamed) carbon stars, so we use their stars, resulting from those selections to recreate the selection region. The dashed region is the colour selection used by \citet{Cruzetal03}. A reddening vector for A$_{V}$=5 is also shown, using the normal \citet{Cardellietal89} interstellar extinction law \citep{BinneyMerrifield98}.}\label{Fi:all_JH_HK_plot_unc_S2}
\end{figure}

Figure \ref{Fi:all_JH_HK_plot_unc_S2}   illustrates the consequences of the different selection techniques made for each sample on their location in the NIR colour-colour plots. The upper right of the plot is populated by 2MASS-selected stars, while those derived from optical imaging -- such as the \citet{TottenIrwin98} and the SDSS \citep{Green13} samples -- are found preferentially toward the blue end. Note that many of the previously published carbon stars in the MW satellites are found in the blue region, due to their discovery on photographic plates.

\begin{figure}
 \centering
 \includegraphics[angle=0,width=85mm]{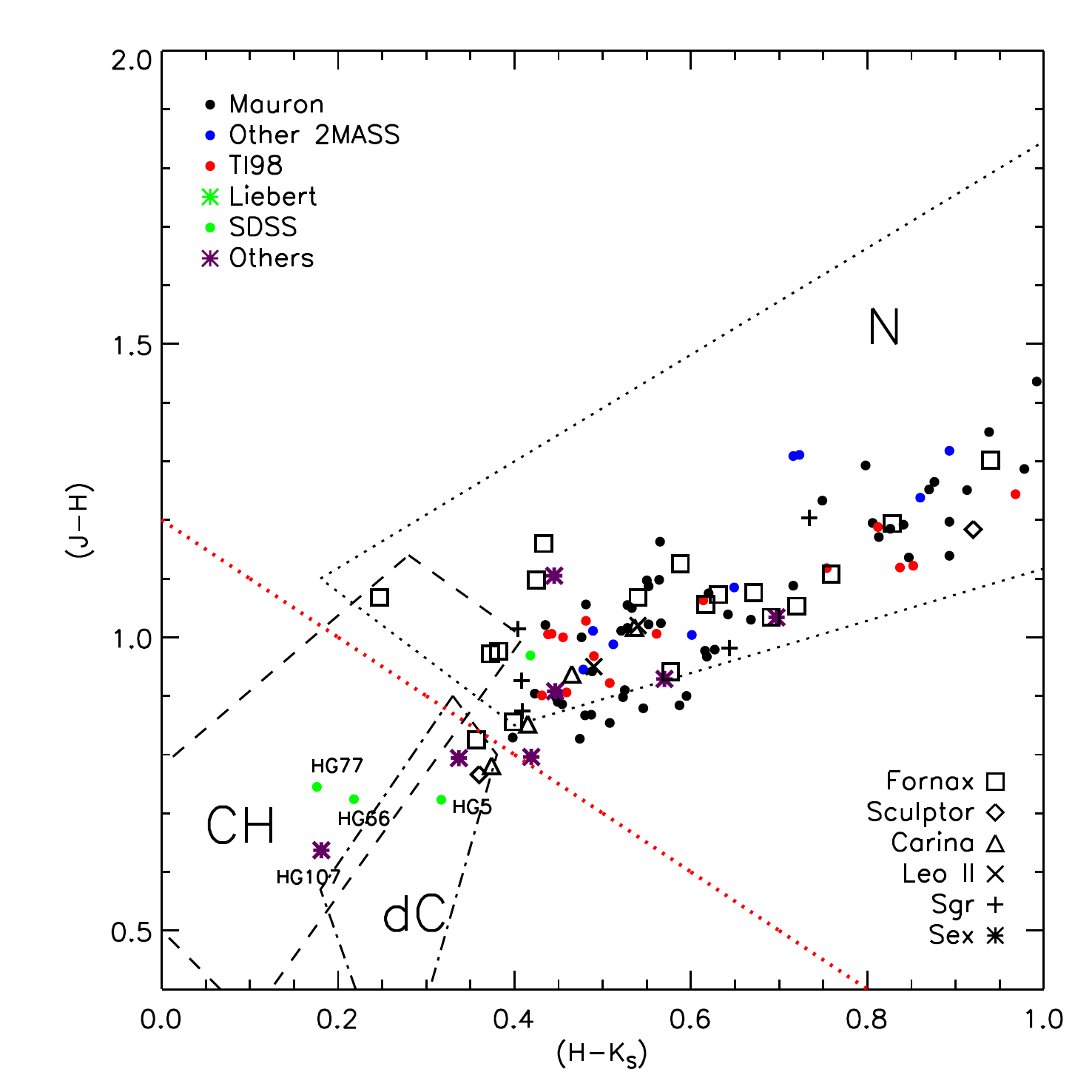}
 \caption{Carbon star colour-colour diagram of $(J-H)$ against $(H-K_{\rm s})$, uncorrected for reddening. Symbols are the same as Fig \ref{Fi:all_JH_HK_plot_unc_S2}. The regions show the locations of N-type, CH-type and dC carbon stars, derived from Figure 3 of \citet{Tottenetal00}. We see that the \citet{Tottenetal00} stars fit within their region defining the N-types, yet the Mauron group stars also fall just below. Also shown is the location of $(J-K_{\rm s}) =  1.2$ (red dotted line), which we use to distinguish AGB stars from our ``second sample" (in \S \ref{noPeriods}.)}\label{Fi:all_JH_HK_plot_unc_catagories}
\end{figure}

Most of the stars are within the area expected for N-type stars. A handful, at the blue end of the distribution, overlap in the region of $(H-K_{\rm s})/(J-H)$ colour where the CH-type carbon stars are also expected (e.g. see Figure \ref{Fi:all_JH_HK_plot_unc_catagories}). Inspection of the spectra for the SDSS sample found one star  (HG77) with a G-band feature, usually considered a characteristic of a CH-type star. But as we discuss  in \S \ref{R_star} below, it is probably a late-type R carbon star. HG107 is an outlier, classified as an N-type by \citet{Gigoyanetal01} (so we include it in our sample) but only with objective prism spectra, so this star should be treated with caution, especially given its location in the NIR colour-colour plot.

Notably three stars from the satellites also sit in the dC region, one each from Fornax (S6), Sculptor (S2) and Carina (S24). The Fornax and Sculptor stars seem to have reasonable light-curves, the Carina star less so. The Fornax and Sculptor stars also have V-band amplitudes of almost one mag, while the Carina star is only 0.38 mag. 
\citet{Zamoraetal09} find a couple of late-R type stars, which are essentially the same as N-type, in the region of the Totten NIR colour plot, that is near the dC region, and off the N-type, near to these MW satellite stars, suggesting that they too may be late-R types.

\subsection{Variability data}
\label{variability_data}

Our star variability data come primarily from Data Release 2 of the Catalina Surveys \citep[e.g.][]{Drakeetal13a}. The main focus of these surveys is the discovery and tracking of near-Earth asteroids, but their multi-epoch imaging of large regions of the sky provide excellent data for the study of the time-varying sky, including variable stars. In total these cover $\sim$33,000 square degrees of sky within the approximate region $-75^{\circ} < $ Dec $ < 65 ^{\circ}$ and more than 10--15$^{\circ}$ from the Galactic plane. The  main bodies of the LMC and SMC are also excluded, due to crowding. The time coverage is greater than seven years.
The whole survey has three components: the Catalina Sky Survey (CSS), the Southern Sky Survey (SSS), and the Mount Lemmon Survey (MLS), which cover the northern sky, southern sky and the ecliptic ($\pm$10$^{\circ}$) respectively. In many cases, stars have data from more than one of these surveys, and in a few cases all three, adding to the extent and quality of the light curves.

The Catalina Surveys data \citep{Drakeetal09,Drakeetal13a,Drakeetal13b} are provided by three telescopes: 1) The 0.68 metre Catalina Schmidt Telescope, which has a 8 deg$^{2}$ field of view, a pixel scale of 2.5 arcsec and which reaches a limit of $V\sim 19-20$ mag. 2) The 1.5 metre Mount Lemmon reflector, with a  1.1 deg$^{2}$ field of view  goes deeper,  to $V \sim 21.5$ mag. And 3) the 0.5 metre Uppsala Schmidt at Siding Spring, Australia, with a 4.2 deg$^{2}$ field of view a 1.8 arcsec/pixel image scale and a depth of $V \sim 19$ mag. 

All three telescopes operate with cooled, 4k$\times$4k back-illuminated unfiltered CCD cameras and use comparable data reduction pipelines. The telescopes observe for 21-24 nights per lunation (avoiding bright nights). A sequence of four exposures of each field are taken, each separated by 10 minutes. The sequences  are repeated, giving a typical 2 to 4 times, per field, during each lunation.
The Catalina Surveys photometry is undertaken with SExtractor\footnote{The Catalina data are unfiltered, but a conversion to the Johnson-Cousins $V$-band is provided by $V = V_{CSS} + 1.07 \times (V-I)^{2} + 0.04 (\sigma = 0.063)$}, and the data are made available through their website\footnote{http://nesssi.cacr.caltech.edu/DataRelease/}.

\begin{figure} 
  \centering
 \includegraphics[angle=0,width=85mm]{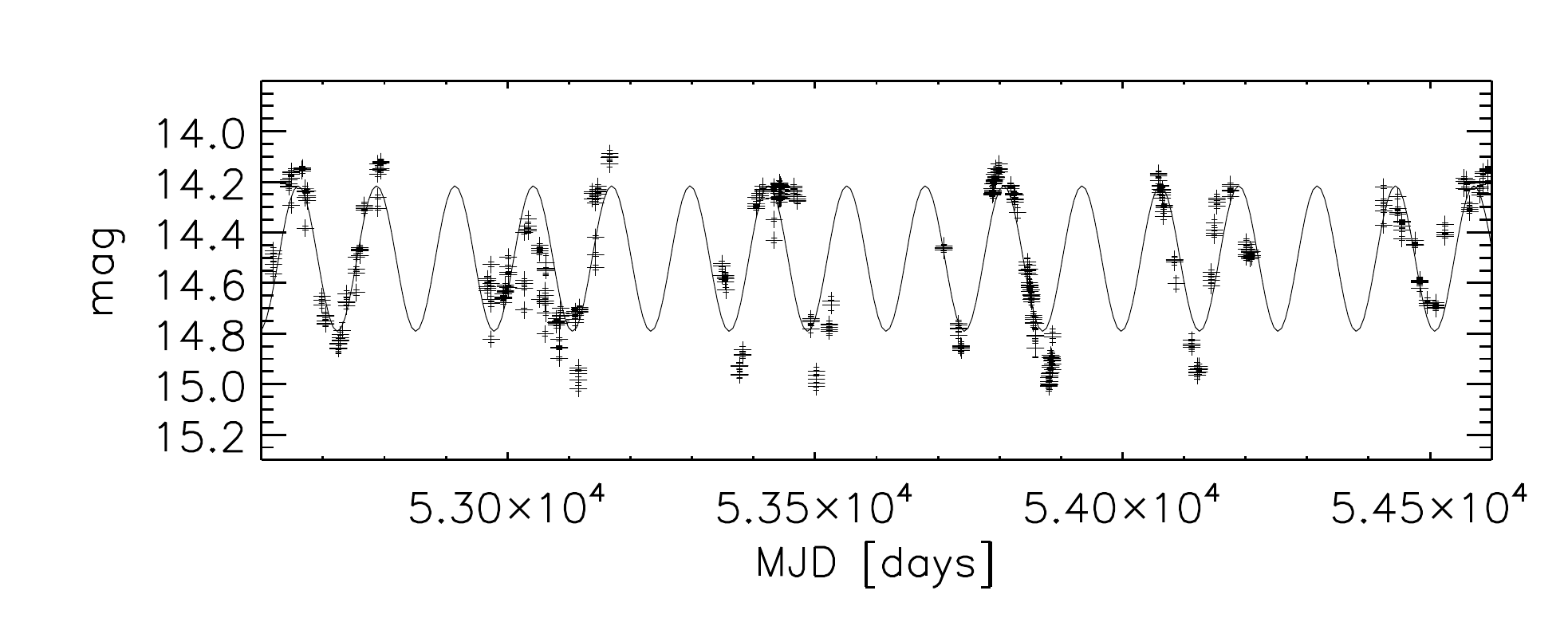}
 \caption{Light curve from the LINEAR survey of HG53.}\label{Fi:m89_linear_lightcurve_plot}
\end{figure}

In one case we also use LINEAR data in addition to the Catalina Surveys. The LINEAR survey \citep{Sesaretal11a} is similar to Catalina, in that its primary aim is the search for Near Earth Objects (NEOs). It operates on two 1 metre telescopes located at the White Sands Missile Range in New Mexico. The camera is unfiltered and has a $\sim$2 deg$^{2}$ field of view and 2.25 arcsec/pixel image scale. However, it is not as deep as Catalina, reaching down to $V < 18$, and the current public LINEAR\footnote{https://astroweb.lanl.gov/lineardb/} database only covers the overlap with the main contiguous Legacy Survey area of the SDSS DR7 footprint. The only example of the use of LINEAR is for HG53, for which the Catalina data could not well fit (see Appendix). The LINEAR fit is comparable to our Catalina data fit, with a period of 128 days and an amplitude of 0.58 mags (Fig. \ref{Fi:m89_linear_lightcurve_plot}).

\section{Distance Determination}
\label{methods}

\subsection{Period determination}

We use the light curves for our sample of carbon stars to derive their periods, from which we determine luminosities and hence distances. Light curve fitting was undertaken with the Period04 code \citep{LenzBreger04}, which has previously been successfully used on the light curves of AGB stars \citep[e.g.][]{Kamathetal10}. The light curves, folded light curves, and power spectra for the MW satellite carbon stars, and out first sample, are shown in the figures in the Appendix to this paper.

Many carbon stars -- estimated at approximately one third of carbon Miras, and an unknown proportion of other carbon-rich variable stars -- exhibit major obscuration events \citep{Whitelocketal06}. \citet{Soszynskietal11} present example light curves for both O and C-rich stars. The C-rich stars are seen to exhibit much greater changes in their mean luminosity, connected with episodes of intensive mass-loss. During these events, the magnitude can drop dramatically.  \citet{Feastetal03} conclude that these are due to the formation of dust clouds along the line of sight; an idea supported by the simulations of  \citet{WoitkeNiccolini05}. 

To account for these obscuration events, we visually inspect all our light curve fits. In a few cases, only parts of the light curve are used for fitting, as obscuration events make a solution for the whole duration implausible. One example is HG4, which has a sudden drop of $\sim$1.5 mags in mean magnitudes at a MJD of $\sim$54830 days, with a slow rise thereafter. Hence, in this case, we only fit the data up to that date. The drop is so sudden that no long secondary fit will fit all the data.

It is also notable that many of our light curves, from spectroscopically-confirmed carbon stars, have regular periods (e.g. HG46). In \citet{Soszynskietal11} such regularity is seen as characteristic of O-rich stars. Although they primarily use the stellar colours to assess chemistry,  \citet{Soszynskietal11} also use light-curve morphology to assign stars to either O-rich or C-rich classes. The Catalina light curve data suggest that one should be cautious in assuming that well-behaved light curves indicate an O-rich star.

 \begin{figure}
 \centering
 \includegraphics[angle=0,width=85mm]{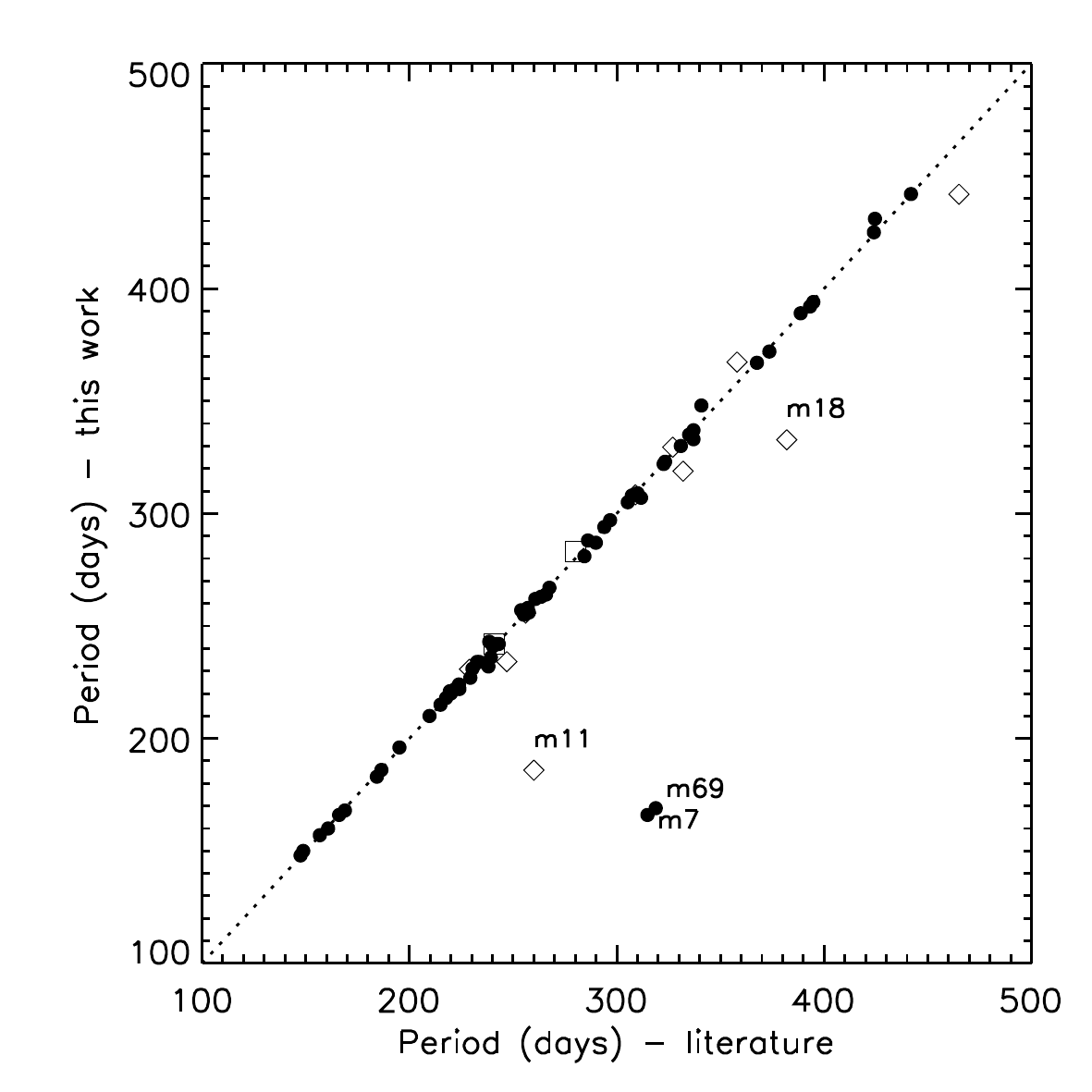}
 \caption{Plot of period from \citet{Mauronetal14} (filled circles), \citet{BattinelliDemers12}  (diamonds) and  \citet{BattinelliDemers13}  (squares) against ours, where objects are common. The dashed line shows where the periods are equal, and is plotted to guide the eye.}\label{Fi:literature_mine_plot_S2}
\end{figure}

\citet{Mauronetal14} list their own period determinations from Catalina data (combined with LINEAR data in some instances), of which 63 are shared with our own ``first sample". There is an excellent agreement between our periods and theirs, despite the different analysis techniques used (Fig. \ref{Fi:literature_mine_plot_S2}). 

An exception is the star HG52 (M7) for which we have a very much shorter period, 166 days (compared to their 314.8 days) on top of a longer secondary period (LSP). Inspection of our fit to the light curve, compared to that of \citet{Mauronetal14} indicates that the shorter period is a better fit to the data.

Another outlier is HG27 (M69) where again we have a period about half that found by \citet{Mauronetal14}. We believe that the light curve, folded light curve and power spectrum (see Appendix) support our shorter period.

There may be artefacts in the period determination for both \citet{Mauronetal14} and ourselves, due to the cadence of the Catalina observations that we both use, so it is also valuable to compare our periods to other data sets. A few stars are also shared between our first sample and that of \citet{BattinelliDemers12, BattinelliDemers13}, and allow us to compare the periods determined. These data allow a good cross-check as they were taken in NIR filters, compared to the  optical band of the Catalina Surveys, and used different light-curve analysis tools. Moreover, their data were observed within the dates (MJD from 55,250 to 55,950) contained within those of the Catalina Surveys data, ensuring a fair comparison.

Again we see a good match except for a couple of outliers, HG70 (M11) and HG98 (M18). Inspection of the \citet{BattinelliDemers12} light curve shows that their data for HG70 (M11) are the least reliable, with few data points, so such a bad match to our period is not unexpected. Similarly, HG98 (M18), has fewer data points than most of their sample, and also reaches down to the faintest magnitudes. Our light curves show our fit is very good for HG98, but more uncertain for HG70.

\subsubsection{Period-luminosity relations in LPVs}

\begin{figure}
 \centering
 \includegraphics[angle=0,width=80mm]{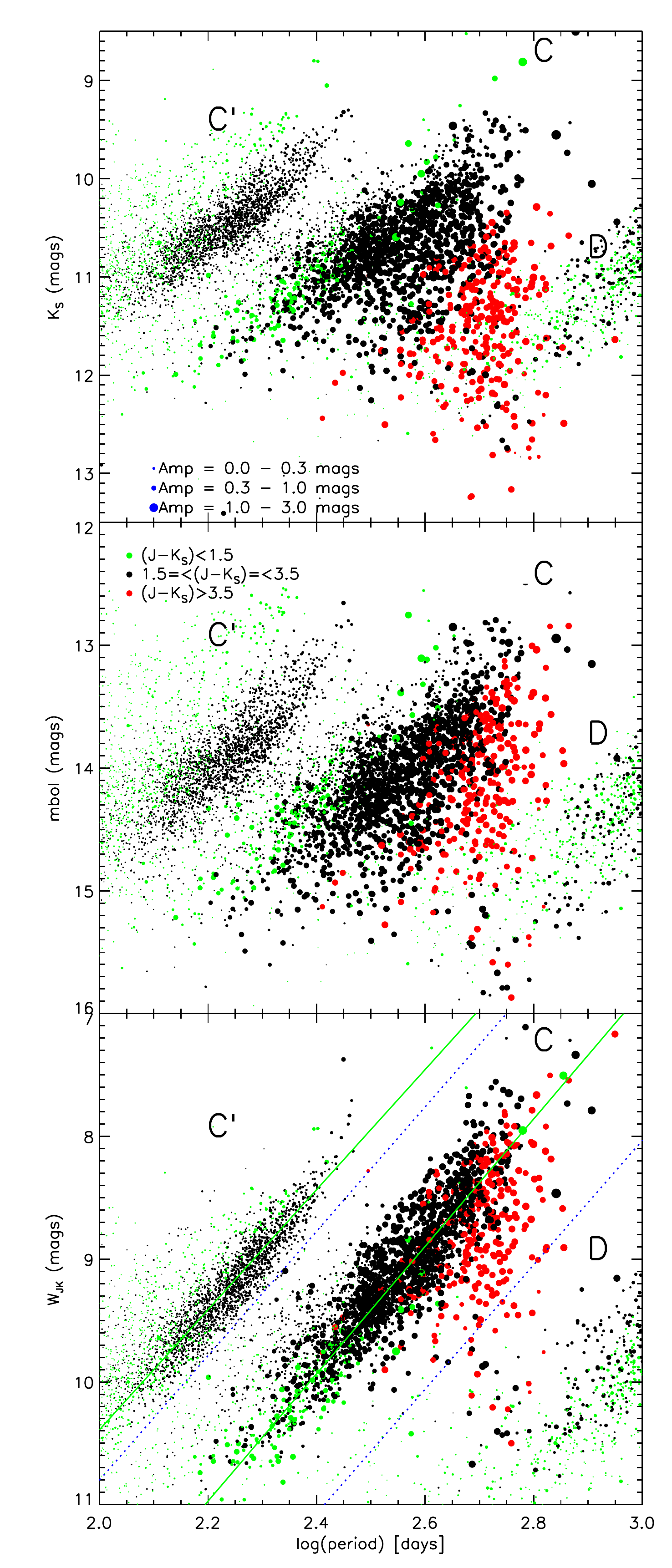}
 \caption{Plots of periods against luminosity for LMC carbon stars, from \citet{Soszynskietal09a} data. Top panel:  Plot of period against $K_{\rm s}$ magnitude. Centre panel: As above, but using the bolometric magnitude, derived from $J$ and $K_{\rm s}$, as given by \citet{Whitelocketal06}. Bottom panel: As above, but using the Wesenheit $W_{JK}$ magnitude. The blue dotted lines separate, by eye, the three sequences: C$^{\prime}$, C and D. The solid green lines are the fits for the sequences C$^{\prime}$, $W_{JK} = -4.80(log[period] - 2.0) + 10.39$,  and C, $W_{JK} = -5.19(log[period] -2.0) + 12.01$ from \citet{Soszynskietal07}. } \label{Fi:multi_soszynski_09_LMC_carbon_2MASS}
\end{figure}

We determine the distances to our ``first sample" stars from their periods. 
The primary period of a long-period variable (LPV) star is a good luminosity, and hence distance, indicator, although it is known that a number of period-luminosity relations exist (e.g. \citealt{Itaetal04a,Fraseretal08}).  Each relation follows a sequence in a plot of log(period) against magnitude (see top panel Figure \ref{Fi:multi_soszynski_09_LMC_carbon_2MASS}). 

Using the notation of \citet{Soszynskietal05} and  \citet{Itaetal04a} there are three sequences of relevance, where the period is greater than about 90 days: C$^{\prime}$, C and D. 
The different sequences in the PL relationship are due to a variety of reasons. In particular the C and C$^{\prime}$ sequences are thought to represent the fundamental and first overtone of pulsation, respectively. The origins of the D sequence are still debated, but may involve binarity \citep{SoszynskiUdalski14}.

Our first task is to allocate our carbon stars to one of the sequences in the P-L relationship. We achieve this by using a large sample of carbon stars with relatively well-known properties (and a known distance) from which we can derive a likely connection between the observable properties of the carbon star (i.e. colour, period, amplitude) and the sequence to which it belongs.
Overtone variables (C$^{\prime}$), for instance,  are expected to have smaller amplitudes than the fundamental (C) mode \citep{WoodSebo96}.

Our main comparison set of LMC carbon stars is the sample of \citet{Soszynskietal05} from the OGLE survey,  which they identify from their position in the Period-W$_{I}$ plot, where $W_{I}$ is a Wesenheit Index \citep{Madore82}, given by:

\begin{equation}
W_{I} = I -1.55 \times (V-I) 
\end{equation}

This is a reddening-free quantity that is defined for any two given wavebands $\lambda1,\lambda2$ with magnitudes $m_{\lambda1},m_{\lambda2}$  as

\begin{equation}
W_{\lambda1,\lambda2} = m_{\lambda2} - R_{\lambda1,\lambda2}(m_{\lambda1} - m_{\lambda2})
\label{eqn:general_wesenheit}
\end{equation}

They find that the location of spectroscopically-confirmed carbon-rich and oxygen-rich AGB stars separate effectively in these plots \citep{Soszynskietal05}.
We use their sample of LMC carbon stars, cross-matched with 2MASS to obtain NIR photometry, to create a series of selections that allow us to assign our sample to certain sequences in the period-luminosity plot. We remind the reader that the also LMC contains much younger populations than expected for the Sgr dSph. Hence we cross-check and refine these cuts by ensuring that they attribute C and C$^{\prime}$ to the MW satellite carbon stars such that they give consistent distances.

Previous applications of the LMC P-L relations have mostly used the $K$-band magnitudes. We, on the other hand, follow-up the suggestion of \citet{Soszynskietal07}, who noted that a Wesenheit index  in the NIR allows for the correction of the reddening due to the circumstellar dust. This is because in the 2MASS colours, it is hard to distinguish between interstellar and circumstellar extinction \citep{Messineoetal05}. Thus, we create a Wesenheit Index for $J$ and $K_{\rm s}$,denoted $(W_{JK})$:

\begin{equation}
W_{JK} = K_{\rm s} - 0.686 \times (J - K_{\rm s})_{0}
\label{eqn:wesenheit}
\end{equation}

Using this value we find that the scatter of the LMC carbon star sequences, shown in the $log(period)-K_{\rm s}$ plot,   tighten up considerably, when we  replace $K_{\rm s}$ with $W_{JK}$ (see bottom panel of Figure \ref{Fi:multi_soszynski_09_LMC_carbon_2MASS}). This allows for a better discrimination of stars between the various sequences, assuming that the scatter in the sequences are mainly due to circumstellar dust.

We also investigated the use of the relations between period and bolometric magnitude \citep{Whitelocketal06}. We plot the LMC PL relationships using their approach and we see that it is not as tight as the Wesenheit index (see centre panel of Fig. \ref{Fi:multi_soszynski_09_LMC_carbon_2MASS}). We believe that this is because the relation given by \citet{Whitelocketal06} requires  averaged values for $J$ and $K$, which they obtained from multi-epoch monitoring for their sample. However, as $(J-K)$ varies throughout the phase of pulsation, our single-epoch 2MASS photometry will likely have the wrong NIR colour to put into the relation. Hence we use the Wesenheit Index for this study, as it is less sensitive to these changes in colour.

If the 2MASS observations occurred during the Catalina Surveys, we might be able to estimate where in the light curve they were made. Unfortunately, the 2MASS data precede the Catalina Surveys by many years, making any such estimate impossible.

\subsection{Determination of P-L sequence for our sample}

We determine the observational features from the LMC data that we can use to assign a sequence to our first sample. This is done step-wise in the following manner.

\begin{figure}  
 \centering
 \includegraphics[angle=0,width=90mm]{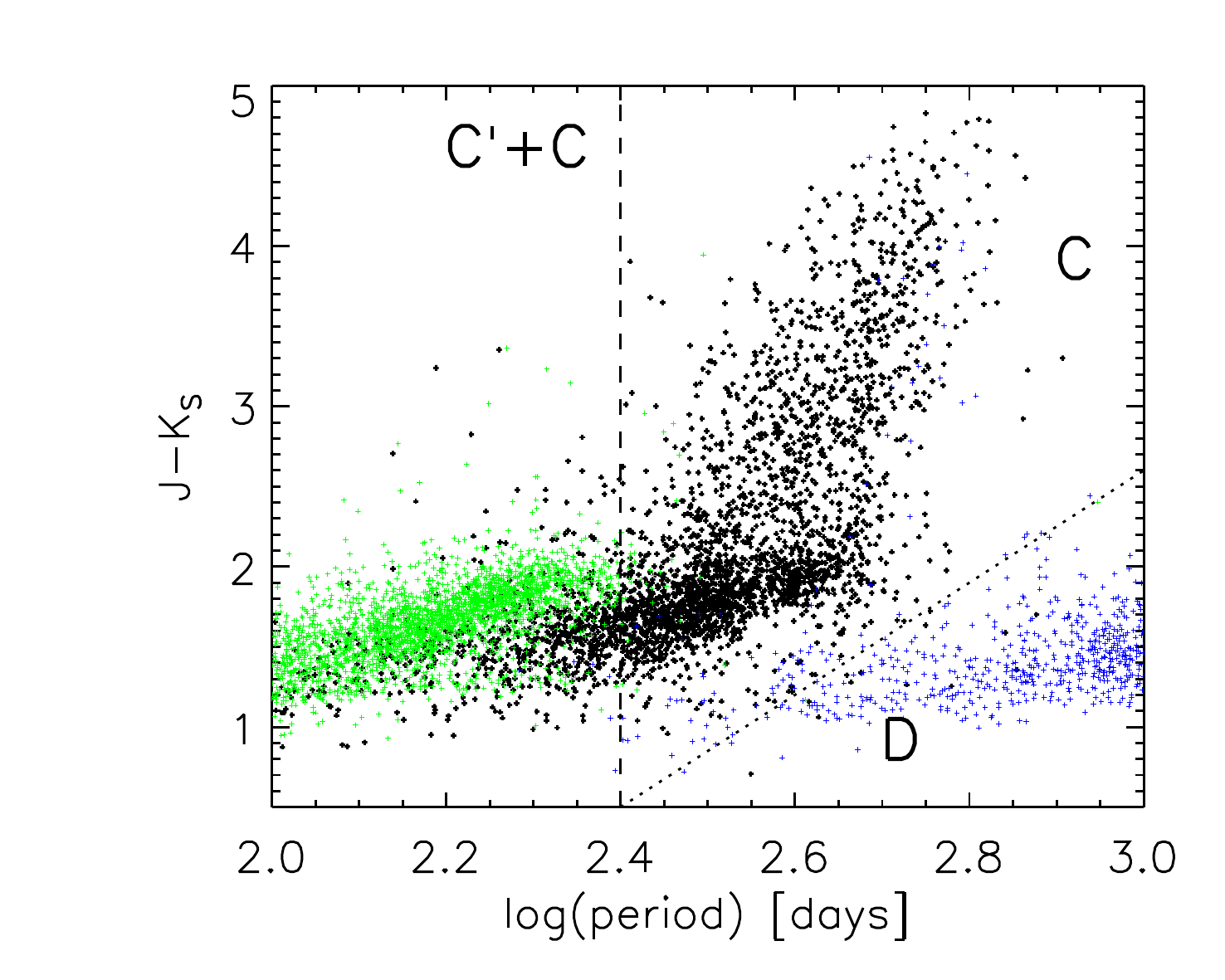}
 \caption{Plot of periods against $(J-K_{\rm s})$ mag for LMC carbon stars, from \citet{Soszynskietal09a} data. The green points represent the C$^{\prime}$ sequence, black points are stars on the C sequence and blue points are D sequence, as derived from the selections shown in Figure \ref{Fi:multi_soszynski_09_LMC_carbon_2MASS}. The region occupied by sequence D stars is primarily below the dotted line: $(J-K_{\rm s}) = 3.5 \times log(Period) - 7.9$, and stars above this line and with a log(period) $>$ 2.4 (dashed line) are assigned to the C sequence.}\label{Fi:soszynski_period_JK_plot_S2}
\end{figure}

Firstly, it is relatively simple to separate out the LMC D sequence stars as they primarily fall below the line (see Figure \ref{Fi:soszynski_period_JK_plot_S2}):

\begin{equation}
 (J-K_{\rm s}) = 3.5 \times log(period) - 7.8
 \end{equation}
 
The few apparent D stars sitting above this line are almost certainly C sequence, as they have a large amplitude, characteristic of the C sequence. One can see a number of very reddened stars in the bottom panel of Fig. \ref{Fi:multi_soszynski_09_LMC_carbon_2MASS} which still fall below the line separating the C and D sequences. Those stars above the line $log(period) > 2.4$, i.e. greater than $\sim$250 days, are  assigned to the C sequence.

 \begin{figure} 
  \centering
 \includegraphics[angle=0,width=90mm]{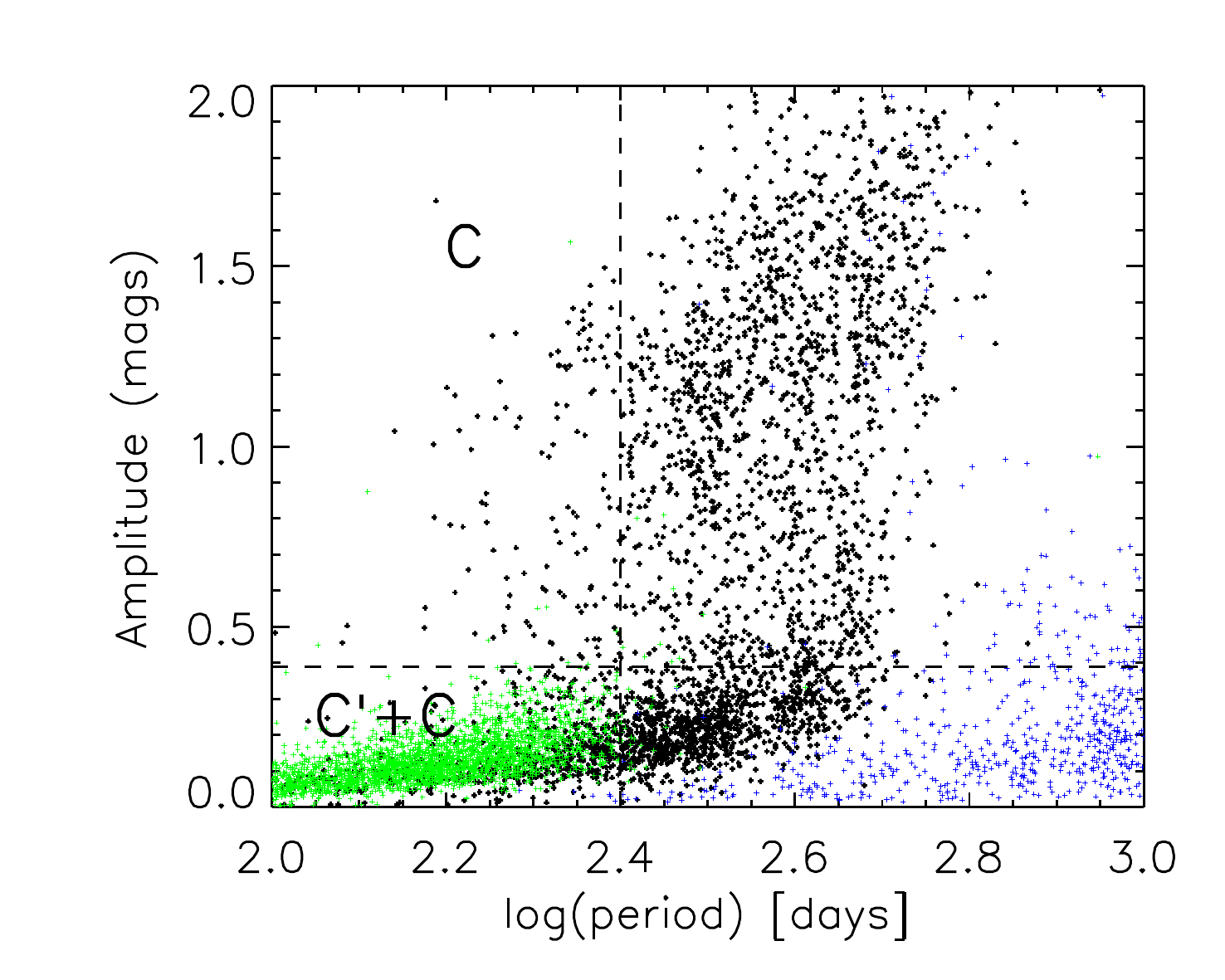}
 \caption{Plot of periods against the I-band amplitude  for LMC carbon stars, with data and symbols as in Figure \ref{Fi:soszynski_period_JK_plot_S2}. The stars with log(period) $<$2.4  (dashed line)  and an amplitude $<$ 0.39 are mainly on the C$^{\prime}$ sequence. }\label{Fi:soszynski_period_amplitude_plot_S2}
\end{figure}

 \begin{figure} 
  \centering
 \includegraphics[angle=0,width=90mm]{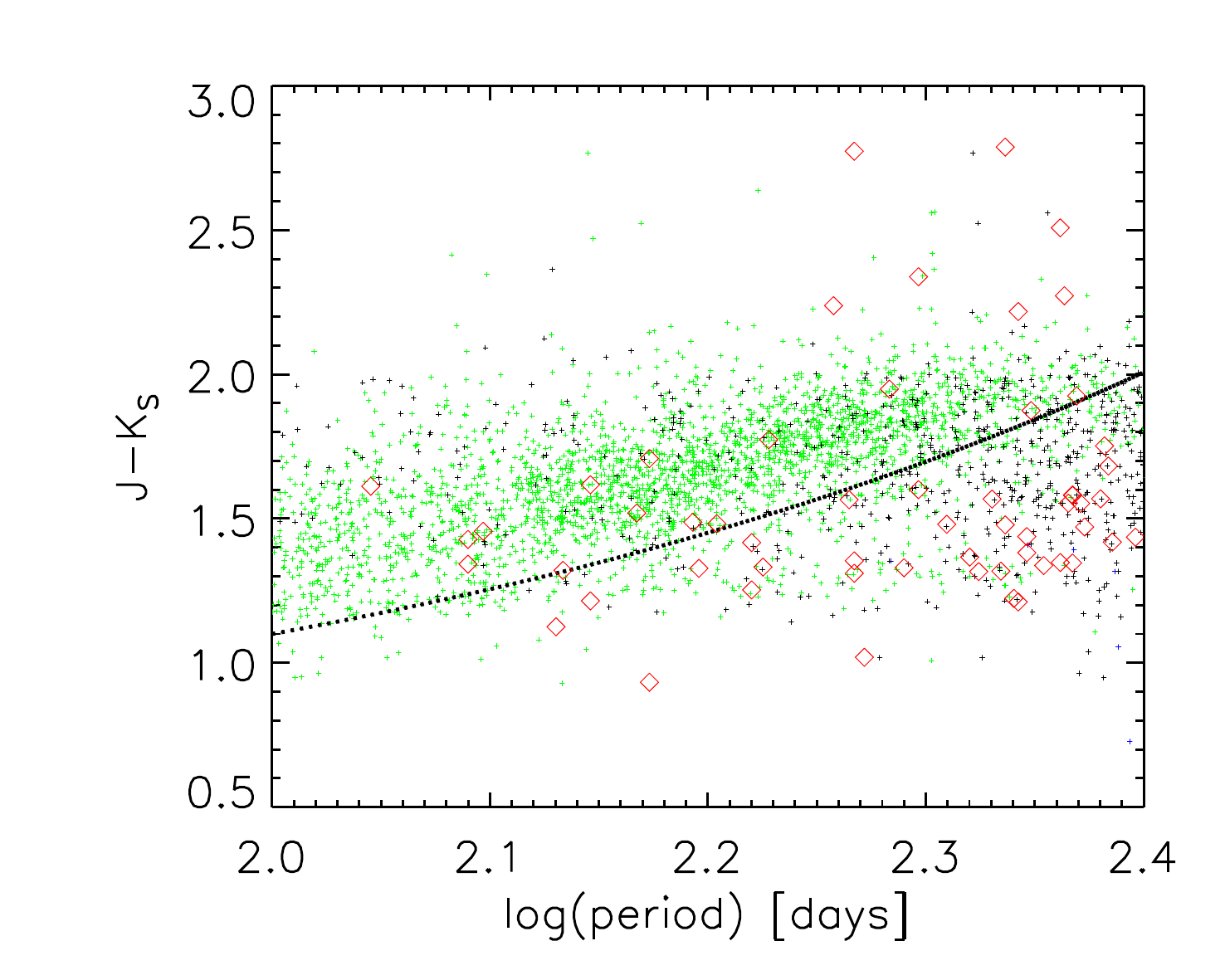}
 \caption{Plot of log(period) against J-K mag for stars with periods $<$ 250 days, with data and symbols as in Figure \ref{Fi:soszynski_period_JK_plot_S2}.  Although with some scatter, stars below the black dotted line (given by $J-K = 0.9 \times period/150 + 0.4)$ are mostly C sequence. Also shown are the first sample stars that lie in this range of periods (red asterisks). }\label{Fi:soszynski_period_JK_plot_short_S2}
\end{figure}

For stars with a  period below 250 days, it is found that many of those with I-band amplitudes $> 0.39$ mags are actually on the C sequence, as are those with $(J-K_{\rm s}) > 2.0$ mag. This value for the amplitude was adjusted to 0.42 mags to ensure that all the calibration stars in our MW satellite sample (Table \ref{tab:satellite_catalog}) have an appropriate distance. This approach also allows us to take into account that the LMC data have amplitudes in the $I$-band, while our data are in the V-band. We cannot, therefore, simply apply the selection that is most appropriate for $I$-band data. Moreover, any correction from I to V-band is almost certainly a function of colour. Using our MW satellite data circumvents this issue to some extent, although we acknowledge that the allocation of sequence adds uncertainties to our results.

For those that remain, the plot of $(J-K_{\rm s})$ against log(period) reveals that many of stars are on the C sequence if they are below the line given by (Fig. \ref{Fi:soszynski_period_JK_plot_short_S2}):

\begin{equation}
 (J-K_{\rm s}) = (0.9 \times period/150) + 0.4  
 \end{equation}
 
 This last selection is the most uncertain. If a star fails these cuts, it is assigned to the C$^{\prime}$ sequence.  Inspection of Figure \ref{Fi:soszynski_period_JK_plot_short_S2} shows significant scatter of C and C$^{\prime}$ stars on either side of this line. Hence we expect that there will be some mis-allocation of C and C$^{\prime}$ sequence stars by our cuts for these short period LPVs. 
 
\begin{figure}  
 \centering
 \includegraphics[angle=0,width=80mm]{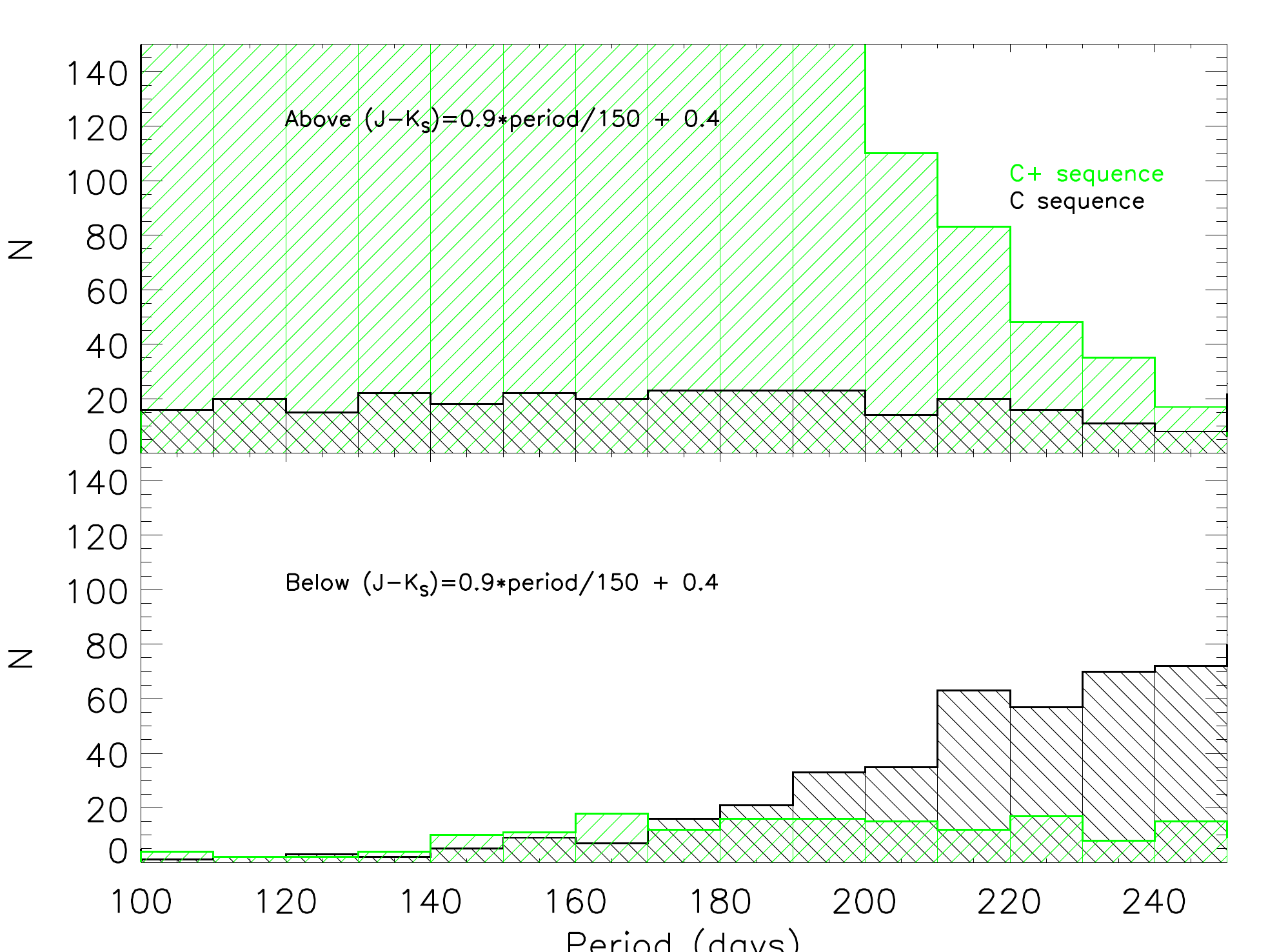}
 \caption{Distribution of C (black hatching) and C$^{\prime}$ (green hatching) stars, with data and colours as in Figure \ref{Fi:soszynski_period_JK_plot_short_S2}, separated into those found above (top panel) and below (bottom panel) the line defined by  $0.9 \times period/150 + 0.4$. The y-axes are both set to 140 counts to ease comparison. }\label{Fi:soszynski_period_JK_histo_short_S2}
\end{figure}

In Figure \ref{Fi:soszynski_period_JK_histo_short_S2}, we plot the number of LMC C sequence and C$^{\prime}$ sequence carbon stars that are found above and below our final selection cut. We see that for a period $<$ 210 days, the cut  selects the C$^{\prime}$ sequence relatively effectively. Above 210 days, the selection of C sequence stars is also reasonable. However, there is considerable mixing of C and  C$^{\prime}$ stars below 210 days.

It is notable that our ``first sample" contains very few stars on the C$^{\prime}$ sequence (see Table \ref{tab:first_sample_catalog}), compared to those in the LMC (see Figure \ref{Fi:multi_soszynski_09_LMC_carbon_2MASS}). The study of LPVs in NGC 147 and NGC 185 \citep{Lorenzetal11} reveals that these two dwarf elliptical galaxies also contain few C$^{\prime}$ stars compared to the C sequence, which they put down to their age. However, crowding probably affected these observations, which is more likely to impact on the smaller amplitude C$^{\prime}$ stars in these relatively distant galaxies. Similarly, our Catalina data are also biased against low amplitude stars, where the photometric errors become proportionally large as compared to the amplitude.

We do not use the period-luminosity relationship based on these LMC carbon stars. Instead, we use the carbon stars of our MW dSph satellite sample to rederive the relation. 

The period-luminosity relations in the literature are typically based on the LMC. But the LMC experienced an increase in star formation $\sim$ 3.5 Gyr ago \citep{Weiszetal13}, whilst the primary population of Sgr is estimated to have an age of $\sim$8 Gyr \citep{Bellazzinietal06b}. Moreover, the stars in the Sgr stream (one of our main objects of investigation) are older and more metal-poor than those in the bound remnant of Sgr \citep{Bellazzinietal06a,Chouetal07}. Thus, we might expect that a calibration based on the LMC may not  be appropriate. By using a range of other MW satellites, we provide a sample with a wider range of star-formation histories. Moreover, as there are Catalina Survey data for 32 LPV carbon stars in these other satellites, compared to only 2 or 3 which are likely in the outer regions of the LMC, we can use a calibration set that  is more comparable with our ``first" sample, having undergone the same procedures as our first sample.

Using our MW satellite stars requires NIR photometry for them, which was obtained from 2MASS \citep{Skrutskieetal06}, except for the two Leo II stars. 
For these, we use the NIR photometry of \citet{Gullieusziketal08} as, at the distance of Leo II, the 2MASS photometry is not reliable.
The photometry is  corrected for extinction and reddening by dust in the MW using  \citet{Schlegeletal98}. As our sample lies away from the Galactic plane, these corrections are mostly small.

\begin{figure}  
 \centering
 \includegraphics[angle=0,width=85mm]{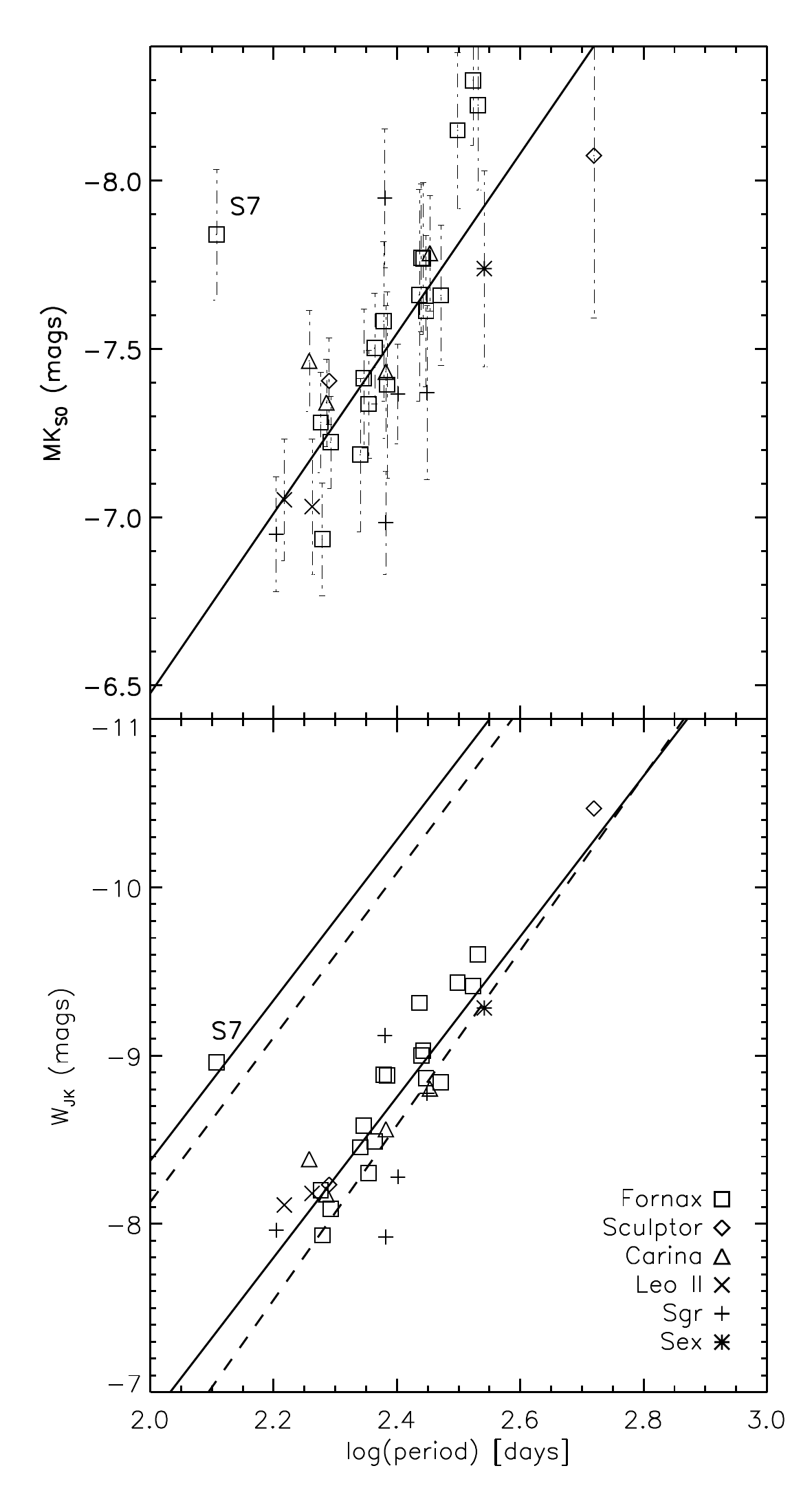}
 \caption{Period-luminosity plots in both the $K_{\rm s}$-band (top)  and $W_{JK}$ (bottom) for the satellites of the MW for the C sequence -- excluding the star S7, in Fornax, which has the shortest period and lies on the C$^{\prime}$ sequence. Top panel: Best linear fit (solid line)  with estimated uncertainties in M$_{K_{S}}$. Bottom panel: The dashed  lines show the same fits to the LMC data as in Figure \ref{Fi:multi_soszynski_09_LMC_carbon_2MASS}. The solid lines show the best linear fit for the stars on the C sequence $(W_{JK} = -4.771 \times log[period] + 2.698)$ -- the one sigma uncertainty on the slope of this fit is 0.42, and of 1.01 mag for the intercept. The upper line has the same slope as the lower,  but  is 1.53 mags brighter (the difference for the LMC data at log[period] = 2.3: see bottom panel of Fig. \ref{Fi:multi_soszynski_09_LMC_carbon_2MASS}).  }\label{Fi:multi_fornax_period_mag_plot}
\end{figure}

We then fit the Wesenheit magnitude $(W_{JK})$ against the periods found for these stars from Catalina (bottom panel of Fig. \ref{Fi:multi_fornax_period_mag_plot}), and determine the best linear fit for the C sequence stars (solid line). As only one star sits on the C$^{\prime}$ sequence, we simply add 1.53 mags to the C sequence fit. For comparison we also plot (dashed lines) the fits for the LMC data as found in Figure \ref{Fi:multi_soszynski_09_LMC_carbon_2MASS}. Note that the PL relations derived from the LMC and the other MW satellites are similar, with a 0.1 mag offset for a period of 300 days, increasing to 0.2 mag for a  short period of 125 days. These values are typically less than the uncertainties driven by the impact of variability.

\subsection{Distance determination and uncertainties on distances}

From these absolute Wesenheit magnitudes, we obtain the distances for all stars in our first (LPV) sample, which are given in the Table \ref{tab:first_sample_catalog}.

One major source of uncertainty in the absolute magnitude, and thus distance, is the variability of the stars. The PL relationships use mean K-band magnitudes. As we do not have information on the phase in which the 2MASS observations were made, they may (in the worst case) be at the maximum or minimum of the variability.

\begin{figure}  
 \centering
 \includegraphics[angle=0,width=80mm]{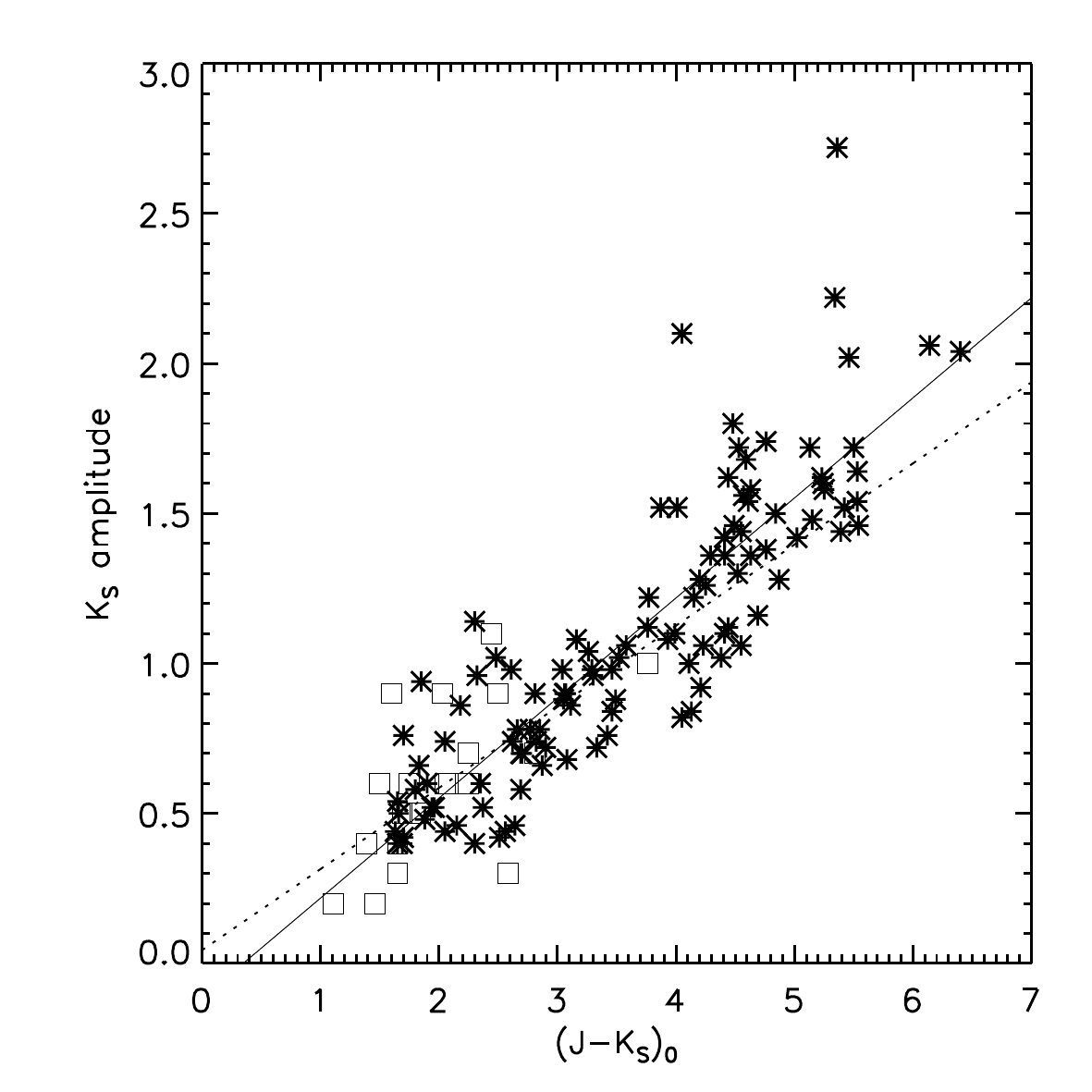}
 \caption{Plot of mean $(J-K_{\rm s})_{0}$ colour against K-band amplitude for the \citet{Whitelocketal06} sample (asterisks), with the best linear fit (solid line). Also shown is the smaller  \citet{BattinelliDemers13} sample (squares) with the best fit to these data (dotted line).}\label{Fi:derive_jk_kamp_fit}
\end{figure}

To estimate the variability in the $K_{\rm s}$-band amplitude of our sample we use their $(J-K)$ colour. This is possible as there is a correlation between the amplitude of $K$-band variability. We use the data  from Table 3 of \citet{Whitelocketal06} -- who had multi-epoch observations covering the whole period of variability -- for all their stars with amplitudes given (excluding those stars for which only limits are given). A linear fit (Fig. \ref{Fi:derive_jk_kamp_fit}) to these data give the following relation:

\begin{equation}
K_{amplitude} = 0.334\times(J-K_{S})_{0} - 0.116
\label{eqn:estimate_Kamp}
\end{equation}

We can check this relation using the sample of \citet{BattinelliDemers13}, who obtained $J$ and $K$ light curves for many of the Mauron group stars, and so overlap our first sample. We see that the fits are comparable, although both show some considerable scatter (Fig. \ref{Fi:derive_jk_kamp_fit}). It is difficult to estimate the impact of this scatter on our distances as it appears irregular; for example, there is a smaller scatter at $(J-K_{\rm s})\sim3.7$ than either blueward or redward of this value.
The distance errors are then derived by taking half of the peak-to-peak amplitude in K magnitudes and propagating these through the distance estimates.

\begin{figure}
 \centering
 \includegraphics[angle=0,width=85mm]{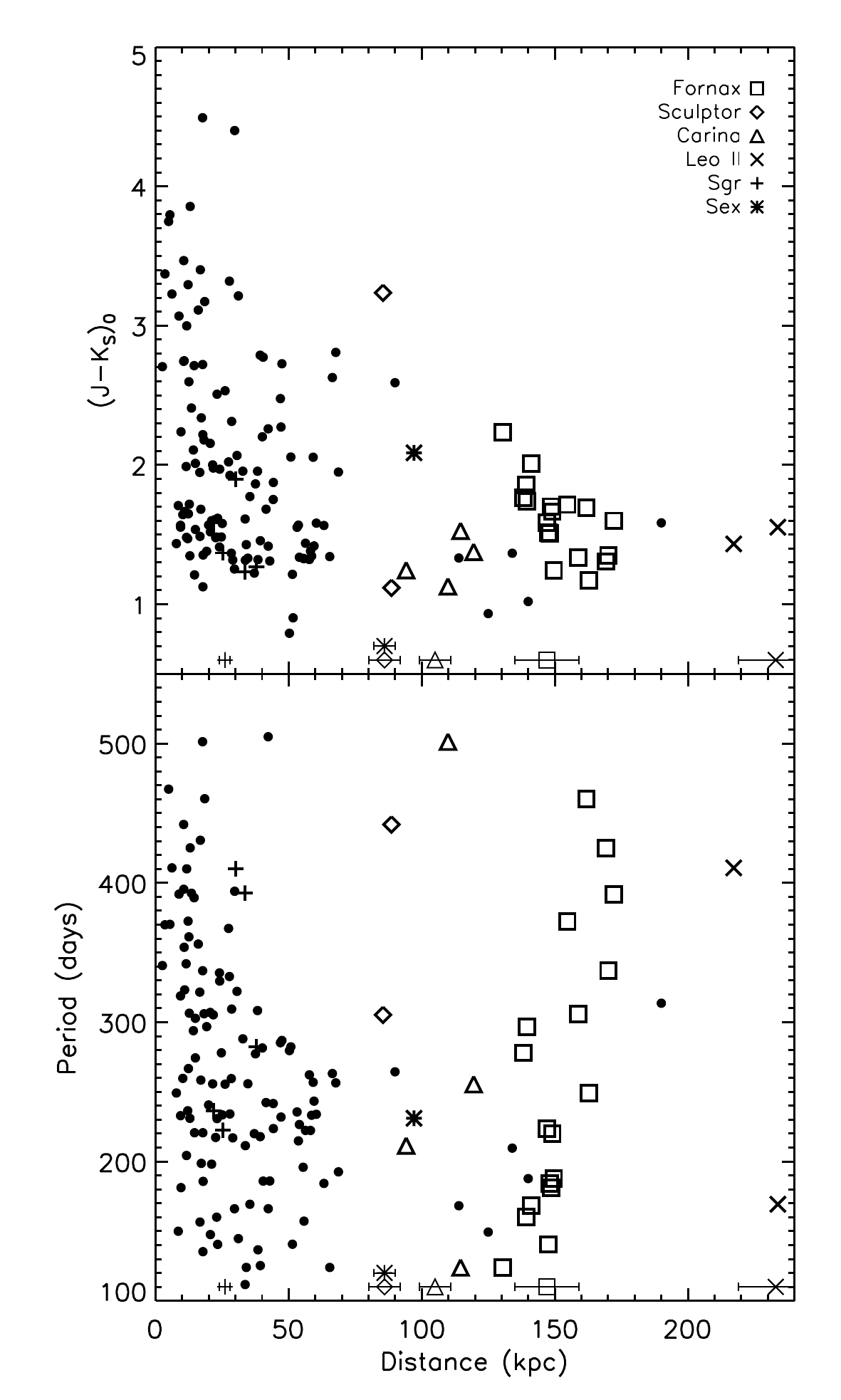}
 \caption{Top: Plot of $(J-K_{\rm s})_{0}$ colour against distance for our ``first" sample of our LPV carbon stars (Tables 2, 3 and 4; black symbols). Bottom: Plot of period against distance. The distances shown for the historical carbon stars in known dwarf galaxies (open symbols) are those derived by our method, and not of the known distance to the host, to illustrate the spread of values and systematic errors in our method. The distances, and their uncertainties, to the host galaxies \citep{McConnachie12} are shown by the same symbols with error bars at the bottom of each panel.}\label{Fi:multi_dist_JK_period_plot_errs_S2}
\end{figure}

The plot of period against distance  (Figure \ref{Fi:multi_dist_JK_period_plot_errs_S2}, bottom panel) shows that the more distant halo carbon stars in our first sample have shorter periods. It has been suggested that the period of an LPV is an indicator of age \citep{Menziesetal11}. If this is correct, the more distant stars are older than those nearer to the MW. One object stands out here, HG118, which has a weak light curve, and which we discuss later (\S \ref{noPeriods}) as one of our least certain stars.

\section{Results}
\label{results}

The main results of this study can be found in Table \ref{tab:first_sample_catalog} where details of the variable carbon stars are given. This includes RA and Dec (J2000), derived NIR values (from 2MASS), our distance estimate, and any radial velocity (converted to Galactic Standard of Rest) as given in the literature. We also give the period and amplitude (in Catalina V magnitudes) found from our best estimate using Period04 fits, and the adopted sequence (C or C$^{\prime}$). Finally, we also give the reference in the literature that provided the most valuable data, and note the halo feature to which we judge the carbon star to most likely be associated. Note that Unc in this table represents ``unclear", where it is not clear to which -- if any -- structure in the star may sit (a colon after the feature name indicates, in the usual manner, some uncertainly in the attribution given).

\subsection{Comparison of the data to the LM10 model}

\begin{figure*}
 \centering
 \includegraphics[angle=0,width=160mm]{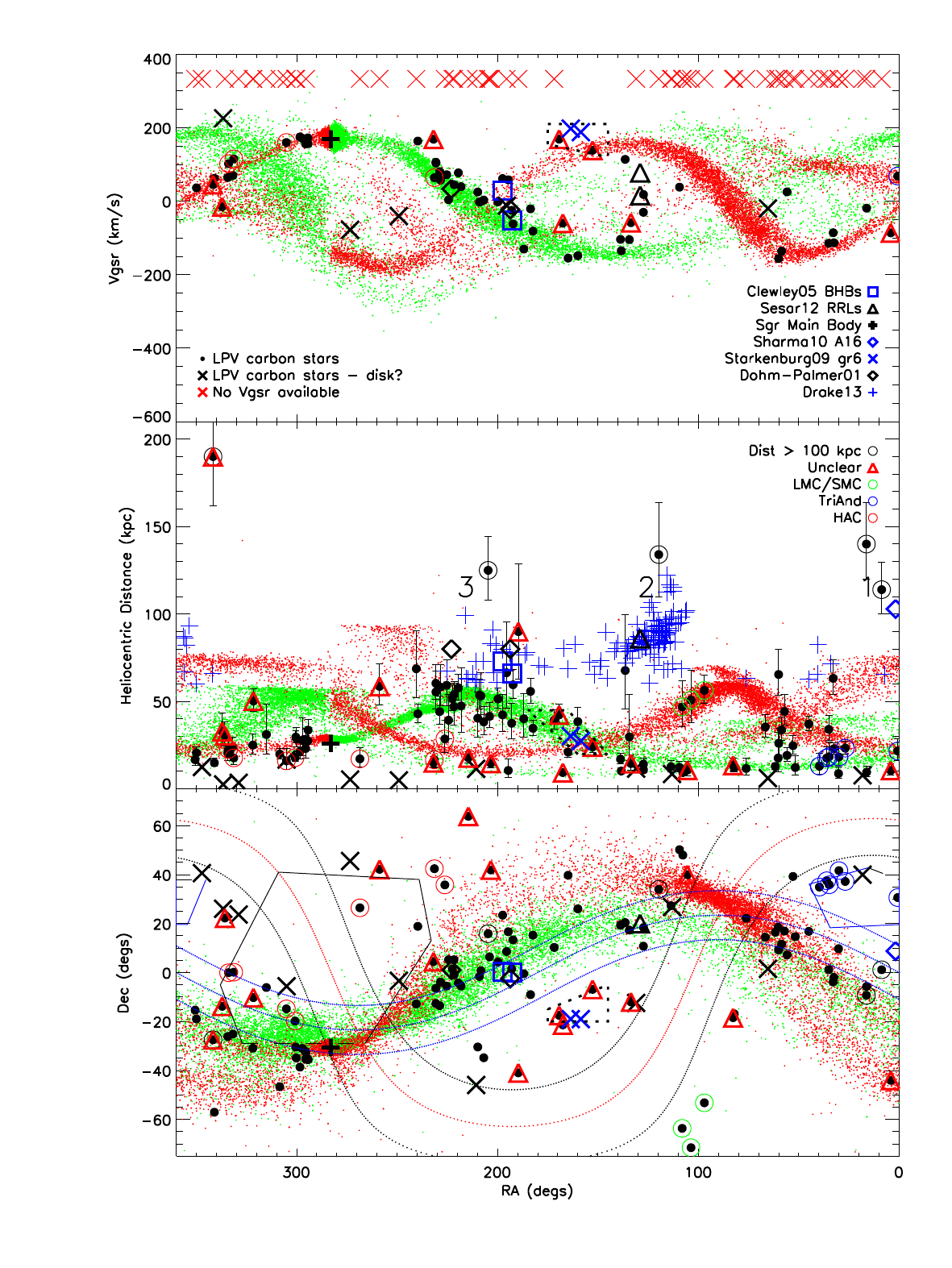}
 \caption{Location of carbon LPVs in relation to the LM10 model (small points) for the last wrap; showing  the leading (green) and trailing (red) arms.  Top panel: RA against the radial velocity (with respect to the Galactic Standard of Rest: Vgsr). Carbon stars without radial velocities are illustrated by red crosses at V$_{GSR}$=333. Middle panel: RA against heliocentric distance. Stars with a heliocentric distance $>$ 100 kpc are marked with outer circles -- it can be seen that these lie exclusively near the Sgr orbital plane. Bottom panel: RA against Dec. The red dotted line is the Galactic Plane and the black dotted lines are $\pm$15$^{\circ}$; within this region of the Galactic Plane, Catalina Survey data are not available The limits in declination also correspond to the Catalina limits. The approximate regions of  TriAnd  (blue polygon) and Hercules-Aquila Cloud (black polygon) are also shown.  The blue dotted curves show the ecliptic, $\pm$10$^{\circ}$, indicating the region of the MLS survey used by \citet{Drakeetal13b}.}\label{Fi:multi_plot_lm10_t1_random_S2}
\end{figure*}

One of the major goals of this study is to investigate whether the distances derived from newly available Catalina light curves will better reveal those carbon stars that belong to the Sgr streams, and those that may be from other substructure. In many cases we have been able to use the location, distance and (where available) the radial velocity of a carbon star to assign it to known substructure. These assignments are given in the final columns of Table \ref{tab:first_sample_catalog}. For many substructures, they are also indicated in Figure \ref{Fi:multi_plot_lm10_t1_random_S2}, except for the overlapping set of candidate structures near the Galactic Anti-centre.

Previous  authors, such as  \citet{Mauronetal04} have compared their halo carbon stars to the older  \citet{Ibataetal01} model, and later papers compare their carbon star data to the model of \citet{Lawetal05}. For the first time we study a large sample of LPV carbon stars comparing it to the LM10 model.

Plots of the location of the carbon stars in comparison to the LM10 model, for location of the sky, distance and radial velocity, are shown in Figure \ref{Fi:multi_plot_lm10_t1_random_S2}. For these plots we use a simple selection to identify stars from our lists that may be associated with the disk. Using the distance, we derive the locations of our stars in the Galactocentric coordinate system. We then highlight stars that are within a radius of \mwRadiusKpc{} kpc from the Galactic Centre and a distance of \mwHalfHeight{} kpc above or below the Galactic Plane. These are indicated as ``Disk" in the tables, and by asterisk symbols in Figure \ref{Fi:multi_plot_lm10_t1_random_S2}. Our values for the size of the Galactic disc region is a conservative one, based on a disc radius of $\sim$15 kpc \citep{Ruphyetal96,Minnitietal11}, but taking into account the known presence of young stars out to 20 kpc \citep{Carraroetal10}. Heliocentric radial velocities, as given in the literature, were also converted to the Galactic Standard of Rest (V$_{GSR}$).

A notable aspect of the top panel of Figure \ref{Fi:multi_plot_lm10_t1_random_S2} is how many stars of the first sample do not have published velocities. Yet we can see how well the velocity data can help identify the stars that are most likely in the Sgr tidal streams. This immediately suggests that a programme of observations to obtain radial velocities for all the carbon stars without such data will  be invaluable. Given the luminosity of these stars, such an effort will not be too demanding.
 
One significant result is that although the majority of our input lists were based on all-sky surveys (e.g. 2MASS), the majority of the carbon stars in our study appear as though they can be attributed to the Sgr streams. This is particularly true for those for which velocity (V$_{GSR}$) data are available, but is also the case for central (RA against heliocentric distance) and bottom (RA against Dec) panels. Many carbon stars trace the leading arm across much of its length, and a group of them at $RA\sim$300$^{\circ}$ are very close the main body of Sgr (which itself lies near the Galactic Plane and hence is not in the Catalina Surveys). In many cases we can allocate the stars to either the leading or trailing arms (green and red points respectively in Figure \ref{Fi:multi_plot_lm10_t1_random_S2}).

The apocentre of the trailing arm in the LM10 model, centred at  $RA\sim$100$^{\circ}$, contains no carbon stars. This is not totally unexpected as much of it lies behind the Galactic Plane. But, even taking this in consideration, there are very few stars in the region of the trailing arm apocentre just north of the Galactic plane. This may suggest that the LM10 model is not a good representation of the trailing arm. Note that all of the (three) carbon stars that appear to sit in the arm at the appropriate distance (middle panel of Figure \ref{Fi:multi_plot_lm10_t1_random_S2}), and that are marked with green open circles, are actually in the southern Galactic hemisphere, and  associated with the LMC (see lower panel). 

 \begin{figure*}
 \centering
 \includegraphics[angle=0,width=180mm]{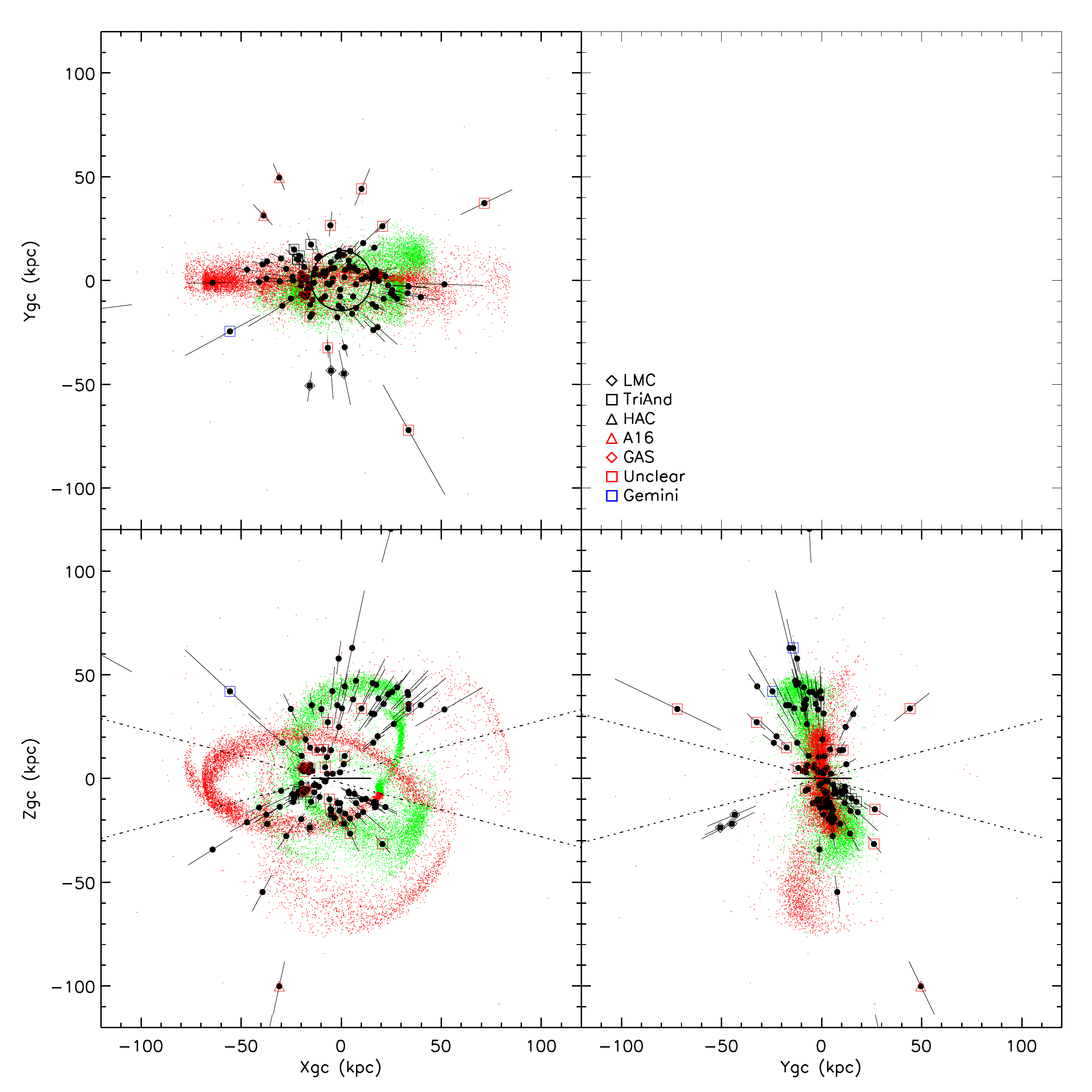}
 \caption{Location of carbon LPVs in relation to the LM10 model (small points) for the last wrap; showing  the leading (green) and  trailing (red) arms, as seen in X, Y and Z coordinates. The Sun is 8.5 kpc from the Galactic Centre. The dotted black lines illustrate the cone of avoidance due to the Catalina Surveys reaching down to $\pm$15$^{\circ}$. Those stars attributed to key halo substructures, in addition to Sgr, are indicated with surrounding symbols. For clarity, we only show the region within 120kpc of the Galactic Centre, hence a few very distant stars are not seen on these plots. Two stars at [X,Z]$\sim$[-50,-50] are probably C sequence stars that have been wrongly classified as C$^{\prime}$, by our selection cuts. }\label{Fi:multi_plot_XYZ_lm10_S2}
\end{figure*}

It is worth commenting on a number of stars. Two stars (HG11 and HG23) stand out as inconsistent with LM10, sitting either side RA = 50$^{\circ}$, and at a distance of $\sim$60 kpc. Both these have properties (period, colour and amplitude) that place them close to the boundary between the C and C$^{\prime}$ sequence in Figure \ref{Fi:soszynski_period_JK_plot_short_S2}. They also have V$_{GSR}$ values (--86 and --156 km/s), however, that are consistent with them sitting in the trailing Sgr arm. Hence, we believe that these are actually C sequence stars, not the C$^{\prime}$ that our selection attributes to them. If this is the case, it demonstrates the caution that we must exercise in using these selections near the boundary.
HG107, which was discussed in \S \ref{properties_of_sample}, at RA$\sim320^{\circ}$ and a distance of 50 kpc is quite likely (given its NIR colours) not an AGB star.

We can also look at the locations of our first sample stars in Galactocentric cartesian coordinates (Fig. \ref{Fi:multi_plot_XYZ_lm10_S2}). Again it is striking that the majority of our sample is consistent with much of the LM10 model. The distant carbon stars that we attribute to A16, the Gemini Arm are at the edge of the plots. It is also easier, in this representation, to see the apparent lack of Sgr stars -- in the X vs. Z plot, just above the Galactic Plane and affecting both the leading and trailing arms. The two stars that we believe may be mis-assigned C$^{\prime}$ can also be seen.

\subsubsection{The Nature of Distant Carbon Stars}

There are six carbon stars with estimated distances that are greater than 100 kpc. Some of these are somewhat blue compared to the rest of the first sample, which arouses suspicion that they may be CH, early-R or dC stars.

\begin{figure}
\centering
\includegraphics[angle=0,width=90mm]{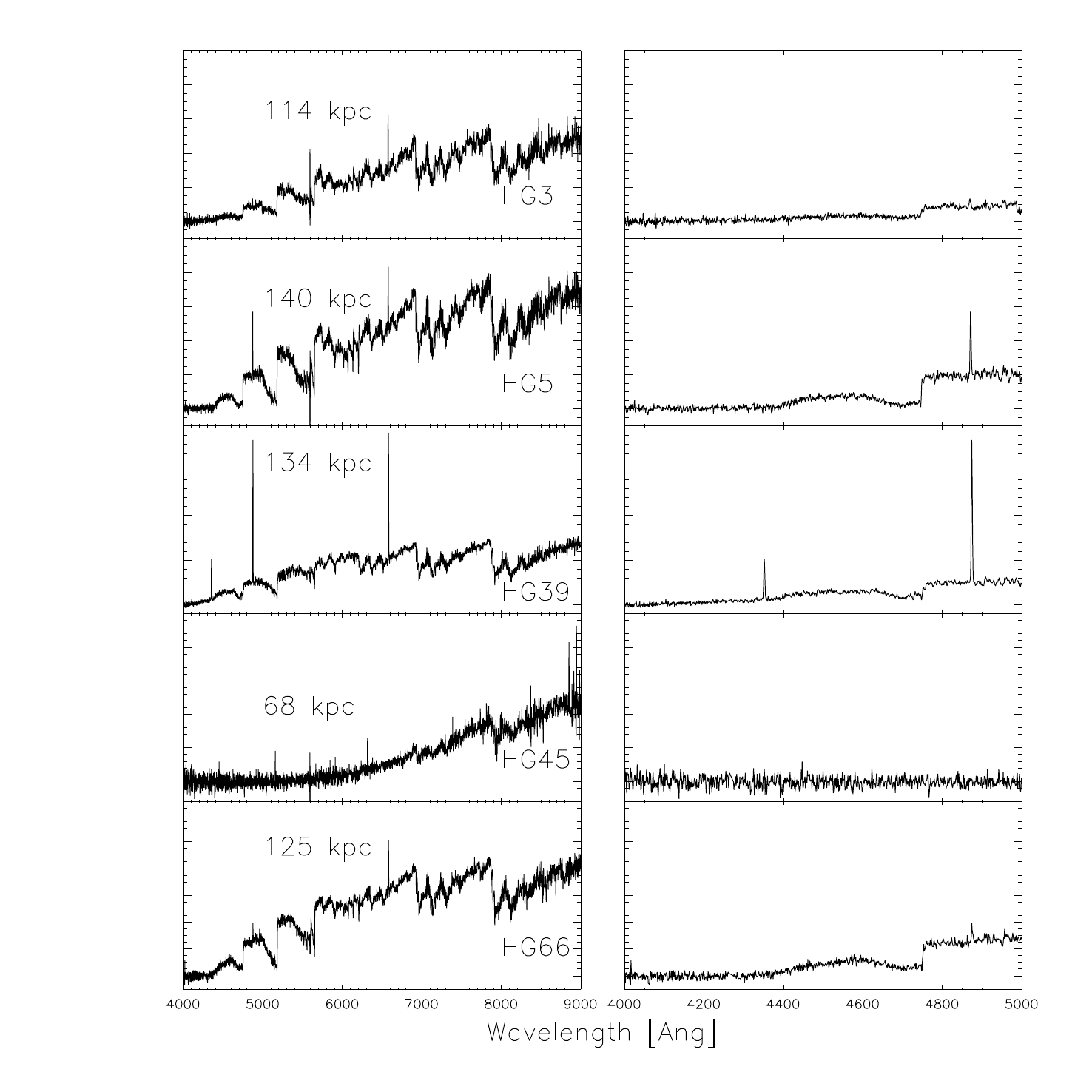}
\caption{Spectra for distant (R $>$100 kpc) carbon stars with SDSS data. The left panels show the full spectrum for 4000 -- 9000\AA, and the right panels show the corresponding region for 4000--5000\AA, where the G-band sits at $\sim$4300\AA.}\label{FI:multi_plot_sdss_spectra_S2}
\end{figure}

The spectra for four of these six stars are available from the SDSS, and we also plot the spectrum for another SDSS star at 68 kpc  (Fig \ref{FI:multi_plot_sdss_spectra_S2}). Four show classic N-type spectra, with little flux below 4500\AA, and all four of the stars more distant than 100 kpc showing Balmer emission. All four except HG66, show very good light curves. Although the spectrum of HG45 has a low signal-to-noise, it is consistent with a carbon star. Its light curve (see Appendix) is also of good quality, reinforcing its claim to be an AGB star.

The distant carbon stars are apparently associated  with other tracers, and all are likely related to structures which are already known. These are labelled on the middle panel of Figure \ref{Fi:multi_plot_lm10_t1_random_S2} as 1, 2 and 3.  
Many other stars of the sample can be attributed to known substructures. We now discuss each of these groups and features in more detail, followed by the closer substructures.

\subsubsection{A16/Pisces Overdensity (Feature 1)}

Feature 1 comprises two carbon stars at a distance of $\sim$100 kpc  at RA $\sim$10--30. These are probably associated with a feature known as A16, detected in M-giants by \citet{Sharmaetal10}, who show it to be a rather large and diffuse structure at a distance of 100 kpc and stretching 20$^{\circ}$ either side of RA = 0. A16 encloses the previously reported and better known ``Pisces Overdensity", a feature  found in RR Lyrae \citep{Watkinsetal09,Sesaretal10b} in the SDSS stripe 82, lying at a distance of $\sim$80 kpc, and by \citet{Drakeetal13b} in Catalina data,  at RA from 350--360$^{\circ}$. The large extent of A16 makes it  more plausible that our carbon stars are associated with this structure. It is possible that HG118 may also be part of A16, but due to its much greater distance ($\sim$180 kpc) and uncertaintly over its nature we do not include it as such.

\subsubsection{Gemini Arm (Feature 2)}

Feature 2 comprises two distant carbon stars (HG39 and the closer HG45) that sit near a feature first found in BHBs by \citet{Newbergetal03}, further studied in  \citet{Ruhlandetal11}, and later denoted the  ``Gemini Arm" by   \citet{Drakeetal13a}. \citet{Belokurovetal14}, argue that the Gemini Arm is actually the true location of the apocentre of the trailing arm of Sgr stream, explaining why that part of the LM10 model has not been seen at the distance expected in the area  north of the Galactic Plane.

The Gemini Arm is particularly apparent in the RR Lyrae of \citet{Drakeetal13b}. They also comment on  several of the most distant RR Lyrae, which they think may  be distinct from the main body of the Gemini Arm. These more distant RR Lyrae lie close -- on the sky -- to HG39, and are also at about the same distance. It is significant that these RR Lyrae come from the MLS component (which reaches much deeper than the CSS and SSS surveys) of the Catalina Surveys, but which is only 20$^{\circ}$ wide, straddling the ecliptic. These distant RR Lyrae stretch across the full width of the MLS strip, and so likely cover a wider region of the sky, possibly including  the position of HG39. 

 \citet{Sesaretal12} report two kinematically distinct, but spatially close, groups of RR Lyrae stars (denoted Cancer A and Cancer B). They both lie at the same location in the sky and distance ($\sim$86 kpc), but differ in radial velocity: the RR Lyrae of Cancer A have a V$_{GSR}$ of 78 kms$^{-1}$  ($\sigma$ = 12 kms$^{-1}$), while those in Cancer B have a V$_{GSR}$ of 16.3 kms$^{-1}$  ($\sigma$ = 15 kms$^{-1}$. The  V$_{GSR}$ of HG45 (114$\pm$10 kms$^{-1}$) might argue against a connection with the Gemini Arm, but it may be related to Cancer A,  given the offset in distance between it and the RR Lyrae.

\subsubsection{Sgr apocentre (Feature 3)}

Feature 3 lies near -- but somewhat more distant than -- the apocentre of the leading arm of the Sgr stream. It comprises only one carbon star (HG66) at a distance of 125 kpc, but is surrounded by a number of other tracers. This group of objects is rather diffuse, and appears to extend from the known apocentre of the Sgr arm to a distance of more than 100 kpc. These stars may be part of the Sgr leading arm proper, or they may be a distinct, more distant feature. 

A hint of an extension to the leading arm was seen in \citet{Belokurovetal06}, and the other tracers suggest that the situation at this location is complex. In the central panel of Figure \ref{Fi:multi_plot_lm10_t1_random_S2}, RR Lyrae from \citet{Drakeetal13b} can be seen at this location. Again -- as noted earlier -- these are from the MLS which has limited sky coverage, and the actual number of RR Lyrae at this distance may well be greater.

Further tracers in this region include two giant stars from \citet{DohmPalmeretal01}, which they suggest may have come from an older wrap of the Sgr streams. There are also BHBs  \citep{Clewleyetal05}, six of which lie in two velocity groups: four at V$_{GSR}$$\sim$20 kms$^{-1}$, and two at $\sim$--50 kms$^{-1}$. \citet{Clewleyetal05} argue this represents stars at the turn-around of an elliptical orbit, although as their BHBs do not lie on the model of \citet{Lawetal05}, they do not argue for an association with Sgr. 

Another carbon star, HG57, also lies at a large distance at this right ascension, but sits far to the south of the other tracers. This star is isolated from any known halo structure, and so we assign as ``Unclear".

\subsubsection{LMC}

Star HG32 is some 7.7$^{\circ}$ on the sky  from the centre of the LMC, compared to the tidal radius for the LMC of $\sim$ 17.1$^{\circ}$  \citep{vanDerMarel02}. It has a distance of  $\sim$50kpc, which is also consistent, and this star is most likely a member of the LMC. Likewise, at 12.2$^{\circ}$, HG35 also a probable member of the LMC. HG31 lies 18.1$^{\circ}$ away from the LMC, just beyond the tidal radius, but is 55 kpc distant and may also be associated. \citet{Munozetal06} found stars 22$^{\circ}$ from the centre of the LMC, near the Carina dwarf, and HG31 lies in the direction between the LMC and Carina. It would be valuable to obtain the velocity of the carbon stars to see they are similar to the giants that \cite{Munozetal06} proposes to be a part of the LMC. \citet{Majewskietal09}  found RGB stars out to the limit of their survey -- 23$^{\circ}$ from the centre of the LMC, which they attribute to a ``classical" halo. 
 
If the carbon stars are indeed part of the LMC is it an interesting result, as it would suggest that either the old LMC halo has an intermediate-age component, or that such stars have been stripped from the LMC disk by interactions  \citep{Munozetal06} -- although the latter scenario is challenged by work which indicates the LMC disk is unperturbed, at least  out to a galactocentric radius of 16 kpc \citep{Sahaetal10}.

\subsubsection{The Sgr ``bundle"}

There is a ``bundle" of LPV carbon stars, very close to each other on the sky, (at RA,  Dec, $\sim$ 300$^{\circ}$,--30$^{\circ}$). These lie on the outer region of the main body of the Sgr dSph, although the main part of the Sgr galaxy itself is close to the Galactic plane, and so not within the Catalina Surveys footprint. These are almost certainly Sgr stars. \citet{McConnachie12} gives the half-light radius of Sgr as $\sim$5.7$^{\circ}$, while \citet{Majewskietal03} derive a King fit with a limiting radius of 30$^{\circ}$, which would include these stars.

\subsubsection{Galactic Anti-centre structures (GAS)}

There are a number of known features in the region of the Galactic Anti-Centre, especially north of the Galactic Plane. These include the Anti-Centre Stream (ACS) of \citet{Grillmairetal08}, and the Monoceros Overdensity \citep{Newbergetal02}. Of these only Monoceros has been detected south of the Plane \citep{Meisneretal12,Slateretal14}.

Two of our stars (HG40, HG41) stand out as having velocities inconsistent with the LM10 model. Both are near the Galactic Plane,  are relatively close to the Sun (11, 14 kpc); and have locations, distances and velocities consistent with  of the Anti-Centre Stream.  Another, HG36, although it lies close on the sky to the trailing arm of Sgr, is too close at 12 kpc to be part of the stream. 
All three are consistent with of the Anti Centre Stream as found in \citet{Lietal12}

One carbon star  in the south, HG29, of the Galactic Plane may  belong to the Monoceros Overdensity. It just lies in the region discussed by  \citep{Slateretal14}, at a distance of $\sim$12 kpc.

\subsubsection{TriAnd}

At a location of approximately RA $\sim$30$^{\circ}$ and  Dec $\sim$35$^{\circ}$, we find a small group of carbon stars (HG1, HG7, HG8, HG12, HG14 and HG15) with a distance of about $\sim$20kpc. This region corresponds to the Triangulum-Andromeda (TriAnd) structure, a relatively dispersed stellar system  found in M giants \citep{RochaPintoetal04}, but which  has  also been detected in MS and MSTO stars \citep{Majewskietal04,Martinetal07}  (blue polygon on lower panel of Figure \ref{Fi:multi_plot_lm10_t1_random_S2}). 

The TriAnd feature, as shown in Figure 2  of \citet{RochaPintoetal04}, has a suggestion of two separate regions, and five of the six candidate TriAnd carbon stars all lie in one of these regions, only HG1 in the second. This may indicate that the two regions have different stellar populations, although the small number of carbon stars involved prevents any firm conclusion. The location of our carbon stars to one region of TriAnd is not due to a lack of Catalina Surveys data at low right ascension. 

\citet{Sheffieldetal14} note that \citet{Martinetal07} report two features,  denoted TriAnd1 and TriAnd2, which they find to have distances of 15--21 and 24--32 kpc respectively, with ages of 6--10 and 10--12 Gyr . The first sample carbon stars are more consistent with TriAnd1, from distance, and this is also consistent with the expectations of an intermediate-age population. Only HG1 of these carbon stars has a published velocity (V$_{GSR}$), 69 kms$^{-1}$, consistent with membership of TriAnd1 as determined by \citet{Sheffieldetal14}.

\subsubsection{Hercules-Aquila Cloud (HAC)}

The Hercules-Aquila Cloud (HAC), discovered by  \citet{Belokurovetal07}, is a large and diffuse feature straddling the Galactic plane (black polygon on lower panel of Figure \ref{Fi:multi_plot_lm10_t1_random_S2}). It stretches $\sim$80$^{\circ}$, centred  on l=40$^{\circ}$,  appears to extend either side of the plane by some 50$^{\circ}$, and lies at a distance of 10--20 kpc.  In the north, \citet{Belokurovetal07} have evidence of a V$_{GSR}$ of roughly centred on 180 kms$^{-1}$. 

RR Lyrae have been found that correspond to the HAC by both \citet{Watkinsetal09} and \citet{Sesaretal10a} in SDSS stripe 82. More recently,  \citet{Simionetal14}, also using RR Lyrae as a tracer, find a strong asymmetry in the HAC. It is very prominent in the Galactic Southern Hemisphere, but barely seen in the North. But they do suggest that this is due to interstellar dust reaching to greater Galactic latitudes in the Norther Hemisphere.

HG91,HG102, HG104, HG110 and HG112  are good candidates for this feature, especially we consider a region of the sky somewhat larger than drawn on Figure \ref{Fi:multi_plot_lm10_t1_random_S2}. Two others, HGHG80 and HG85 might be associated, and we tentatively allocate them to the HAC.

\subsubsection{Starkenburg Group 6}

HG52 (RA,Dec $\sim$ 170$^{\circ}$,--18$^{\circ}$) lies off the LM10 model -- as seen on the sky -- far to its south. This star has a distance of $\sim$40 kpc and a velocity of 169 kms$^{-1}$.  It is close to one of the pairs of K-giants (their group 6) from \citet{Starkenburgetal09} on the sky, and has a similar distance and velocity (see short-dashed box in all panels of Figure \ref{Fi:multi_plot_lm10_t1_random_S2}). The Starkenburg pairs are good candidates to be tracing halo substructure due to the improbability of random halo stars sharing location, distance and velocity. The presence of HG52 supports the reality of the group.

HG48 may also be associated with the Starkenburg group 6, as it too has an unusual radial velocity that places it close to both this group and to HG52.

\subsection{Notes on specific stars}

During our analysis, a number of individual stars stood out as worthy of comment:

\subsubsection{HG77 -- SDSS J144945.37+012656.2}
 \label{R_star}
\begin{figure}
\centering
\includegraphics[angle=0,width=85mm]{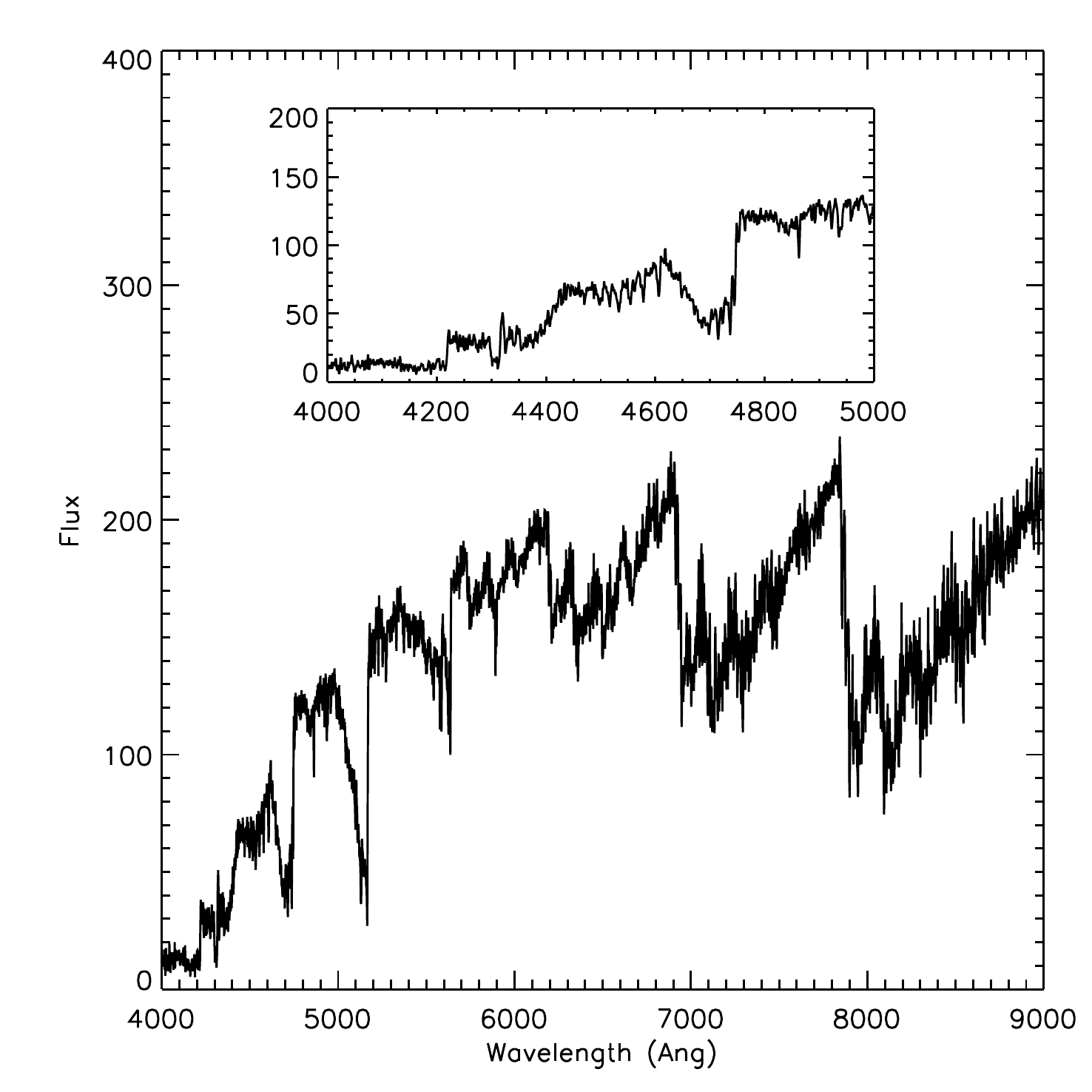}
\caption{Spectrum for star HG77, from SDSS data. The inset shows the blue region of the spectrum, and the CH feature at $\sim$4300\AA . }\label{Fi:sgr_r_star_spectrum_plot}
\end{figure}

One star [HG77] is unusual in that it has a strong CH feature in its spectrum (Figure \ref{Fi:sgr_r_star_spectrum_plot}), is quite blue, with a $(J-K_{\rm s})_{0}$ colour of  0.9 mag, and a short (88 day) period.
Although with a CH feature, this star is probably a C-R star. Comparison with the criteria listed by \citet{Goswami05} -- the sharper Na D feature and  strong CN lines -- are indicative of a C-R star, as is the weakness of H$_{\alpha}$.  

Our belief that this is a late-R type AGB, rather than a CH-type star, is strengthened by its location (assuming it is an AGB star). It position on the sky, and its derived distance (52 kpc) place it in the apocentre of the Sgr leading arm, accompanied by many other carbon stars. 

\subsubsection{Star  S25}

The star  S25,  found in Sextans \citep{Mauronetal04}, has been previously reported by \citet{Sakamotoetal12}, using their own photometry, which has fewer data points than in the Catalina Surveys. We analysed this star in more detail, as a carbon star with such a long period (and hence not old) is not expected in this dwarf galaxy, which is metal-poor and primarily composed of an ancient stellar population \citep{Leeetal09}.
The light curve and initial fit of  S25 are also shown in the Appendix, to be consistent with the full sample. However, it was sufficiently of  interest to warrant further analysis.

 \begin{figure}
 \centering
 \includegraphics[angle=0,width=85mm]{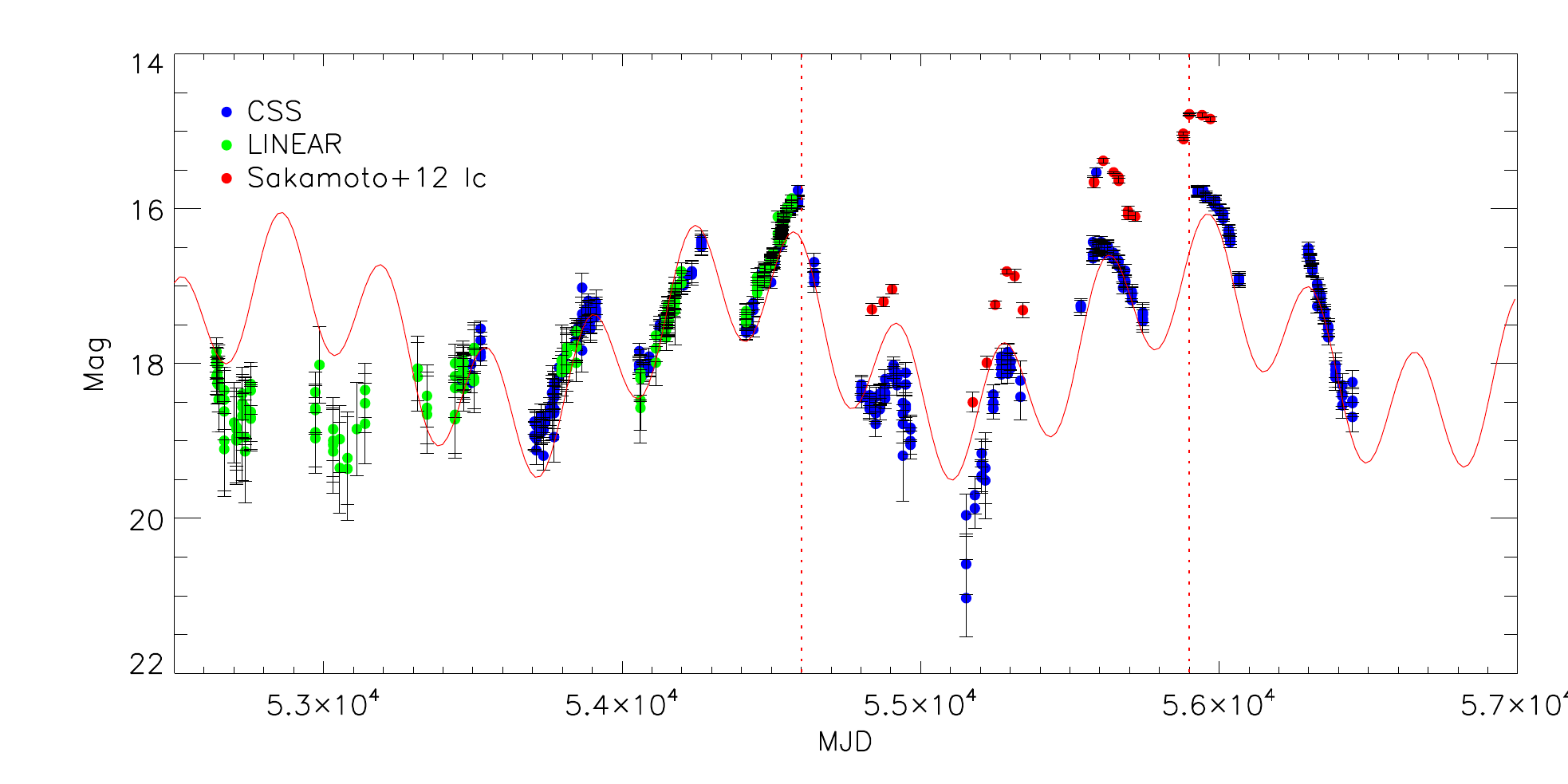}
 \caption{Plot of all LINEAR and CSS data for the star  S25, with a model curve that is the best fit for all the Catalina data (red). The red vertical dashed lines bound the obscuration event. We also show the I-band data from \citet{Sakamotoetal12}.}\label{Fi:sextans_MAG04_allLCs_art_plot_P4}
\end{figure}

The extensive Catalina data have been supplemented LINEAR observations, to reveal some of the behaviour prior to the Catalina data being taken. The LINEAR and CSS data overlap between MJD = 53800 and 54600, and within the range MJD= 54400 and 54600 they both have small errors in both samples. We use this region to determine the offset between the two systems as 0.1 magnitudes. This correction is then applied to the CSS data for the subsequent analysis.

The light curve (Figure \ref{Fi:sextans_MAG04_allLCs_art_plot_P4}) shows it suffers from an obscuration event in the the region from MJD 54,600--55,900 days (between the red vertical lines), with a drop of over four magnitudes. 
A fit to all the Catalina Surveys data (the red curve in Figure \ref{Fi:sextans_MAG04_allLCs_art_plot_P4}) seems to explain the obscuration event  as actually being part of a longer secondary period. However, we can see that this fit does not match the LINEAR data well. Thus we believe that this is a genuine obscuration event from significant mass-loss.

Even we exclude the 'dip', with its larger uncertainties, the total range in amplitude of the combined LINEAR and Catalina Surveys data is still over 3.8 magnitudes. This compares to a range of $\sim$3 mags for S1, (M29 in the Mauron numbering system).  \citet{Sloanetal09} found that M29 in the Sculptor dSph is undergoing substantial mass-loss, and if  S25 is similar it will provide an additional example of the production of carbonaceous dust in a metal-poor environment. Hence, S25 is worthy of observational follow-up. Additional photometry will add to the light-curve data to better understand the nature of the long-term changes in luminosity, i.e. whether they are periodic or not.

\subsubsection{HG2-42: [TI98]1450-1300}

\begin{figure}
 \centering
 \includegraphics[angle=0,width=80mm]{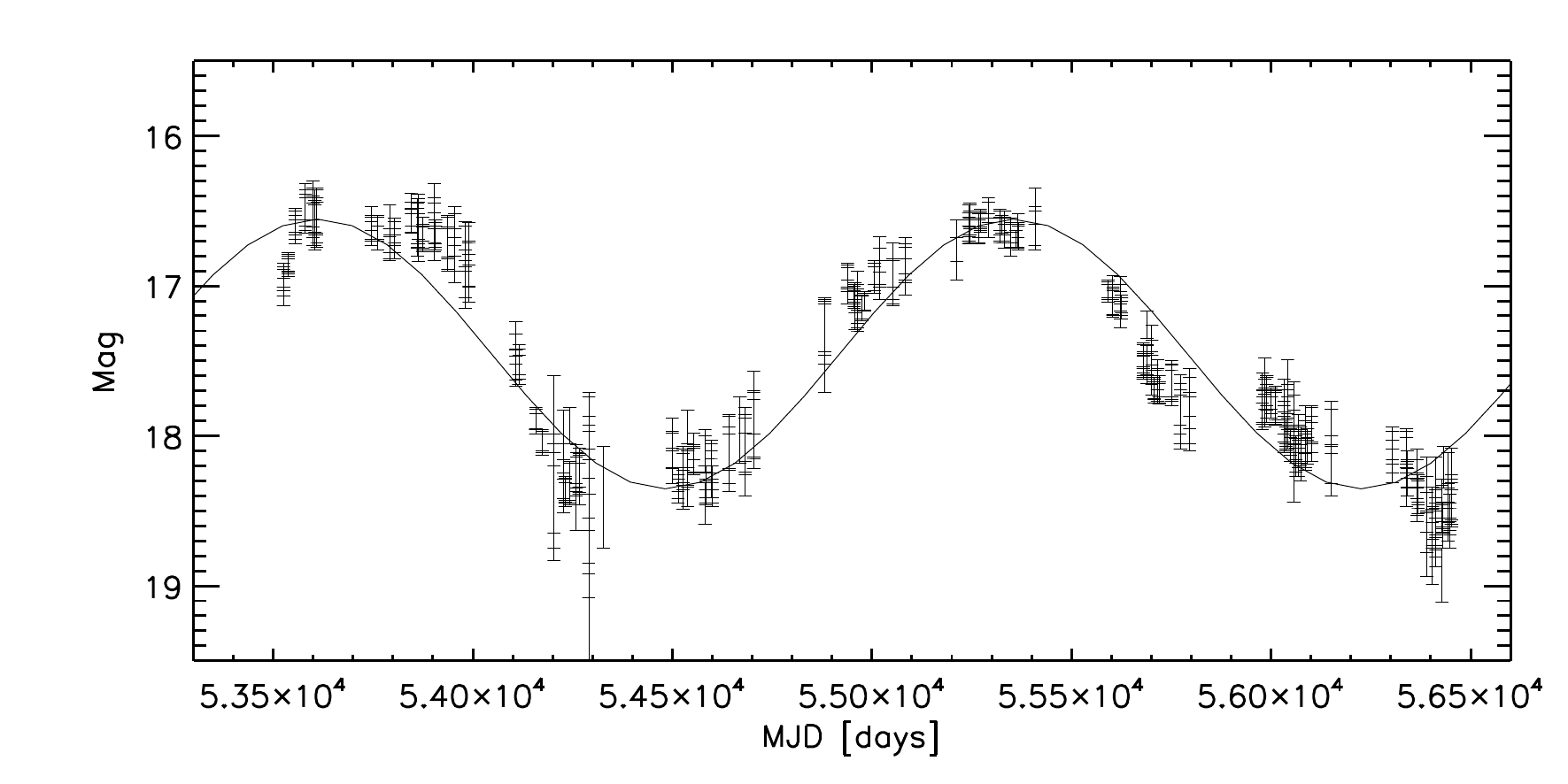}
 \caption{Light curve for HG2-42 with the best Period04 model fit, with a period of 1744 days and an amplitude of 1.8 mag.}\label{Fi:TI981450_1300_LC_plot}
\end{figure}

HG2-42, originally from \citet{TottenIrwin98}, has a  very long period of 1638 days and an amplitude of  1.8 magnitudes (See Figure  \ref{Fi:TI981450_1300_LC_plot}).
We exclude from our first sample, as this period lies far from the region in which we could expect the period-luminosity relation to hold. Hence it is not in  Table \ref{tab:first_sample_catalog}, but we place it in Table  \ref{tab:second_sample_catalog}, using its luminosity to estimate a distance. 

It has been suggested that stars having periods longer than $\sim$1000 days are  massive enough for hot bottom burning (HBB) to occur \citep{Whitelocketal13}. This will result in  carbon being consumed and the carbon star turning into an O-rich star \citep{Feast09}. The carbon stars with the longest period in the LMC are  just below 1000 days \citep{Whitelocketal03}. So this carbon star is quite unexpected. 

\citet{Nannietal13} have models in which a very low metallicity star may become C-rich if it is massive ($M> 5M_{\odot}$), due to a very efficient HBB in which the ON cycle is activated, leading to the depletion of oxygen \citep{Siess10}. However, the V$_{GSR}$ of HG2-42 places it in the Sgr leading arm, and the distance derived using NIR photometry, is also consistent with a location in the Sgr leading arm. Although circumstantial, this does suggest that it may not be a very massive star, but a normal AGB star with a very long period.

This idea may find some support  from \citet{Soszynskietal09a} who find a group of 10 Miras with P$>$1000 and amplitudes in the I-band of $~$4mag. As they exhibit 'regular' variability,  they assume they are O-rich, not C-rich. However,  we found that is not the case for shorter period stars (in that many of our carbon stars show very regular light curves), so maybe there are carbon stars among this group of 10. By extension there may be a population of very long period carbon Miras that have hitherto not be found due to their scarcity.

\section{Carbon stars without periods}
\label{noPeriods}

For many stars in our input lists (see \S \ref{data}) we do not have periods. This can occur for a number of reasons. In many cases the star lies outside the Catalina Surveys footprint, in others the Catalina light curves have too few (or too noisy)  data-points to obtain a period fit, or show no obvious variability. And finally, in some cases, the light curves exhibit behaviour that  prevents a period fit being made. This is often the case for very bright stars in which the Catalina data are saturated. For these stars we use the more common method of distance estimation from NIR photometry alone. 

We were encouraged to do so when we found that the distances obtained by Mauron et al. using $K_{\rm s}$-band magnitudes  and $(J-K_{\rm s})$ colours were relatively consistent with those derived by using periods. This consistency may, of course, only be the case for those AGB carbon stars which have well-behaved light curves, or at least are shown to be variable. Therefore although we include the stars without periods in this section of the paper, we caution against drawing very strong conclusions from them. We include a list of them here, to encourage further observations of those with weak light-curves, or none at all (i.e. those not in the Catalina footprint)

We remove stars classified as CH-type by \citet{TottenIrwin98}, and those in the list of CH-stars of \citet{Bartkevicius96}, as these will not fit the relation that is true of AGB stars. Of course in some of our sources, an object is reported as a carbon star, without further details as to the sub-type, so one must be further cautious of the results from the second sample. To address this problem, we also exclude those stars for which $(J-K_{\rm s})_{0} < 1.2$, as such blue colours are typical of CH-stars (e.g. see Fig. \ref{Fi:all_JH_HK_plot_unc_catagories}), unless they are noted as C(N) by \citet{GigoyanMickaelian12}. Finally, we remove stars for which $K_{\rm s} <  6$, as these will all be in the disc. This new sample, which we term the ``second sample" is shown in Table \ref{tab:second_sample_catalog}.

We now estimate the distances of the second sample stars following the approach of Mauron et al. They exploit relations between absolute $K_{\rm s}$ --band magnitude and J and K photometry, derived from a sample of LMC carbon stars.
The equation(s) used by Mauron et al. to estimate distances are not given, so we re-derive them from the data given in their papers. We use the reddening corrected NIR photometry and derived absolute $K_{\rm s}$-band magnitudes from \citet{Mauronetal07b} and \citet{Mauron08} to obtain  two fits:

\begin{equation}
\begin{split}
 M_{K_{S}} = 1.550\times(J - K_{S})_{0} - 4.927    \mbox{ for } 1.2 < (J - K_{S})_{0} < 1.8 
 \end{split}
 \label{eqn:jk_to_Mk_1}
\end{equation}    
\begin{equation}
\begin{split}
 M_{K_{S}} = 0.414\times(J - K_{S})_{0} - 8.429    \mbox{ for }  (J - K_{S})_{0} > 1.8 
 \end{split}
 \label{eqn:jk_to_Mk_2}
\end{equation}

\begin{figure}
 \centering
 \includegraphics[angle=0,width=85mm]{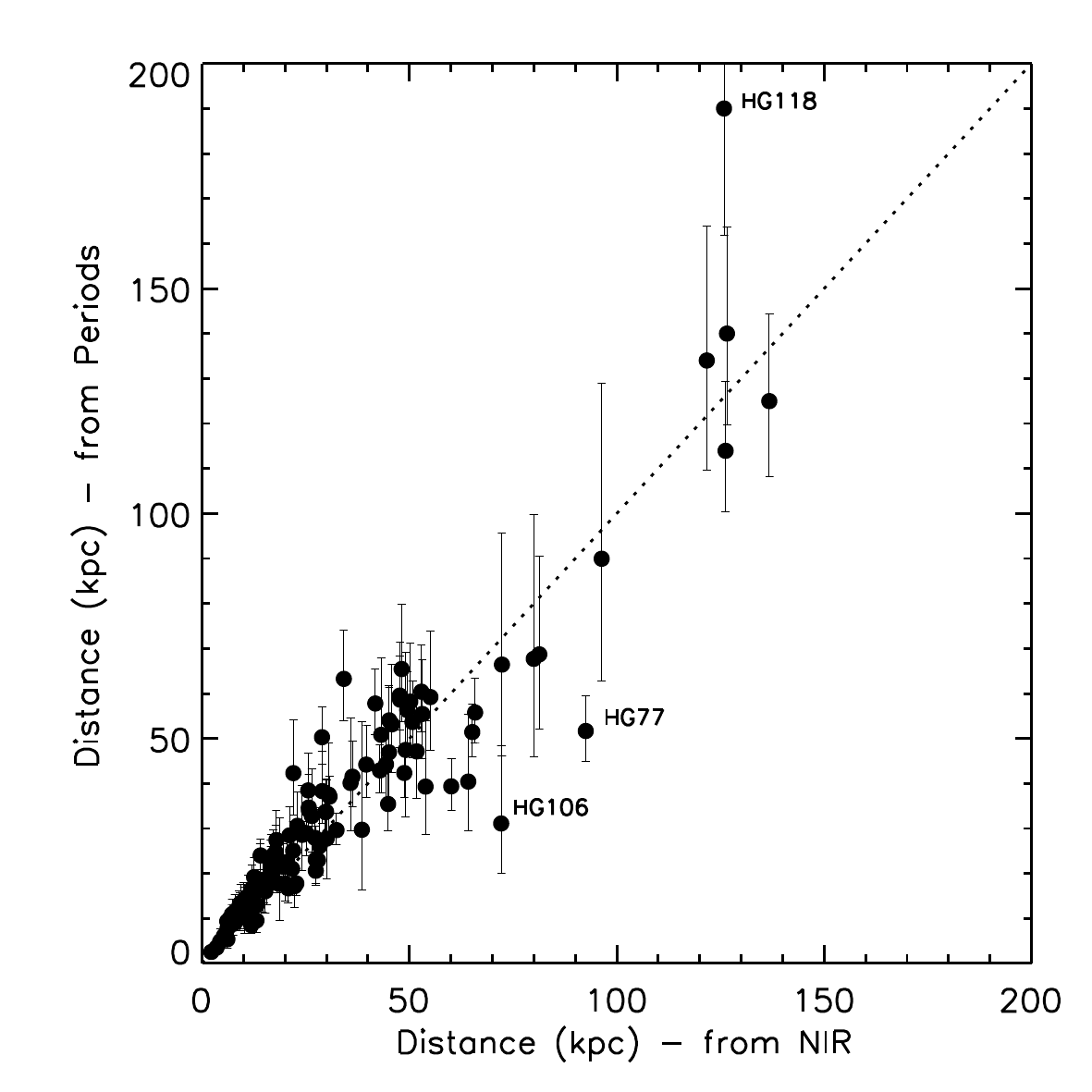}
 \caption{Plot of our distances, derived from variable periods, against their distance determination from NIR photometry. There is a reasonable correlation between the two distance measures for this sub-sample. The line of equality (dotted) is shown to guide the eye. }\label{Fi:plot_period_NIR_distances}
\end{figure}

We can test these in more detail by comparing the distances obtained from periods, against those from NIR photometry alone, for our first sample (see Fig. \ref{Fi:plot_period_NIR_distances}). Given the distance errors the correlation is remarkably good. A few outliers are outside of the estimated distance uncertainties: HG118, HG77 and HG106. The first two of these have given us problems before: HG118 has a weak light curve and was found to be an outlier in Fig. \ref{Fi:multi_dist_JK_period_plot_errs_S2}, HG77 is one of the bluer stars with a very short period. HG106 also has a weak light curve. 

Equations \ref{eqn:jk_to_Mk_1} and \ref{eqn:jk_to_Mk_2} are then used to derive  estimates for $M_{K_{\rm s}}$ and then distances for the stars in our second sample. This approach effectively assumes that the carbon  stars are all on the C sequence. Hence any that are, in fact, C$^{\prime}$ or D will have incorrect distances. But without period and amplitude data, it is difficult to know which sequence our sample stars belong to. The exception to this occurs if the $(J-K_{\rm s})$ colour is greater $\sim$2, in which case it is reasonable to assume that it lies on the C sequence (see Figure \ref{Fi:soszynski_period_JK_plot_S2}). Some stars in this sample have published velocities, which are also given in Table \ref{tab:second_sample_catalog}, and can increase our confidence of whether they lie in known halo substructures. 

The good correlation in the two methods of distance estimation seen in Fig. \ref{Fi:plot_period_NIR_distances} may, of course, be a result of the nature of sub-sample of carbon stars that are found in the first sample. That is, the first sample have variability properties that make the use of NIR photometry appropriate. Many stars in the second sample, however, show either no, or erratic, variability. For this reason, we repeat our note of caution from above. Although we believe that the distances are indicative for the sample as a whole, individual stars may have estimated distances that are incorrect.
Due the additional uncertainties of this sample, we do not attribute the second sample stars to substructure as done for the first (LPV) sample.

\begin{figure*}
 \centering
 \includegraphics[angle=0,width=160mm]{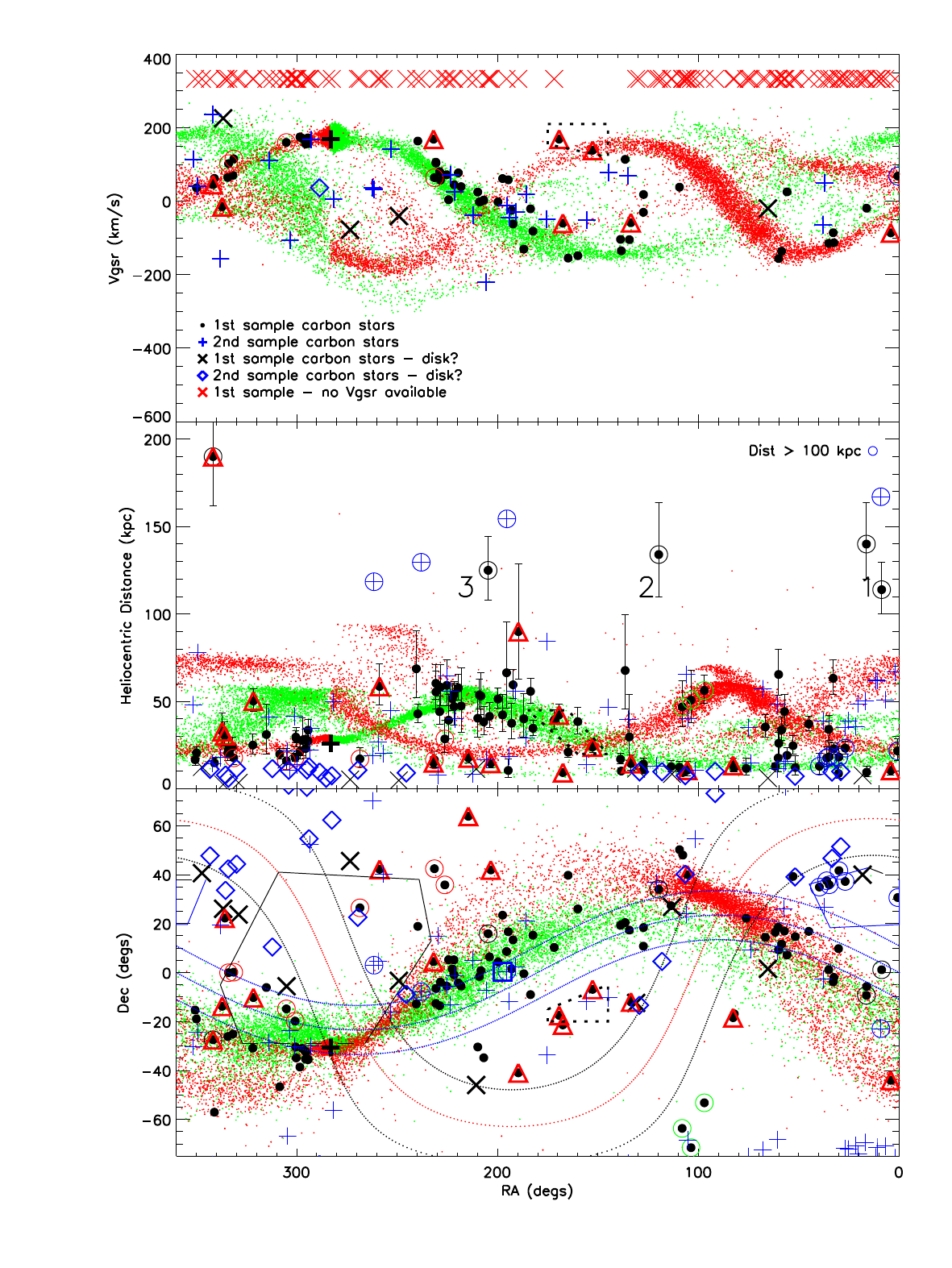}
 \caption{A similar plot as Figure \ref{Fi:multi_plot_lm10_t1_random_S2}, but including the second sample (with no periods). The symbols indicating possible feature membership of the first sample are the same as those shown in Figure \ref{Fi:multi_plot_lm10_t1_random_S2}. The RR Lyrae data of \citet{Drakeetal13b} are excluded to for clarity. }\label{Fi:multi_plot_lm10_t1_second_random_S2}
\end{figure*}

Looking at the position of this second sample in velocity, distance and location (Figure \ref{Fi:multi_plot_lm10_t1_second_random_S2}) a few results are immediately apparent. 
Firstly, many more stars (shown as asterisk symbols in the figure) than our ``first" sample have positions that place them close to the MW disk. This is because many of these stars are too luminous to obtain good light-curves in the Catalina Surveys; they become saturated, making curve fitting impossible.

We also see many stars that are almost certainly in the LMC (either side of RA, Dec $\sim$ 90$^\circ$, --70$^\circ$) and  SMC (surrounding RA, Dec $\sim$ 15$^\circ$, --70$^\circ$). Most of these have no light-curves as they are in the region excluded from the Catalina Surveys due to crowding in the imaging. Others are beyond the southern declination limit of the Surveys. The second sample carbon stars nearby to the SMC reach to a projected sky distance of 4.3$^{\circ}$, equivalent to $\sim$4.7 kpc, a value consistent with the recent results of \citet{NoelGallart07}, who found evidence of an intermediate-age population up to 6.5 kpc from the centre of the SMC. So these stars are consistent.

One interesting feature are the several ``second sample" carbon stars at distances $>$100 kpc, and that lie close to the features identified in the first sample. One lies close to the A16/Pisces Overdensity, at a distance of over 150 kpc, in ``Feature 1".  Three second sample stars lie beyond the apocentre of the Sgr leading arm,  from 100--160 kpc in ``Feature 3"; two on the LM10 model (on the sky), and the third is not far from it. 

Finally, HG2-26 is particularly interesting. Its location on the sky (RA, Dec $\sim$ 106$^\circ$, 24$^\circ$), and its estimated distance of 65 kpc would place it in the region of the trailing arm in which no other tracers have been found. It is in the Catalina data, but with too few data points to obtain a good light curve. If we use the same distance uncertainty used for the first sample, based on the $(J-K_{\rm s})$ colour, this star could be at a distance from $\sim$47--90 kpc. The lower value would be consistent with the LM10 model, while the upper value would locate it in the Gemini Arm (which of course may be the true Sgr trailing arm). The Catalina data for this star range over 3 magnitudes in the V-band, suggesting it also suffers from episodes of obscuration, and that the uncertainty in the distance may be greater still. 

\section{Discussion}
\label{discussion}

In the following discussion we focus on the results from the first sample (Table \ref{tab:first_sample_catalog}), rather than those for which distances have only been obtained from magnitude (Table \ref{tab:second_sample_catalog}), due to the far greater uncertainties associated with second sample. 

\subsection{Sgr carbon stars}

A major finding of this study is that  the majority of the first (LPV)  sample can be understood as being part of the Sgr streams, being consistent in location, distance and velocity (where available). This highlights the importance of the Sgr accretion in the history of the MW halo, especially as seen through intermediate-age tracers, such as carbon stars. This result suggests that there has been little recent merger activity of massive dwarf galaxies onto the MW, except for the Sgr dwarf, and any progenitors of the features discussed above. Of course, our LPV sample in particular, avoids the Galactic Plane, so any accretion events occurring in the Plane would not be detected. 

This finding is consistent with the view that the MW has had a quiescent history \citep{Hammeretal07}. Specifically,  accretion into the field halo by satellites with extended SFH has been unimportant for the last $\sim$8 Gyr \citep{Wyse08}. However, older, more metal-poor satellites may well have been accreted, or are still being so, in the more distant halo.

However, we find -- as many others have using a variety of tracers -- that the LM10 model is not a good fit to the data along its full spatial extent. There does indeed appear to be a lack of carbon stars in the trailing arm north of the Galactic Plane, where the model suggest a relatively high stellar density (and as noted above). Only HG2-26 is plausible as being consistent with the LM10, but  -- as noted earlier -- it could equally be part of the Gemini Arm. The lack of the trailing arm in the region predicted by the LM10 model has been reported before. For example, \citet{Drakeetal13a} have also found that their RR Lyrae show little of the LM10 trailing arm at RA = 110$^{\circ}$ at a distance of $\sim$40 kpc. 

But our data are also ambiguous with regards to the leading arm south of the Plane. Although we have carbon stars that we attribute to Sgr in this region, they are few. Moreover, many of these may in fact be part of the Monoceros Overdensity. Such a result, as with the trailing arm in the north, is not totally new. Other tracers have either not been detected in the leading arm \citep[e.g.][]{Sesaretal11b,Ruhlandetal11,Sirkoetal04,Carrelletal12}, or have been only  detected to a minor extent \citep{Shietal12}, in this region of the sky.

\subsection{Distant carbon stars}
\label{distant_stars}

Another result  is the presence of a handful of carbon stars at very large distances, greater than 100 kpc. Feature 1 is almost certainly part of the Pisces Overdensity, and Feature 2 is very likely tied up with the ``Gemini Arm".  Feature 3 is more difficult to explain. As these stars lie on the Sgr orbital plane, one plausible explanation is that they arise from older wraps of the Sgr streams, lost on previous orbits, especially if the Gemini Arm is indeed the actual apocentre of the Sgr trailing arm.

The idea that the most distant carbon stars may be older wraps of Sgr is consistent with \citet{Shietal12} who argues that it lost an older, more metal-poor, component in the older wraps, and then the younger more metal-rich component in the youngest wraps (the arms shown in Figures \ref{Fi:multi_plot_lm10_t1_random_S2} and \ref{Fi:multi_plot_lm10_t1_second_random_S2}). For example,  HG39 has a period of 210 days and  HG66 of 149 days. The shorter periods are indicative of a greater age than the majority of the sample.

An alternative scenario  \citep{Carlinetal12b}  suggests that both the Pisces Overdensity and the Gemini Arm can be best explained as the tidal features of the progenitor of the Virgo Overdensity; a feature that is located at a distance of $\sim$ 15 kpc, over a large area of sky of the SDSS footprint above the Galactic Plane. Curiously, we have no obvious candidate carbon stars for this structure, whereas we have examples in both the Pisces Overdensity and Gemini Arm.

An obvious issue concerns the reliability of the distance determination for the handful of very distant carbon stars. One concern might be that they lie in the plane of the Sgr orbit because they may actually be in the main arms of the  streams (i.e. up to 60 kpc) and are just projected to larger distances by failings in our approach. However, these stars have well-behaved light curves and bluish colours. As  discussed above, the spectra of four of the most distant stars are N-type. We doubt that the NIR data have been affected by obscuration events that may confound the P-L relationships; there is no evidence that these suffer erratic and dramatic changes in their light-curves (as seen in S25), in which the 2MASS NIR data would not be representative of the period covered by the Catalina Surveys light curves. The long duration of the Catalina Surveys, some seven years, is invaluable in assessing this.

By contrast, as we have seen above, other carbon stars were difficult to assign cleanly to the C or C$^{\prime}$ sequences, and some proportion of them may be closer, if they are indeed C$^{\prime}$ stars. However, as we are interested in using these stars as potential tracers of distant structure, this issue can be addressed through follow-up observations. Deep imaging in the regions near these objects  should reveal any halo features at the indicated distances.

\subsection{Other Substructure}

In addition to the many carbon stars in Sgr, and the distant stars which are in halo structures which may or may also not be associated with Sgr, we found LPV carbon stars in other known substructures. In particular, we find a clear group in Triangulum-Andromeda feature, and several that are likely in the GAS. Less clear are the stars we attribute to the HAC, primarily as the true extent of  this feature is still uncertain.

The most interesting feature, for our purposes, are the two carbon stars near to the Starkenburg group 6. Even though we assign these carbon stars as ``Unclear", the grouping in velocity, distance and location strongly suggest that this is a genuine halo feature. Follow-up imaging in this region should provide a definitive answer. This is the best candidate for ``new" substructure.

Finally, there are several isolated carbon stars which are not close to any known substructure. These may be contaminants (e.g. non AGB stars; possibly HG107, for example) that have fallen into our sample due to our generous selection approach, or they may indicate unknown halo sub-structure. What is remarkable is that there are so few. This is even more so when one considers that some of the stars that are not easily attributed to a substructure are also those with the most uncertain light curves (e.g. HG64 and HG72).

\subsection{Period distribution of the carbon stars}

One of our  conclusions, that the majority of the carbon stars are associated with Sgr,  is quite different from that of \citet{BattinelliDemers12}. They argue that the halo carbon stars were too young to be from the Sgr streams,  based on the distribution of periods for their sample. They argued that their sample has too many 'shorter' period LPVs than would be expected if the Sgr dSph were a relatively old to intermediate-age system.  

\begin{figure}  
 \centering
 \includegraphics[angle=0,width=85mm]{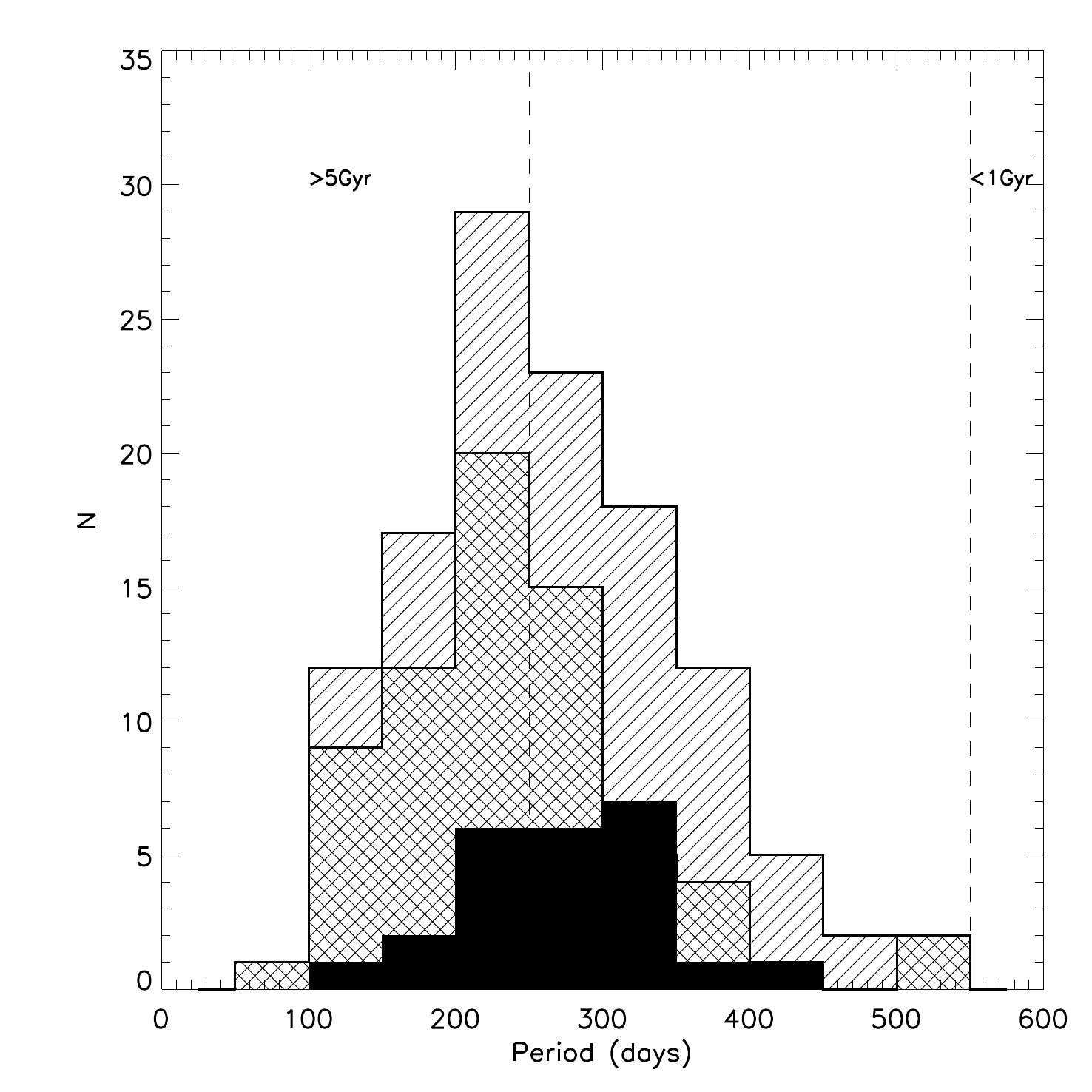}
 \caption{Histogram of periods for all our sample (hatched), those attributed to Sagittarius (labelled 'Sgr' and 'Sgr:' in the tables; cross-hatched), and those with a declination  less than --20$^{\circ}$ (solid fill), comparable to the sample of \citet{BattinelliDemers12}. The vertical dashed lines delineate the ages adopted by  \citet{BattinelliDemers12}. }\label{Fi:histo_sample_periods_P4_S2}
\end{figure}

But we suggest that their conclusion is  based on a spatially biased sample. In Figure \ref{Fi:histo_sample_periods_P4_S2}, we plot the distribution of periods for our first sample (single hatched), and those we attribute to Sgr (cross-hatched). The peak for the Sgr stars is at 200 -- 250 days, not the 250 -- 350 days found by \cite{BattinelliDemers12}. However, they had an implicit selection, due to observational constraints, of a declination less than  --20$^{\circ}$. If we apply this same selection to the first sample (solid fill), we too find a shift in the peak to the longer periods, 300--350 days, consistent with \cite{BattinelliDemers12}. This shift in period distribution is most likely due to the large proportion in this sample of carbon stars that lie close the main body of Sgr. There is known to be an age-metallicity gradient along the streams \citep{Chouetal07,Shietal12}, so the  stars near Sgr have been recently stripped from the main body and so represent a younger population than the bulk of the Sgr stream stars.

\begin{figure}
 \centering
 \includegraphics[angle=0,width=85mm]{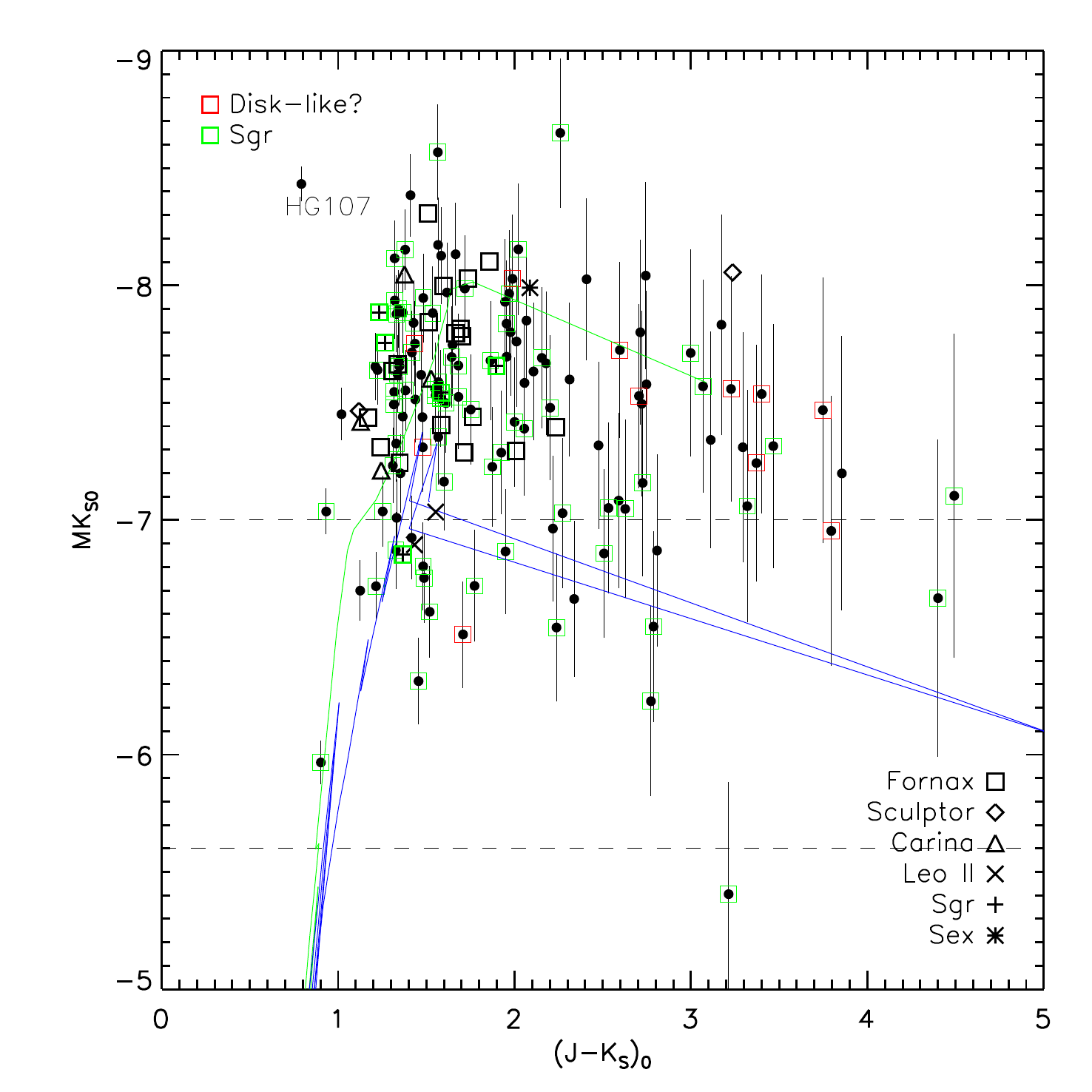}
 \caption{Plot of absolute $K_{\rm s}$-band magnitude against (J-K$_{S}$) colour. Red squares indicate stars that lie in our 'disklike' region (see text), and green squares indicate stars that are in Sgr. Also shown are isochrones from the tool CMD 2.5 \citep{Marigoetal08}, for a metallicity (Z) of 0.004 ([Fe/H] = --0.7) and ages of 1.5 (green) and 2.0 (blue) Gyr, using the bolometric corrections of \citet{Loidletal01} and dust compositions of 0.85 AMC (amorphous carbon) +  0.15 SiC for C-rich stars and 0.6 AlOx+0.4 silicate for the O-rich stars \citep{Groenewegen06}. The horizontal dashed lines shows the tip of the red giant branch (TRGB) value of $M_{K_{\rm s}}$ for [Fe/H] = 0.0 (upper) and [Fe/H] = --2.5 (lower) from \citet{Ferraroetal00}. The outlier HG107 is labelled.}\label{Fi:all_JK_Mk_disky_plot_P4_S2}
\end{figure}

However, the age of the carbon stars is unclear. The short periods of many of the carbon stars should indicate an older age.  A plot of $M_{K_{\rm s}}$ against $(J-K_{\rm s})$ colour (Figure \ref{Fi:all_JK_Mk_disky_plot_P4_S2}) shows where the LPV samples sit in relation to the isochrones of  \citet{Marigoetal08}\footnote{http://stev.oapd.inaf.it/cgi-bin/cmd}. The majority of the sample lies between the isochrones (with a metallicity of [Fe/H] = -- 0.7, an average value for the Sgr body and arms,  \citealt{Chouetal07}) for 1.5 and 2.0 Gyr, suggesting that the sample is relatively young. Alternatively, we could be seeing the consequences of the wide range in metallicity ($\sim$ 1 dex) observed in Sgr and its arms.  The total sample looks very similar in distribution to the carbon stars in the Fornax dSph, which had a fairly continuous star formation, from 14 Gyr ago until as recently as 0.25 Gyr \citep{GrebelStetson99,deBoeretal12}. It is possible that the process of accretion triggered star-formation in the Sgr dSph over the last few Gyr created both a spread in metallicities and ages, or that it contained a range of populations at the time of accretion.

It is not obvious why this inconsistency between the ages implied by the period of variability and the colour-magnitude diagram occurs, although there are still unsolved issues in  AGB evolution, and the models may require further refinement. Recently \citet{Girardietal13} found that, from their models, the numbers of TP-AGB stars are ``boosted" by a factor of $\sim$2 for ages from 1.57 to 1.66 Gyr. This may help explain the the presence of so many of our sample (and the MW satellite carbon stars)  seen in Figure \ref{Fi:all_JK_Mk_disky_plot_P4_S2}, and which sit between the 1.5 and 2.0 Gyr isochrones.

 \section{Summary}
 \label{summary}
 
We have revisited the use of carbon stars as tracers of halo substructure, with a particular focus on the Sagittarius  dSph tidal streams. 

Carbon stars on the AGB have always promised the opportunity to trace intermediate-age stellar populations to large distances in the MW halo\footnote{However, the need for caution is highlighted by the recent findings of \citet{Feastetal13} who found a carbon Mira  with a period of 551 days in the globular cluster Lynga 7. Despite its long period, it must be old, and they suggest it is the result of a stellar merger.}. 
Any such features will likely be the remnants of disrupted satellite galaxies. The potential of these stars has been difficult to realise as they can have a wide range of luminosities, making distance determination imprecise. By using variable carbon stars, we exploit  period-luminosity relations to obtain new estimates of their distances. Our own sample suffers from the inherent uncertainty in our NIR photometry, due to it being single epoch only, but in many cases these distance uncertainties are comparable to those found in the use of other tracer populations.

This study been made possible by the availability of the recently released Catalina Surveys Data Release 2 light curves. These cover a large part of the sky and have data over a duration of seven years, allowing analysis of stars whose periods can be more than a year. The relatively large number of carbon stars that we have assembled from the literature means that those with inadequate light-curves can be excluded from our ``first sample", such that we have confidence in the results from this sample of LPV carbon stars. 

Almost all  of the stars in our ``first sample" can be allocated to  known substructures. The majority can be reasonably assigned to the tidal streams of Sagittarius. Other features that possess AGB carbon stars are the Hercules-Aquila Cloud, the Galactic Anti-centre streams, the LMC and Triangulum-Andromeda. 

We also find that the LM10 model is consistent with many of our Sgr carbon stars. However, we also find two possible regions of failure in the model. There is a distinct lack of carbon stars in the trailing arm north of the Galactic Plane (except for the highly-schrouded HG2-26 and its associated large uncertainty in estimated distance), and evidence of the leading arm south of the Plane is also weak. Due to their location near to the Galactic plane (with the consequent crowding and high extinction), it is difficult to draw very strong conclusions, but these apparent carbon star deficiencies remain a challenge for the model. The regions involved are also confused by the presence of other known substructures: the Anti-Centre Streams, Monoceros, and the Hercules-Aquila Cloud. Future observations to obtain velocities for many of these carbon stars may be valuable in identifying their parent. As crucial will be data that can probe closer to the Galactic Plane than hitherto, as it is in this region that the fate of both the leading and trailing arms should become clearer.

Despite two decades of study, the Sgr accretion still poses many questions. The bifurcation is still unexplained, and the Gemini Arm has been suggested as related to either the Sgr dSph or to the Virgo Overdensity. It is no surprise that \citet{Koposovetal13} remark that the `` Sagittarius (Sgr) dwarf galaxy remains a riddle, wrapped in a mystery, inside an enigma".

Carbon stars provide new probes in the far halo which may point to solutions to these problems. They are too few in number to say much on their own. But if the distant halo carbon stars are representatives of a wider stellar population then deep imaging of the regions around these stars may reveal new stellar substructure, and its nature if present. In particular, imaging of the regions around the very distant carbon stars, and those few that lie away from known substructure would be invaluable. 

Notably, we find that the Trianglum-Andromeda feature (TriAnd) stands out prominently in our data, and we strengthen the likelihood that the Starkenburg group 6 indicates a {\it bone fide} halo feature. Very few of our first sample stars are located in regions where no known substructure is thought to exist.  One aim of this paper is to identify possible targets for follow-up, hence we have taken a generous approach to the AGB star classification, as they add little overhead to follow-up, yet may indicate substructure which will show up in deep imaging. The small number of stars with no attribution facilitates this. 

We have also highlighted several stars which possess unusual properties. Further observations of these objects (e.g. photometric monitoring or high-resolution spectroscopy) might throw new light on them, and thus on the carbon star population more generally.

In conclusion, we suspect that almost all of major  intermediate-age substructure in the Milky Way  have been found already. There may, of course,  be  features hiding behind the region of the Galactic Plane or towards the Galactic Poles, which are not properly covered by our study. The carbon star population has also shown that it can still surprise, with the complex nature of these objects still able to provide examples that fit no easy classification.  But the the luminosity of the AGB population still makes them one of the best probes of the MW  halo.

\section*{Acknowledgments}

This work was supported by Sonderforschungsbereich SFB 881 "The Milky Way System" 
(subprojects A2 and A3) of the German Research Foundation (DFG).

The Catalina Sky  Surveys are funded by the National Aeronautics and Space Administration under Grant No. NNG05GF22G issued through the Science Mission Directorate Near-Earth Objects Observations Program.  The CRTS survey is supported by the U.S. National Science Foundation under grants AST-0909182.  The LINEAR program is funded by the National Aeronautics and Space Administration at MIT Lincoln Laboratory under Air Force Contract FA8721-05-C-0002. 
This publication makes use of data products from the Two Micron All Sky Survey, which is a joint project of the University of Massachusetts and the Infrared Processing and Analysis Center/California Institute of Technology, funded by the National Aeronautics and Space Administration and the National Science Foundation.
This research has also made very extensive use of the Simbad and Vizier databases operated at CDS, Strasbourg, France. 

We gratefully acknowledge the anonymous referee, whose close reading and insightful comments improved both the form and content of this paper.

\clearpage

\input{first_sample_latex_table_S2_V2.txt}

\newpage
\input{ second_sample_latex_table_S2_V2.txt}

\clearpage
\newpage
\input{satellite_latex_table_S2_V2.txt}

\clearpage

\label{lastpage}

\end{document}

%% file: first_sample_latex_table_S2_V2.txt
\begin{table*}
 \centering
\begin{minipage}{160mm}
 \caption{Catalogue of variable carbon stars. The coordinates are given in decimal degrees (J2000), the (reddening corrected) NIR magnitudes and colour are from 2MASS data. The periods and amplitudes (in Catalina V band) of variability are derived from fits using Period04.}\label{tab:first_sample_catalog}
 \vspace{2pt}
\renewcommand{\arraystretch}{1.3}
\begin{tabular}{@{}lrrrrrrrrrlll@{}}
\hline
 ID &  RA  & Dec  & K$_{S0}$ & (J-K$_{S}$)$_{0}$ & MK$_{S0}$ &  R$_{\odot}$ & V$_{GSR}$ & Period & Amp & Seq & Ref & Feature\\
   &  degs & degs &  &  &  & kpc & km/s & days & mags & & & \\
\hline
HG1 & 0.7535 & 30.6399 & 8.88 & 1.98 & -7.78 & 
$          22\substack{+           7\\-           5}$ &           69 & 
         305 & 0.88 & C & 1 & TriAnd \\
HG2 & 4.2325 & -44.0113 & 7.07 & 2.74 & -8.04 & 
$          11\substack{+           4\\-           3}$ &          -86 & 
         442 & 2.18 & C & 17 & Unc \\
HG3 & 8.7696 & 1.1465 & 13.27 & 1.33 & -7.00 & 
$         114\substack{+          15\\-          14}$ & ... &          168 & 
0.60 & C & 18 & A16 \\
HG4 & 16.2505 & -5.6677 & 8.36 & 2.24 & -6.52 & 
$          10\substack{+           4\\-           3}$ &          -19 & 
         181 & 0.71 & C & 1 & Sgr: \\
HG5 & 16.3375 & -9.2952 & 13.28 & 1.02 & -7.44 & 
$         140\substack{+          24\\-          20}$ & ... &          188 & 
1.06 & C & 2 & A16: \\
HG6 & 18.2353 & 39.9958 & 6.71 & 1.44 & -7.74 & 
$           8\substack{+           1\\-           1}$ & ... &          249 & 
0.60 & C & 18 & Disk \\
HG7 & 26.9012 & 37.2081 & 8.87 & 1.62 & -7.95 & 
$          23\substack{+           4\\-           4}$ & ... &          140 & 
0.18 & C' & 18 & TriAnd \\
HG8 & 30.0370 & 41.6299 & 8.63 & 2.18 & -7.64 & 
$          18\substack{+           7\\-           5}$ & ... &          306 & 
0.96 & C & 3 & TriAnd \\
HG9 & 30.2339 & 9.7598 & 7.15 & 3.07 & -7.55 & 
$           9\substack{+           4\\-           3}$ & ... &          392 & 
1.82 & C & 18 & Sgr \\
HG10 & 32.5503 & -1.9609 & 10.00 & 1.48 & -6.79 & 
$          23\substack{+           4\\-           3}$ &         -113 & 
         160 & 0.93 & C & 18 & Sgr \\
HG11 & 32.8786 & -3.8289 & 10.44 & 1.57 & -8.56 & 
$          63\substack{+          11\\-           9}$ &          -86 & 
         184 & 0.22 & C' & 16 & Sgr \\
HG12 & 34.8622 & 35.8497 & 8.50 & 3.17 & -7.81 & 
$          18\substack{+           8\\-           6}$ & ... &          460 & 
1.46 & C & 19 & TriAnd \\
HG13 & 35.0979 & 1.1775 & 9.82 & 1.43 & -7.83 & 
$          34\substack{+           8\\-           6}$ &         -115 & 
         124 & 0.21 & C' & 1 & Sgr \\
HG14 & 36.1332 & 37.4924 & 8.74 & 2.72 & -7.48 & 
$          18\substack{+           6\\-           5}$ & ... &          337 & 
1.58 & C & 19 & TriAnd \\
HG15 & 39.7703 & 34.9187 & 8.37 & 3.86 & -7.18 & 
$          13\substack{+           7\\-           5}$ & ... &          425 & 
1.84 & C & 19 & TriAnd \\
HG16 & 45.0472 & 16.8278 & 10.21 & 1.22 & -7.55 & 
$          37\substack{+           4\\-           4}$ & ... &          220 & 
0.50 & C & 18 & Sgr \\
HG17 & 51.7497 & 14.6658 & 8.12 & 3.29 & -7.22 & 
$          12\substack{+           7\\-           4}$ & ... &          372 & 
1.63 & C & 4 & GAS: \\
HG18 & 52.8899 & 39.2472 & 9.02 & 1.48 & -7.90 & 
$          25\substack{+           6\\-           5}$ & ... &          278 & 
0.25 & C & 14 & Sgr \\
HG19 & 55.8700 & 7.1830 & 8.26 & 1.38 & -8.10 & 
$          19\substack{+           4\\-           4}$ &           25 & 
         297 & 0.97 & C & 1 & Sgr: \\
HG20 & 57.1172 & 16.9510 & 11.00 & 1.87 & -7.13 & 
$          44\substack{+          10\\-           8}$ & ... &          224 & 
1.06 & C & 18 & Sgr \\
HG21 & 58.5677 & 11.6057 & 10.15 & 1.32 & -7.40 & 
$          34\substack{+           7\\-           6}$ &         -136 & 
         211 & 0.18 & C & 1 & Sgr \\
HG22 & 59.9831 & 9.3178 & 10.03 & 2.53 & -6.97 & 
$          26\substack{+           9\\-           6}$ & ... &          255 & 
1.51 & C & 19 & Sgr \\
HG23 & 60.0554 & 9.2858 & 11.18 & 1.34 & -7.81 & 
$          65\substack{+          14\\-          12}$ &         -156 & 
         124 & 0.17 & C' & 1 & Sgr: \\
HG24 & 60.2905 & 18.4690 & 9.13 & 4.49 & -6.98 & 
$          18\substack{+          15\\-           8}$ & ... &          501 & 
2.14 & C & 4 & Sgr: \\
HG25 & 61.7035 & 16.4718 & 7.90 & 1.35 & -7.51 & 
$          13\substack{+           2\\-           2}$ & ... &          231 & 
0.25 & C & 18 & Sgr \\
HG26 & 65.3635 & 1.4871 & 6.38 & 3.23 & -7.52 & 
$           6\substack{+           3\\-           2}$ &          -20 & 
         411 & 1.81 & C & 1 & Disk \\
HG27 & 66.6616 & 14.4210 & 11.03 & 1.77 & -6.54 & 
$          35\substack{+           7\\-           6}$ & ... &          169 & 
1.20 & C & 18 & Sgr \\
HG28 & 76.2175 & 22.1761 & 7.63 & 3.00 & -7.58 & 
$          12\substack{+           6\\-           4}$ & ... &          410 & 
2.07 & C & 5 & Sgr \\
HG29 & 82.1406 & -16.8624 & 7.79 & 1.47 & -7.60 & 
$          12\substack{+           3\\-           2}$ & ... &          236 & 
0.36 & C & 3 & GAS: \\
HG30 & 82.7216 & -18.4236 & 7.62 & 2.41 & -8.01 & 
$          13\substack{+           4\\-           3}$ & ... &          393 & 
1.81 & C & 20 & Unc \\
HG31 & 97.0253 & -53.1848 & 11.24 & 1.44 & -7.49 & 
$          56\substack{+           9\\-           7}$ & ... &          222 & 
0.60 & C & 20 & LMC \\
HG32 & 103.6610 & -71.5165 & 10.95 & 2.06 & -7.55 & 
$          51\substack{+          17\\-          13}$ & ... &          282 & 
0.89 & C & 6 & LMC \\
HG33 & 105.4540 & 39.9977 & 7.58 & 2.75 & -7.55 & 
$          11\substack{+           5\\-           3}$ & ... &          354 & 
1.22 & C & 14 & Unc \\
HG34 & 107.7400 & 47.9719 & 7.33 & 1.55 & -7.51 & 
$           9\substack{+           2\\-           1}$ & ... &          233 & 
0.47 & C & 18 & Sgr \\
HG35 & 108.0770 & -63.6360 & 11.04 & 2.48 & -7.29 & 
$          47\substack{+          15\\-          11}$ & ... &          285 & 
1.64 & C & 20 & LMC \\
HG36 & 109.4010 & 50.1782 & 7.71 & 1.65 & -7.73 & 
$          12\substack{+           3\\-           3}$ &           38 & 
         267 & 1.29 & C & 1 & GAS: \\
HG37 & 113.1360 & 26.7875 & 8.14 & 1.71 & -6.50 & 
$           9\substack{+           2\\-           2}$ & ... &          150 & 
1.26 & C & 14 & Disk \\
HG38 & 113.6000 & 27.3198 & 7.53 & 1.72 & -7.97 & 
$          13\substack{+           4\\-           3}$ & ... &          306 & 
0.79 & C & 3 & Sgr: \\
HG39 & 119.7330 & 33.9772 & 13.19 & 1.37 & -7.43 & 
$         134\substack{+          30\\-          24}$ & ... &          210 & 
1.20 & C & 2 & Gemini: \\
HG40 & 127.3130 & 18.3853 & 7.06 & 1.67 & -8.12 & 
$          11\substack{+           2\\-           2}$ &           18 & 
         323 & 1.35 & C & 17 & GAS: \\
HG41 & 127.3710 & 10.7734 & 8.13 & 2.11 & -7.62 & 
$          14\substack{+           4\\-           3}$ &          -30 & 
         294 & 1.00 & C & 17 & GAS: \\
HG42 & 130.9810 & -12.4019 & 8.02 & 1.48 & -7.30 & 
$          12\substack{+           3\\-           2}$ & ... &          204 & 
0.40 & C & 6 & Disk \\
\hline
\end{tabular}
\end{minipage}
\end{table*}
\addtocounter{table}{-1}
\begin{table*}
 \centering
\begin{minipage}{160mm}
 \caption{Continued}.
 \vspace{2pt}
\renewcommand{\arraystretch}{1.3}
\begin{tabular}{@{}rrrrrrrrrrlll@{}}
\hline
 ID &  RA  & Dec  & K$_{S0}$ & (J-K$_{S}$)$_{0}$ & MK$_{S0}$ &  R$_{\odot}$ & V$_{GSR}$ & Period & Amp & Seq & Ref & Feature\\
   &  degs & degs &  &  &  & kpc & km/s & days & mags & & & \\
\hline
HG43 & 133.5780 & -12.0151 & 8.01 & 2.71 & -7.79 & 
$          14\substack{+           5\\-           4}$ &          -58 & 
         389 & 1.13 & C & 17 & Unc \\
HG44 & 134.3580 & 17.3477 & 10.70 & 4.40 & -6.66 & 
$          30\substack{+          24\\-          13}$ &         -105 & 
         394 & 1.20 & C & 4 & Sgr \\
HG45 & 136.4430 & 20.4105 & 12.28 & 2.81 & -6.86 & 
$          68\substack{+          32\\-          22}$ &          114 & 
         256 & 1.58 & C & 2 & Gemini: \\
HG46 & 138.3830 & 19.5729 & 7.35 & 1.64 & -7.68 & 
$          10\substack{+           2\\-           2}$ &         -135 & 
         260 & 1.25 & C & 16 & Sgr \\
HG47 & 138.7720 & 19.2938 & 8.49 & 1.68 & -7.65 & 
$          17\substack{+           3\\-           3}$ &         -104 & 
         258 & 1.63 & C & 16 & Sgr \\
HG48 & 152.6540 & -6.8538 & 8.52 & 1.41 & -8.37 & 
$          24\substack{+           4\\-           3}$ &          139 & 
         335 & 1.35 & C & 17 & Unc \\
HG49 & 160.0140 & 26.0206 & 9.81 & 1.32 & -8.11 & 
$          38\substack{+           8\\-           7}$ &         -148 & 
         137 & 0.22 & C' & 1 & Sgr \\
HG50 & 164.8490 & 39.7349 & 9.45 & 1.60 & -7.16 & 
$          21\substack{+           4\\-           3}$ &         -155 & 
         198 & 0.84 & C & 16 & Sgr \\
HG51 & 167.4990 & -21.3671 & 6.68 & 1.57 & -8.16 & 
$           9\substack{+           2\\-           1}$ &          -60 & 
         319 & 1.54 & C & 16 & Unc \\
HG52 & 169.3290 & -17.4876 & 11.21 & 1.42 & -6.91 & 
$          42\substack{+           6\\-           5}$ &          169 & 
         166 & 0.86 & C & 16 & Unc \\
HG53 & 171.7730 & 10.2612 & 11.66 & 1.46 & -6.30 & 
$          39\substack{+           6\\-           5}$ & ... &          125 & 
0.64 & C & 19 & Sgr \\
HG54 & 182.3540 & 15.2717 & 9.82 & 1.33 & -7.87 & 
$          35\substack{+           5\\-           4}$ &          -81 & 
         256 & 0.73 & C & 16 & Sgr \\
HG55 & 183.5710 & -9.0139 & 11.86 & 1.33 & -6.86 & 
$          56\substack{+           8\\-           7}$ &          -21 & 
         157 & 1.04 & C & 17 & Sgr: \\
HG56 & 186.9170 & -0.4642 & 10.54 & 2.20 & -7.47 & 
$          40\substack{+          15\\-          11}$ &         -130 & 
         281 & 1.76 & C & 1 & Sgr \\
HG57 & 189.6220 & -40.9361 & 12.69 & 2.59 & -7.04 & 
$          90\substack{+          39\\-          27}$ & ... &          264 & 
1.83 & C & 15 & Unc \\
HG58 & 192.2700 & 13.3432 & 11.16 & 1.42 & -7.71 & 
$          60\substack{+           9\\-           8}$ &          -62 & 
         243 & 0.89 & C & 16 & Sgr \\
HG59 & 192.9580 & 1.5005 & 10.19 & 1.86 & -7.67 & 
$          38\substack{+          11\\-           9}$ &          -25 & 
         277 & 1.39 & C & 1 & Sgr \\
HG60 & 194.6400 & 16.6701 & 7.81 & 3.47 & -7.30 & 
$          11\substack{+           6\\-           4}$ &           58 & 
         395 & 2.00 & C & 1 & Sgr: \\
HG61 & 195.4830 & 8.6087 & 12.06 & 2.63 & -7.04 & 
$          66\substack{+          29\\-          20}$ & ... &          263 & 
1.21 & C & 2 & Sgr \\
HG62 & 197.4730 & 23.4279 & 9.48 & 2.26 & -8.65 & 
$          42\substack{+          12\\-           9}$ &           62 & 
         505 & 1.69 & C & 17 & Sgr \\
HG63 & 199.7550 & 4.3392 & 11.84 & 1.21 & -6.71 & 
$          51\substack{+           6\\-           5}$ &           -2 & 
         141 & 0.20 & C & 17 & Sgr \\
HG64 & 203.3310 & 41.9142 & 8.17 & 1.21 & -7.65 & 
$          15\substack{+           3\\-           2}$ & ... &          221 & 
0.74 & C & 14 & Unc \\
HG65 & 203.9880 & 6.3986 & 10.57 & 1.68 & -7.52 & 
$          42\substack{+           8\\-           7}$ & ... &          242 & 
1.46 & C & 19 & Sgr \\
HG66 & 204.6730 & 15.8689 & 13.45 & 0.93 & -7.03 & 
$         125\substack{+          19\\-          17}$ & ... &          149 & 
0.61 & C & 2 & Sgr: \\
HG67 & 206.8460 & -34.7898 & 10.08 & 1.95 & -7.82 & 
$          38\substack{+           9\\-           7}$ &            3 & 
         308 & 0.85 & C & 17 & Sgr: \\
HG68 & 208.2560 & 0.7874 & 11.07 & 1.55 & -7.54 & 
$          53\substack{+          13\\-          11}$ &           -1 & 
         236 & 0.64 & C & 1 & Sgr \\
HG69 & 209.0100 & -1.6073 & 11.29 & 1.57 & -7.34 & 
$          54\substack{+           9\\-           8}$ &           -2 & 
         215 & 0.68 & C & 16 & Sgr \\
HG70 & 209.8360 & -30.3940 & 11.81 & 2.77 & -6.21 & 
$          40\substack{+          15\\-          11}$ &           25 & 
         186 & 0.92 & C & 16 & Sgr: \\
HG71 & 210.6260 & -45.9354 & 7.27 & 1.99 & -8.00 & 
$          11\substack{+           4\\-           3}$ & ... &          342 & 
0.52 & C & 3 & Disk \\
HG72 & 214.5320 & 63.8186 & 9.55 & 1.13 & -6.69 & 
$          18\substack{+           3\\-           3}$ & ... &          135 & 
0.64 & C & 14 & Unc \\
HG73 & 218.1200 & -5.5217 & 11.22 & 2.73 & -7.14 & 
$          47\substack{+          22\\-          15}$ &           40 & 
         287 & 1.38 & C & 1 & Sgr \\
HG74 & 219.4930 & -4.2267 & 10.87 & 1.32 & -7.91 & 
$          58\substack{+           8\\-           7}$ &           77 & 
         262 & 0.56 & C & 17 & Sgr \\
HG75 & 221.6300 & -0.9168 & 11.04 & 1.34 & -7.61 & 
$          54\substack{+           7\\-           7}$ &           45 & 
         227 & 1.43 & C & 17 & Sgr \\
HG76 & 221.6840 & 5.2108 & 11.34 & 2.27 & -7.02 & 
$          47\substack{+          13\\-          10}$ & ... &          232 & 
1.06 & C & 19 & Sgr \\
HG77 & 222.4390 & 1.4490 & 12.60 & 0.90 & -5.95 & 
$          52\substack{+           8\\-           7}$ & ... &           88 & 
0.16 & C & 2 & Sgr \\
HG78 & 224.3620 & 5.2677 & 11.43 & 2.79 & -6.53 & 
$          39\substack{+          15\\-          11}$ &            4 & 
         218 & 1.34 & C & 17 & Sgr \\
HG79 & 225.2790 & -5.5283 & 11.47 & 2.06 & -7.36 & 
$          59\substack{+          15\\-          12}$ &           72 & 
         257 & 0.86 & C & 16 & Sgr \\
HG80 & 226.2300 & 35.7993 & 9.68 & 2.31 & -7.59 & 
$          29\substack{+          11\\-           8}$ & ... &          309 & 
1.70 & C & 12 & HAC: \\
HG81 & 228.4230 & -3.8967 & 11.27 & 1.38 & -7.52 & 
$          58\substack{+          13\\-          11}$ &           58 & 
         222 & 0.74 & C & 1 & Sgr \\
HG82 & 228.7960 & -13.5412 & 10.76 & 1.75 & -7.43 & 
$          44\substack{+           9\\-           7}$ &           81 & 
         242 & 1.72 & C & 16 & Sgr \\
HG83 & 230.6520 & -6.4262 & 11.39 & 1.33 & -7.30 & 
$          55\substack{+          12\\-          10}$ &          104 & 
         196 & 0.85 & C & 1 & Sgr \\
\hline
\end{tabular}
\end{minipage}
\end{table*}
\addtocounter{table}{-1}
\begin{table*}
 \centering
\begin{minipage}{160mm}
 \caption{Continued}. 
 \vspace{2pt}
\renewcommand{\arraystretch}{1.3}
\begin{tabular}{@{}rrrrrrrrrrlll@{}}
\hline
 ID &  RA  & Dec  & K$_{S0}$ & (J-K$_{S}$)$_{0}$ & MK$_{S0}$ &  R$_{\odot}$ & V$_{GSR}$ & Period & Amp & Seq & Ref & Feature\\
   &  degs & degs &  &  &  & kpc & km/s & days & mags & & & \\
\hline
HG84 & 230.6850 & -12.6305 & 11.38 & 1.58 & -7.47 & 
$          60\substack{+          10\\-           9}$ &          106 & 
         234 & 0.37 & C & 17 & Sgr \\
HG85 & 231.4140 & 42.4141 & 8.17 & 1.95 & -7.92 & 
$          17\substack{+           5\\-           4}$ &           64 & 
         322 & 2.40 & C & 1 & HAC: \\
HG86 & 231.8480 & 4.4744 & 8.11 & 2.01 & -7.75 & 
$          15\substack{+           4\\-           3}$ &          169 & 
         303 & 0.90 & C & 17 & Unc \\
HG87 & 239.6760 & 18.8796 & 10.93 & 1.31 & -7.22 & 
$          43\substack{+           6\\-           5}$ &          164 & 
         186 & 0.35 & C & 16 & Sgr: \\
HG88 & 240.4270 & -12.8290 & 12.32 & 1.95 & -6.80 & 
$          69\substack{+          22\\-          17}$ & ... &          193 & 
0.91 & C & 10 & Sgr: \\
HG89 & 249.1320 & -3.3937 & 5.98 & 3.75 & -7.35 & 
$           5\substack{+           3\\-           2}$ &          -40 & 
         467 & 1.89 & C & 19 & Disk \\
HG90 & 258.7400 & 42.1733 & 11.16 & 1.35 & -7.67 & 
$          59\substack{+          13\\-          11}$ & ... &          233 & 
0.26 & C & 13 & Unc \\
HG91 & 268.5050 & 26.4534 & 9.51 & 2.34 & -6.64 & 
$          17\substack{+           7\\-           5}$ & ... &          199 & 
1.45 & C & 3 & HAC: \\
HG92 & 273.3720 & 45.5220 & 6.70 & 3.80 & -6.94 & 
$           5\substack{+           3\\-           2}$ &          -78 & 
         370 & 2.22 & C & 19 & Disk \\
HG93 & 294.2910 & -35.5042 & 10.14 & 1.61 & -7.41 & 
$          34\substack{+           6\\-           5}$ &          157 & 
         112 & 0.16 & C' & 17 & Sgr \\
HG94 & 294.3920 & -35.5440 & 9.04 & 2.02 & -8.07 & 
$          27\substack{+           7\\-           5}$ &          173 & 
         367 & 0.61 & C & 17 & Sgr \\
HG95 & 295.3690 & -32.5606 & 9.39 & 1.37 & -7.84 & 
$          28\substack{+           6\\-           5}$ & ... &          260 & 
0.32 & C & 12 & Sgr \\
HG96 & 295.5790 & -35.3270 & 9.96 & 2.51 & -6.78 & 
$          23\substack{+           7\\-           6}$ &          156 & 
         231 & 1.58 & C & 16 & Sgr \\
HG97 & 295.5890 & -32.1840 & 9.94 & 1.92 & -7.24 & 
$          28\substack{+           6\\-           5}$ &          163 & 
         234 & 1.12 & C & 16 & Sgr \\
HG98 & 297.2110 & -30.9750 & 10.16 & 3.32 & -7.01 & 
$          28\substack{+          13\\-           9}$ &          171 & 
         333 & 1.62 & C & 16 & Sgr \\
HG99 & 298.3760 & -38.6000 & 9.24 & 2.00 & -7.39 & 
$          21\substack{+           5\\-           4}$ &          175 & 
         256 & 1.53 & C & 16 & Sgr \\
HG100 & 300.2200 & -34.8657 & 9.77 & 1.32 & -7.52 & 
$          29\substack{+           6\\-           5}$ & ... &          217 & 
1.31 & C & 3 & Sgr \\
HG101 & 300.4330 & -30.4129 & 10.32 & 1.25 & -7.00 & 
$          30\substack{+           4\\-           3}$ & ... &          166 & 
0.88 & C & 20 & Sgr \\
HG102 & 300.7660 & -19.8179 & 9.06 & 1.35 & -7.15 & 
$          18\substack{+           2\\-           2}$ & ... &          186 & 
1.48 & C & 19 & HAC \\
HG103 & 305.0020 & -5.5973 & 8.59 & 3.40 & -7.51 & 
$          17\substack{+           8\\-           5}$ & ... &          431 & 
1.73 & C & 19 & Disk \\
HG104 & 305.1150 & -14.8243 & 8.68 & 3.11 & -7.32 & 
$          16\substack{+           7\\-           5}$ &          160 & 
         356 & 1.78 & C & 16 & HAC \\
HG105 & 308.4490 & -46.6057 & 8.91 & 1.57 & -7.58 & 
$          20\substack{+           3\\-           3}$ & ... &          241 & 
0.93 & C & 20 & Sgr \\
HG106 & 315.0780 & -6.1150 & 12.06 & 3.21 & -5.39 & 
$          31\substack{+          17\\-          11}$ & ... &          145 & 
1.45 & C & 10 & Sgr: \\
HG107 & 321.5260 & -10.2295 & 10.07 & 0.79 & -8.41 & 
$          50\substack{+           7\\-           6}$ & ... &          280 & 
0.13 & C & 14 & Unc \\
HG108 & 321.8190 & -30.8660 & 9.48 & 1.58 & -7.50 & 
$          25\substack{+           6\\-           5}$ & ... &          233 & 
1.62 & C & 3 & Sgr: \\
HG109 & 328.8620 & 23.7038 & 5.48 & 3.37 & -7.22 & 
$           4\substack{+           2\\-           1}$ & ... &          370 & 
1.16 & C & 18 & Disk \\
HG110 & 331.3110 & 0.1457 & 9.28 & 2.22 & -6.94 & 
$          18\substack{+           5\\-           4}$ &          115 & 
         221 & 1.32 & C & 16 & HAC \\
HG111 & 331.7240 & -25.1080 & 8.94 & 1.97 & -7.95 & 
$          24\substack{+           6\\-           5}$ &           70 & 
         330 & 1.60 & C & 16 & Sgr \\
HG112 & 333.9050 & -0.0498 & 9.33 & 1.48 & -7.42 & 
$          23\substack{+           5\\-           4}$ &          102 & 
         217 & 0.19 & C & 1 & HAC: \\
HG113 & 334.2910 & -26.1176 & 8.87 & 2.15 & -7.68 & 
$          21\substack{+           5\\-           4}$ &           64 & 
         307 & 1.50 & C & 16 & Sgr \\
HG114 & 335.7550 & 22.2824 & 9.58 & 2.07 & -7.83 & 
$          31\substack{+           8\\-           6}$ & ... &          322 & 
1.39 & C & 18 & Unc \\
HG115 & 336.5810 & 26.0606 & 4.51 & 2.71 & -7.51 & 
$           3\substack{+           1\\-           1}$ &          226 & 
         341 & 1.38 & C & 1 & Disk \\
HG116 & 337.0440 & -13.7730 & 9.88 & 1.96 & -7.68 & 
$          33\substack{+          10\\-           8}$ &          -15 & 
         288 & 1.32 & C & 1 & Unc \\
HG117 & 340.9600 & -57.0232 & 7.99 & 1.54 & -7.88 & 
$          15\substack{+           2\\-           2}$ &           62 & 
         274 & 0.60 & C & 17 & Sgr \\
HG118 & 341.6190 & -27.4490 & 13.27 & 1.58 & -8.12 & 
$         190\substack{+          33\\-          28}$ &           45 & 
         314 & 0.35 & C & 17 & Unc \\
HG119 & 347.1460 & 40.5928 & 7.76 & 2.60 & -7.69 & 
$          13\substack{+           5\\-           4}$ & ... &          361 & 
1.59 & C & 3 & Disk \\
HG120 & 349.8980 & -18.9401 & 9.96 & 1.52 & -6.60 & 
$          21\substack{+           3\\-           3}$ &           37 & 
         148 & 0.89 & C & 16 & Sgr \\
HG121 & 350.5460 & -15.3045 & 9.36 & 1.49 & -6.74 & 
$          17\substack{+           4\\-           3}$ & ... &          157 & 
0.88 & C & 11 & Sgr \\
\hline
\end{tabular}
\bigskip
 
The references are numbered as follows: 1. \citet{TottenIrwin98}, 2. \citet{Green13}, 3. \citet{Reidetal08}, 4. \citet{Liebertetal00}, 5. \citet{LeeChen09}, 6. \citet{Cruzetal03}, 7. \citet{Moodyetal97}, 8. \citet{Groenewegenetal97}, 9. \citet{Totten98}, 10. \citet{Kirkpatricketal08}, 11. \citet{Christliebetal01}, 12. \citet{Cruzetal07}, 13. \citet{MeusingerBrunzendorf01}, 14. \citet{GigoyanMickaelian12}, 15. \citet{Gizis02}, 16. \citet{Mauronetal04}, 17. \citet{Mauronetal05}, 18. \citet{Mauronetal07b}, 19. \citet{Mauron08}, 20. \citet{Mauronetal14}.
\end{minipage}
\end{table*}

%% file: second_sample_latex_table_S2_V2.txt
\begin{table*}
 \centering
\begin{minipage}{160mm}
 \caption{Catalogue of second sample carbon stars. As for Table 1, the coordinates are given in decimal degrees (J2000), the (reddening corrected) NIR magnitudes and colour are from 2MASS data. Errors on distances are $\sim$30$\%$}\label{tab:second_sample_catalog}
 \vspace{2pt}
\renewcommand{\arraystretch}{1.3}
\begin{tabular}{@{}lrrrrrrrl@{}}
\hline
 ID &  RA  & Dec & K$_{S0}$ & (J-K$_{S}$)$_{0}$ & MK$_{S0}$ & Distance & V$_{GSR}$ & Reference\\
   &  degs & degs &  &  &  & kpc & km/s  &  \\
\hline
HG2-1 & 2.0708 & -75.6242 & 11.60 & 2.18 & -7.53 &       67 & ... & 12 \\
HG2-2 & 6.9679 & -70.8753 & 11.04 & 2.27 & -7.49 &       51 & ... & 12 \\
HG2-3 & 9.1348 & -22.9142 & 14.14 & 1.32 & -6.98 &      167 & ... & 19 \\
HG2-4 & 11.0646 & -71.2611 & 11.36 & 2.02 & -7.59 &       62 & ... & 12 \\
HG2-5 & 11.5033 & -75.3533 & 11.29 & 1.83 & -7.67 &       62 & ... & 12 \\
HG2-6 & 16.9675 & -69.3600 & 10.97 & 2.18 & -7.53 &       50 & ... & 12 \\
HG2-7 & 20.4304 & -74.0628 & 11.08 & 2.60 & -7.35 &       48 & ... & 12 \\
HG2-8 & 21.6450 & -70.6631 & 10.34 & 2.43 & -7.42 &       36 & ... & 12 \\
HG2-9 & 25.2263 & -72.1475 & 10.30 & 1.87 & -7.66 &       39 & ... & 3 \\
HG2-10 & 27.0325 & -71.9225 & 10.31 & 2.43 & -7.42 &       35 & ... & 12 \\
HG2-11 & 29.1083 & 51.4225 & 7.60 & 1.60 & -7.40 &       10 & ... & 12 \\
HG2-12 & 33.4667 & 46.6868 & 8.02 & 1.28 & -6.91 &       10 & ... & 14 \\
HG2-13 & 37.1960 & 26.7951 & 8.59 & 1.47 & -7.21 &       14 &           50 & 1
 \\
HG2-14 & 37.8151 & -2.7185 & 11.52 & 0.96 & -6.42 &       39 &          -64 & 1
 \\
HG2-15 & 51.8808 & 39.0806 & 7.00 & 1.53 & -7.30 &        7 & ... & 14 \\
HG2-16 & 56.6607 & 75.7879 & 8.78 & 1.46 & -7.19 &       16 & ... & 3 \\
HG2-17 & 57.5445 & 26.0841 & 8.75 & 1.37 & -7.05 &       14 & ... & 18 \\
HG2-18 & 60.4306 & 8.7029 & 8.11 & 1.45 & -7.18 &       11 & ... & 19 \\
HG2-19 & 60.6383 & -68.2731 & 11.00 & 2.43 & -7.42 &       48 & ... & 6 \\
HG2-20 & 67.8925 & -72.5356 & 11.16 & 1.92 & -7.63 &       57 & ... & 6 \\
HG2-21 & 74.0112 & 9.3719 & 8.55 & 1.52 & -7.29 &       15 & ... & 18 \\
HG2-22 & 74.9562 & -75.1550 & 10.25 & 2.30 & -7.48 &       35 & ... & 6 \\
HG2-23 & 91.6438 & 73.1742 & 7.84 & 1.44 & -7.16 &       10 & ... & 18 \\
HG2-24 & 101.5750 & 54.5260 & 8.05 & 1.37 & -7.06 &       10 & ... & 18 \\
HG2-25 & 105.2837 & -68.3150 & 10.51 & 2.22 & -7.51 &       40 & ... & 6 \\
HG2-26 & 105.9281 & 23.9374 & 11.67 & 2.47 & -7.41 &       65 & ... & 4 \\
HG2-27 & 106.5324 & 40.2004 & 7.05 & 1.57 & -7.36 &        8 & ... & 14 \\
HG2-28 & 118.1038 & 4.5663 & 7.71 & 1.51 & -7.26 &       10 & ... & 3 \\
HG2-29 & 120.9702 & 36.7454 & 8.80 & 1.20 & -6.78 &       13 & ... & 14 \\
HG2-30 & 129.3208 & -13.3539 & 7.69 & 1.51 & -7.26 &       10 & ... & 6 \\
HG2-31 & 134.9821 & -77.8848 & 10.50 & 2.23 & -7.51 &       40 &           69 & 
17 \\
HG2-32 & 144.7517 & -10.3629 & 10.71 & 1.75 & -7.63 &       47 &           78 & 
1 \\
HG2-33 & 155.5614 & -11.8608 & 8.47 & 1.49 & -7.24 &       14 &          -51 & 1
 \\
HG2-34 & 175.4265 & -33.6926 & 12.16 & 1.64 & -7.47 &       84 &          -49 & 
17 \\
HG2-35 & 185.7478 & 21.0906 & 9.97 & 1.56 & -7.35 &       29 &           19 & 7
 \\
HG2-36 & 190.9055 & 2.3584 & 11.44 & 1.53 & -7.29 &       56 &          -29 & 1
 \\
HG2-37 & 194.2376 & -11.7719 & 9.39 & 1.31 & -6.95 &       19 &          -44 & 1
 \\
HG2-38 & 195.3270 & 0.4974 & 13.51 & 1.62 & -7.44 &      154 &          -12 & 19
 \\
HG2-39 & 205.6118 & -7.2565 & 9.23 & 1.27 & -6.89 &       17 &         -220 & 9
 \\
HG2-40 & 212.1464 & 5.1075 & 7.25 & 1.60 & -7.41 &        9 &          -37 & 1
 \\
HG2-41 & 221.2036 & -1.1824 & 10.74 & 1.52 & -7.28 &       40 &           26 & 1
 \\
HG2-42 & 223.4306 & -13.2182 & 11.19 & 2.12 & -7.55 &       56 &           72 & 
1 \\
HG2-43 & 225.2789 & -5.5274 & 11.47 & 2.06 & -7.58 &       65 & ... & 10 \\
\hline
\end{tabular}
\end{minipage}
\end{table*}
\addtocounter{table}{-1}
\begin{table*}
 \centering
\begin{minipage}{160mm}
 \caption{continued.}
 \vspace{2pt}
\renewcommand{\arraystretch}{1.3}
\begin{tabular}{@{}lrrrrrrrl@{}}
\hline
 ID &  RA  & Dec & K$_{S0}$ & (J-K$_{S}$)$_{0}$ & MK$_{S0}$ & Distance & V$_{GSR}$ & Reference \\
   &  degs & degs &  &  &  & kpc & km/s  &  \\
\hline
HG2-44 & 229.6677 & 14.9842 & 7.33 & 1.44 & -7.15 &        8 & ... & 20 \\
HG2-45 & 237.9550 & -7.8469 & 13.49 & 3.27 & -7.07 &      130 & ... & 10 \\
HG2-46 & 245.4011 & -8.8886 & 7.74 & 1.39 & -7.08 &        9 & ... & 3 \\
HG2-47 & 253.1154 & -16.1274 & 11.30 & 1.31 & -6.96 &       45 &          142 & 
17 \\
HG2-48 & 256.7116 & 40.2096 & 9.67 & 1.25 & -6.86 &       20 & ... & 14 \\
HG2-49 & 259.4814 & 4.6020 & 10.17 & 1.31 & -6.96 &       27 & ... & 18 \\
HG2-50 & 261.4760 & 3.0070 & 13.37 & 1.33 & -7.00 &      118 &           32 & 19
 \\
HG2-51 & 262.1073 & 70.1418 & 9.01 & 2.50 & -7.39 &       19 &           37 & 16
 \\
HG2-52 & 269.5650 & 22.5980 & 8.05 & 1.41 & -7.11 &       11 & ... & 18 \\
HG2-53 & 281.7095 & -56.2341 & 11.01 & 1.64 & -7.47 &       50 &            5 & 
17 \\
HG2-54 & 282.4620 & 62.2910 & 7.48 & 1.29 & -6.92 &        8 & ... & 19 \\
HG2-55 & 285.2664 & 87.0623 & 7.26 & 1.12 & -6.67 &        6 & ... & 14 \\
HG2-56 & 288.6007 & -78.3782 & 7.37 & 2.14 & -7.54 &       10 &           37 & 
17 \\
HG2-57 & 292.9105 & -30.0417 & 10.25 & 1.63 & -7.46 &       35 &          168 & 
17 \\
HG2-58 & 293.3575 & 52.3578 & 9.30 & 2.42 & -7.43 &       22 & ... & 3 \\
HG2-59 & 293.8285 & 54.6649 & 8.22 & 1.53 & -7.29 &       13 & ... & 3 \\
HG2-60 & 294.8768 & 75.6946 & 7.76 & 1.61 & -7.42 &       11 & ... & 18 \\
HG2-61 & 299.6685 & 77.7570 & 8.65 & 3.30 & -7.07 &       14 & ... & 18 \\
HG2-62 & 301.1035 & -30.1142 & 10.78 & 1.54 & -7.31 &       41 & ... & 20 \\
HG2-63 & 303.3270 & -23.6980 & 8.08 & 1.38 & -7.06 &       11 &         -106 & 
16 \\
HG2-64 & 303.9993 & 76.5857 & 7.91 & 1.49 & -7.23 &       11 & ... & 18 \\
HG2-65 & 304.5785 & -66.8494 & 8.98 & 1.80 & -7.68 &       22 & ... & 20 \\
HG2-66 & 312.0750 & 10.4440 & 7.81 & 2.20 & -7.52 &       12 & ... & 20 \\
HG2-67 & 313.7274 & -28.4824 & 10.81 & 1.51 & -7.27 &       41 &          111 & 
16 \\
HG2-68 & 329.9533 & 44.3110 & 7.68 & 1.90 & -7.64 &       12 & ... & 14 \\
HG2-69 & 333.9954 & 42.3795 & 7.74 & 5.35 & -6.21 &        6 & ... & 14 \\
HG2-70 & 335.4107 & 33.5997 & 7.22 & 2.72 & -7.30 &        8 & ... & 14 \\
HG2-71 & 337.9708 & 19.2946 & 9.20 & 1.38 & -7.06 &       18 &         -157 & 1
 \\
HG2-72 & 341.9084 & -78.4548 & 8.17 & 1.75 & -7.63 &       14 &          237 & 
17 \\
HG2-73 & 343.1504 & 47.6903 & 7.92 & 1.56 & -7.34 &       11 & ... & 12 \\
HG2-74 & 349.3378 & -24.1952 & 12.26 & 1.47 & -7.20 &       78 &           41 & 
16 \\
HG2-75 & 351.3807 & -30.1822 & 10.99 & 2.44 & -7.42 &       48 &          114 & 
16 \\
\hline
\end{tabular}
\end{minipage}
\end{table*}

%% file: satellite_latex_table_S2_V2.txt
\begin{table*}
 \centering
\begin{minipage}{160mm}
  \caption{Variable carbon stars in Galactic satellites. Distance to the hosts are Sculptor: 86 kpc,  Fornax: 147 kpc, Carina: 105 kpc, Leo II: 233 kpc, Sextans: 86 and Sgr: 26 kpc \citep{McConnachie12}. As for Table 1, the coordinates are given in decimal degrees (J2000), the (reddening corrected) NIR magnitudes and colour are from 2MASS data. The periods and amplitudes (in Catalina V band) of variability are derived from fits using Period04.}\label{tab:satellite_catalog}
 \vspace{2pt}
\renewcommand{\arraystretch}{1.3}
\begin{tabular}{@{}lrrrrrrrrll@{}}
\hline
 ID &  RA  & Dec  & K$_{S0}$ & (J-K)$_{S0}$ & MK$_{S0}$ & R$_{\odot}$  & Period & Amp & Seq. & Host \\
   &  degs & degs &  &    &   & kpc & days & & & \\
\hline
S1 & 14.9736 & -33.6418 & 11.60 & 3.24 & -8.06 &       85 &      524 & 1.90 & C
 & Scu \\
S2 & 14.9956 & -33.4764 & 12.27 & 1.13 & -7.46 &       89 &      195 & 0.96 & C
 & Scu \\
S3 & 39.7211 & -34.8224 & 13.45 & 2.02 & -7.30 &      141 &      242 & 1.09 & C
 & For \\
S4 & 39.7375 & -34.7760 & 13.08 & 1.71 & -7.78 &      149 &      277 & 0.46 & C
 & For \\
S5 & 39.7377 & -34.7969 & 13.56 & 1.26 & -7.31 &      150 &      189 & 0.30 & C
 & For \\
S6 & 39.8138 & -34.2523 & 13.62 & 1.18 & -7.44 &      163 &      196 & 0.95 & C
 & For \\
S7 & 40.0039 & -34.3787 & 13.00 & 1.52 & -7.84 &      148 &      128 & 0.18 & C'
 & For \\
S8 & 40.0056 & -34.3386 & 13.34 & 1.35 & -7.66 &      159 &      231 & 0.47 & C
 & For \\
S9 & 40.0114 & -34.5303 & 13.18 & 1.61 & -8.00 &      172 &      296 & 0.27 & C
 & For \\
S10 & 40.0278 & -34.3895 & 12.62 & 1.87 & -8.10 &      139 &      340 & 0.73 & C
 & For \\
S11 & 40.0426 & -34.5561 & 12.55 & 1.52 & -8.31 &      148 &      334 & 1.37 & C
 & For \\
S12 & 40.0651 & -34.5675 & 13.23 & 1.70 & -7.81 &      162 &      280 & 0.46 & C
 & For \\
S13 & 40.0741 & -34.4599 & 13.18 & 2.24 & -7.39 &      130 &      273 & 0.96 & C
 & For \\
S14 & 40.0831 & -34.5527 & 13.91 & 1.36 & -7.24 &      170 &      190 & 0.83 & C
 & For \\
S15 & 40.1043 & -34.4829 & 13.51 & 1.32 & -7.64 &      169 &      226 & 0.45 & C
 & For \\
S16 & 40.1301 & -34.4789 & 13.07 & 1.67 & -7.79 &      149 &      275 & 0.76 & C
 & For \\
S17 & 40.2177 & -34.6231 & 13.66 & 1.73 & -7.29 &      155 &      219 & 0.36 & C
 & For \\
S18 & 40.2223 & -34.2037 & 13.26 & 1.77 & -7.44 &      138 &      239 & 0.85 & C
 & For \\
S19 & 40.2648 & -34.8015 & 12.69 & 1.75 & -8.03 &      139 &      315 & 1.19 & C
 & For \\
S20 & 40.2847 & -34.5978 & 13.43 & 1.59 & -7.41 &      147 &      222 & 0.31 & C
 & For \\
S21 & 100.3062 & -50.9068 & 12.66 & 1.27 & -7.21 &       94 &      181 & 0.17 & 
C & Car \\
S22 & 100.4230 & -50.9689 & 12.34 & 1.40 & -8.05 &      119 &      284 & 0.66 & 
C & Car \\
S23 & 100.4353 & -51.0058 & 12.69 & 1.55 & -7.60 &      115 &      241 & 0.41 & 
C & Car \\
S24 & 100.5431 & -50.9400 & 12.78 & 1.15 & -7.42 &      110 &      193 & 0.38 & 
C & Car \\
S25 & 153.8581 & -2.0757 & 11.95 & 2.10 & -7.99 &       97 &      348 & 1.41 & C
 & Sex \\
S26 & 168.3034 & 22.1872 & 14.79 & 1.44 & -6.89 &      217 &      164 & 0.42 & C
 & LeoII \\
S27 & 168.3359 & 22.1878 & 14.81 & 1.56 & -7.03 &      234 &      183 & 0.82 & C
 & LeoII \\
S28 & 288.9922 & -30.5467 & 9.73 & 1.94 & -7.66 &       30 &      281 & 0.47 & C
 & Sgr \\
S29 & 289.7724 & -30.9072 & 9.16 & 1.63 & -7.54 &       22 &      240 & 1.88 & C
 & Sgr \\
S30 & 290.9395 & -34.5209 & 9.74 & 1.28 & -7.89 &       34 &      252 & 1.26 & C
 & Sgr \\
S31 & 292.2203 & -29.5899 & 10.16 & 1.42 & -6.85 &       25 &      160 & 0.55 & 
C & Sgr \\
S32 & 295.7120 & -33.5759 & 10.14 & 1.33 & -7.76 &       38 &      241 & 0.92 & 
C & Sgr \\
\hline
\end{tabular}
\end{minipage}
\end{table*}